\documentclass[twoside,11pt]{article}

\usepackage{blindtext}
\usepackage{appendix}

\usepackage[abbrvbib, preprint]{jmlr2e}

\usepackage{algorithm}
\usepackage[noend]{algorithmic}
\usepackage{placeins}
\usepackage{booktabs}

\usepackage{natbib, graphicx, url}
\usepackage{mathrsfs}

\renewcommand{\P}[1]{\mathbb{P}\left[#1\right]}
\newcommand{\Pbig}[1]{\mathbb{P}\big[#1\big]}

\usepackage{xcolor}
\definecolor{darkgreen}{rgb}{0.0, 0.5, 0.0}

\newtheorem{prop}{Proposition}

\usepackage{bbm}
\newcommand{\I}[1]{\mathbbm{1}\left[#1\right]}

\usepackage{mathtools}

\newcommand{\defeq}{\vcentcolon=}

\renewcommand{\epsilon}{\varepsilon}

\usepackage{lastpage}
\firstpageno{1}

\begin{document}

\title{Conformal Frequency Estimation using Discrete Sketched Data with Coverage for Distinct Queries}

\author{%
 \name Matteo Sesia\\
 \addr Departments of Data Sciences and Operations, and of Computer Science\\
 University of Southern California,
 Los Angeles, California, USA \\
 \texttt{sesia@marshall.usc.edu}
 \AND
 \name Stefano Favaro \\
 \addr Department of Economics and Statistics \\
 University of Torino and Collegio Carlo Alberto,
 Torino, Italy \\
 \texttt{stefano.favaro@unito.it}
 \AND
 \name Edgar Dobriban \\
 \addr Departments of Statistics and Data Science, and of Computer and Information Science \\
 University of Pennsylvania,
 Philadelphia, Pennsylvania, USA \\
 \texttt{dobriban@wharton.upenn.edu}
}

\editor{My editor}

\maketitle

\begin{abstract}
  This paper develops conformal inference methods to construct a confidence interval for the frequency of a queried object in a very large discrete data set, based on a sketch with a lower memory footprint. This approach requires no knowledge of the data distribution and can be combined with any sketching algorithm, including but not limited to the renowned count-min sketch, the count-sketch, and variations thereof.
After explaining how to achieve marginal coverage for exchangeable random queries, we extend our solution to provide stronger inferences that can account for the discreteness of the data and for heterogeneous query frequencies, increasing also robustness to possible distribution shifts.
These results are facilitated by a novel conformal calibration technique that guarantees valid coverage for a large fraction of distinct random queries.
Finally, we show our methods have improved empirical performance compared to existing frequentist and Bayesian alternatives in simulations as well as in examples of text and SARS-CoV-2 DNA data.
\end{abstract}


\section{Introduction}

\subsection{Estimating frequencies from sketched data} \label{sec:intro}

Estimating the frequency of a queried object given a lossy reduced representation, or {\em sketch}, of a large discrete data set
is a classical problem ~\citep[e.g.,][etc]{misra1982finding,charikar2002finding}.
This task is relevant in diverse fields including machine learning~\citep{shi2009hash}, cybersecurity~\citep{schechter2010popularity}, natural language processing~\citep{goyal2012sketch}, privacy~\citep{cormode2018privacy}, and biology~\citep{zhang2014these}.
For example, in biology, researchers may want to efficiently count the occurrences of a contiguous sequence of nucleotides within a large DNA database, as that can help identify common motifs that are associated with evolutionary relatedness between different organisms or are involved in important regulatory processes \citep{saavedra2020mining}.

Sketching tends to be motivated either by memory limitations, as large numbers of distinct symbols may otherwise be computationally expensive to analyze~\citep{zhang2014these}, or by privacy constraints when dealing with sensitive data~\citep{kockan2020sketching}.
Several sketching algorithms can provide compressed data representations that enable accurate approximations of the frequency of any object \citep{cormode2020small}. Classical approaches are based on random hashing \citep{cormode2020small}, but some recent works have proposed more sophisticated machine learning-driven algorithms that can automatically adapt to the features of the data distribution in order to optimize the data compression \citep{hsu2019learning,jiang2019learning,aamand2019learned,bertsimas2021frequency}.

An important statistical problem in the context of sketching is to quantify the uncertainty of frequency queries, as exact recovery of the latter is typically unfeasible due to some loss of information during the data compression.
Prior works took a number of very different routes to address this topic, ranging from data-conditional and Bayesian methods to the bootstrap
 \citep{cormode2020small,ting2018count, cai2018bayesian, dolera2021bayesian}.
This paper presents a novel conformal inference method \citep{vovk2005algorithmic}.
As we will explain, our approach is principled and offers some notable advantages, starting from the ability to obtain informative inferences without any parametric assumptions about the distribution of the sketched data. Further, a key strength of our approach is that it can provide rigorous uncertainty estimates for any sketching algorithm, including the classical count-min sketch (CMS) \citep{cormode2005improved}, its non-linear variations \citep{estan2002new}, the count-sketch (CS) \citep{charikar2002finding}, and even more complex learning-based techniques \citep{bertsimas2021frequency}. 
As we shall see, different sketching algorithms can lead to more or less accurate frequency queries for different types of data, and therefore the flexibility of our methods will be practically useful.

After reformulating the problem so that standard split conformal inference can be applied,
developing our methodology requires overcoming several challenges.
%  Firstly, conformal inference was originally developed to estimate predictive uncertainty in a very different type of supervised learning tasks, and connecting it to our problem introduces some additional computational and memory costs compared to traditional sketching. 
% Fortunately, these costs tend to be negligible compared to the scale of the data sets involved.
First, standard conformal inference techniques provide relatively weak statistical guarantees, which are less satisfactory than usual in the context of answering frequency queries about discrete data. Indeed, if some objects in the data are much more frequent than others, standard statistical coverage guarantees can be satisfied even by meaningless inferences that are only valid for the most common queries.
We address this limitation by proposing two methodological improvements that provide conformal inferences whose validity holds separately for queries with different frequencies, and for all distinct objects in a possibly large set of queries.
Further, we prove that our methods are more robust to distribution shifts compared to standard conformal inferences, which rely on the relatively strong assumption of data exchangeability.

\subsection{Problem statement and preview of our contributions} \label{sec:problem-statement}

We now present a simplified version of our problem statement and data observation model; see Section \ref{sec:methods-marginal} for the complete version.
Consider $m$ data points $Z_1,\ldots,Z_m \in \mathscr{Z}$, taking values in a discrete and possibly infinite dictionary $\mathscr{Z}$. 
We consider the setting where $m$ is very large, and $\mathscr{Z}$ is possibly also large; thus exact computations with  $Z_1,\ldots,Z_m$ are infeasible.
Instead, the data are processed via an arbitrary {\it sketching} function $\phi : \mathscr{Z}^m \to \mathcal{C}$  that produces a reduced representation of these data with lower memory footprint, where $\mathcal{C}$  consists for instance of $L$ discrete counters, so that  $\mathcal{C} = \mathbb{N}^L$ with $L \ll m$.
A well-known example of $\phi$ is the CMS \citep{cormode2005improved}, reviewed in Appendix~\ref{app:cms}.
The methods developed in this paper can be applied in combination with the CMS or with any other sketching function. However, the choice of $\phi$ is important in practice because it affects the efficiency of the data compression and the informativeness of our inferences, as it will become clear in Sections~\ref{sec:experiments}--\ref{sec:app}.

\begin{figure}[!htb]
\begin{center}
\includegraphics[scale=1]{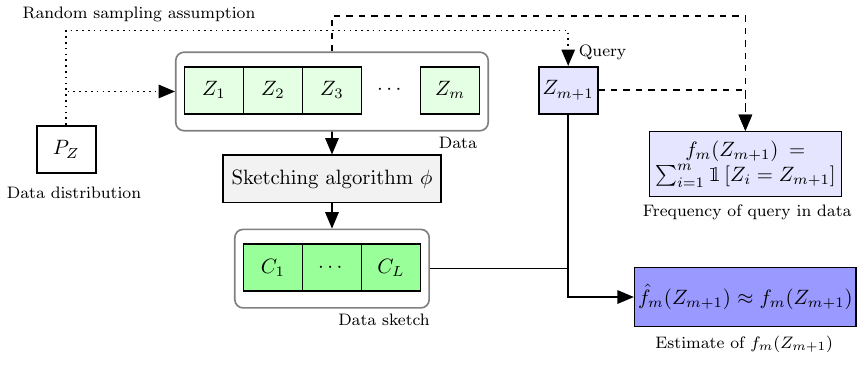}
\caption{Schematic visualization of the problem of estimating the empirical frequency of a queried object in a large data set, given a sketched representation of the latter.}
\label{fig:sketch-diagram}
\end{center}
\end{figure}

In general, our target of inference is the number of occurrences (or empirical frequency) of a given object (or query) $z \in \mathscr{Z}$ in the data set $Z_1,\ldots, Z_m$:
\begin{align} \label{eq:f-true}
  f_m(z) \defeq \sum_{i=1}^{m} \I{Z_i =z}.
\end{align}
Of course, since $Z_1,\ldots, Z_m$ are not available for direct computations, the exact value of $f_m(z)$ is not known.
Instead, we aim to approximate these values for an appropriate $z$ using the sketch.
Specifically, we seek an informative {\em confidence interval} for $f_m(z)$ that enjoys precise statistical guarantees in finite samples, as previewed next.
As a starting point, we assume that the query, $z=Z_{m+1}$, is a random draw from some distribution $P_Z$, sampled exchangeably with  $Z_1,\ldots,Z_m$.
See Figure~\ref{fig:sketch-diagram} for a schematic visualization of this problem.
%As mentioned before, exact recovery of $f_m(z)$ is typically unfeasible because the sketch may lose some information compared to the original data set.

The exchangeability of $(Z_1,\ldots$, $Z_{m+1})$, 
which will be relaxed later in the paper, 
imposes additional conditions compared to some classical analyses of sketching algorithms \citep{cormode2020small}. 
Such analyses typically treat the data as arbitrary---and thus can also handle non-stationary streams or adversarial cases. 
However, we believe our  exchangeability  condition is often realistic, for instance in applications where the data are processed in a random order; see Sections~\ref{sec:app} and~\ref{sec:discussion} for examples.
Treating the data as an approximately i.i.d.~sample from some distribution has been suggested before in the context of sketching~\citep{ting2018count, cai2018bayesian, aamand2019learned,dolera2021bayesian}, but our perspective involves key novelties. First, we assume only exchangeability, not independence. Second, we allow $P_Z$ to be arbitrary and unknown. Third, our results apply to any sketching algorithm.

Section~\ref{sec:conformal-review} reviews the relevant conformal inference background. Then, Section~\ref{sec:methods-marginal} connects conformal inference to our problem and explains how to construct a confidence interval\footnote{Since $f_m(Z_{m+1})$ is also random, it is technically speaking a prediction interval, not a confidence interval. However, we still refer to it as a confidence interval to keep the terminology consistent with prior work.} $[\hat{L}_{m,\alpha}(Z_{m+1}), \hat{U}_{m,\alpha}(Z_{m+1})]$ for $f_m(Z_{m+1})$  with guaranteed {\em marginal coverage},
\begin{align} \label{eq:marginal-coverage}
  \Pbig{\hat{L}_{m,\alpha}(Z_{m+1}) \leq f_m(Z_{m+1}) \leq \hat{U}_{m,\alpha}(Z_{m+1})} \geq 1-\alpha,
\end{align}
at the desired level $\alpha \in (0,1)$.
Such a coverage property is called {\em marginal} because it involves a probability taken with respect to the randomness in both the data and the query.
Its interpretation is as follows: the confidence interval will cover $f_m(Z_{m+1})$ for at least a fraction $1-\alpha$ of data points $Z_1,\ldots,Z_m$ and future queries $Z_{m+1}$.

Marginal coverage is not trivial to achieve with a reasonably short interval, but it is also not fully satisfactory
because our problem involves discrete data that are likely to include many repeated observations of the same objects.
Unfortunately, inferences satisfying \eqref{eq:marginal-coverage} are not necessarily reliable for a sufficient proportion of {\em distinct} or {\em unique} queries, which is what we would ideally like to guarantee.
To the contrary, confidence intervals with marginal coverage are likely to have lower coverage for rarer queries, as illustrated by the following thought experiment.
Imagine a distribution $P_Z$ with support on $\mathscr{Z} = \{0,1,2,\ldots,10^{100}\}$, such that $\P{Z_i = 0} = 0.95$ and $\P{Z_i=z} = 0.05/(|\mathscr{Z}|-1)$ for all $z \in \mathscr{Z} \setminus \{0\}$ and $i \geq 1$.
Marginal coverage at level 95\% would be satisfied even by a non-informative confidence interval that always contains the true frequency for a new query if $Z_{m+1}=0$ and is empty otherwise. However, those inferences are incorrect for all but one possible query.
This issue motivates the development of methods with stronger coverage guarantees.

In Section~\ref{sec:frequency-coverage}, we begin to address the limitations of marginal coverage by presenting a method for constructing confidence intervals that are valid for both rarer and more common random queries, taking inspiration from Mondrian conformal inference for classification~\citep{vovk2003mondrian}.
Section~\ref{sec:coverage-unique} extends these ideas by developing and studying a novel construction of conformal confidence intervals with guaranteed coverage for a large fraction of distinct/unique queries in a possibly redundant test set.
This method is related to the works of \cite{dunn2018distribution} and \cite{park2022pac} on conformal inference for hierarchical models and meta-learning, but the specific notion of coverage proposed here had not been investigated before.
Coverage for a large fraction of distinct queries implies that less frequent queries are given a higher weight. For instance, the 
example above, we expect that out of $M=1000$ test examples $950$ are equal to zero and the others are all distinct.
Then, covering 95\% of the uniques means that we expect to cover approximately $0.95\cdot 951 \approx 903$ distinct queries. Clearly, this is more informative than an interval that covers only zero. 

% Coverage for a large fraction of distinct queries is import1ant, 
% because identical query values are treated as equivalent in our framework.

Exchangeability has a broad scope, and in certain cases it can be ensured by permuting the data---as in the experiments described in this paper. However, in practice, when the data come from a real-time stream---such as a sensor monitoring the weather, internet traffic, etc.---systematic distribution shifts can occur that make test data dissimilar from training data.
Motivated by this problem, we will show that our proposed method also leads to increased robustness to distribution shifts, which allows some relaxation of the exchangeability assumptions and thus broadens the relevance of our results to more applications, possibly to online data streams~\citep{cao2023meta}.

Finally, Sections~\ref{sec:experiments}--\ref{sec:app} present several experiments and illustrations of our methods, using both synthetic data from realistic power-law distributions and two empirical data examples. The latter concern 16-mers in SARS-CoV-2 DNA sequences and  2-grams in English literature.
We consider the classical CMS \citep{cormode2005improved}, the CMS-CU \citep{estan2002new}, the CS \citep{charikar2002finding}, and non-random sketches based on data-driven hash functions \citep{bertsimas2021frequency}.
We compare our methods, according to different performances metrics, to existing uncertainty estimation techniques developed for CMS sketches, including bootstrap and Bayesian approaches \citep{cormode2020small,ting2018count,cai2018bayesian, dolera2021bayesian}.
In addition to being more flexible, as we are not limited to working with the CMS, our methods tend to outperform the existing benchmarks even when the latter are applicable, producing shorter confidence intervals with more consistent coverage.
Further, we verify that our method aiming for coverage of unique elements has a higher robustness to distribution shifts compared to the simpler approach targeting marginal coverage. Additional experiments are discussed in the appendix.
Section~\ref{sec:discussion} concludes with a discussion and some ideas for future work.

\subsection{Related work}

There exist many algorithms for computing approximate frequency queries given a reduced-memory sketch; some are based on random hashing \citep{fan2000summary,goyal2011lossy, pitel2015count,cormode2020small}, while others may involve complex learning algorithms \citep{bertsimas2021frequency}.
Several works have also studied the problem of quantifying uncertainty in this context, but we are the first to propose a conformal inference approach that is not limited to a specific sketching algorithm.
In fact, to the best of our knowledge, the related prior research has focused on the CMS algorithm \citep{cormode2005improved}.
The classical uncertainty estimation strategies treated the data as fixed and leveraged only the randomness in the hash functions of the CMS \citep{cormode2005improved}, which we review in Appendix~\ref{app:cms}. While that approach can lead to rigorous confidence bounds for the unknown empirical frequencies under minimal assumptions, the results are often too conservative to be practically useful \citep{ting2018count}.

This is why more recent works treated the data as random and either derived frequentist inferences using re-sampling techniques~\citep{ting2018count} or calculated a Bayesian posterior distribution for the frequency of the queried object starting from a prior model for the sketched data \citep{cai2018bayesian,dolera2021bayesian,beraha2023random}.
Our work is closer to~\cite{ting2018count}, as we seek frequentist probabilistic guarantees while treating the data as random, but our solution is very different. The method of \cite{ting2018count} is limited to the CMS, whereas we use conformal inference and can handle any sketching algorithm, including non-linear and learning-based ones \citep{estan2002new,hsu2019learning,aamand2019learned,bertsimas2021frequency}.
Such flexibility is useful because different sketching algorithms may allow more efficient data compression and more accurate frequency estimates depending on the data distribution~\citep{aamand2019learned}.

Conformal inference was pioneered by Vovk and collaborators~\citep{saunders1999transduction,vovk2005algorithmic} and brought to the statistics spotlight by works such as~\cite{lei2013distribution,lei2014distribution,lei2018distribution}.
Although primarily conceived for supervised prediction~\citep{vovk2009line, vovk2015cross,lei2014distribution,romano2019conformalized,izbicki2019flexible,park2021pac,qiu2022distribution}, conformal inference
has found other applications including outlier
and anomaly
detection~\citep{bates2021testing,kaur2022idecode,li2022pac,liang2022integrative}, causal inference~\citep[e.g.,]{lei2021conformal}, and survival analysis~\citep{candes2021conformalized}.
We mention here that
the ideas in conformal prediction have deep roots in statistics, dating back at least to the pioneering works of \protect\cite{Wilks1941}, \protect\cite{Wald1943}, \protect\cite{scheffe1945non}, and \protect\cite{tukey1947non,tukey1948nonparametric}; see also \cite{geisser2017predictive}.

\subsection{Relation to shorter conference paper}

The potential of conformal inference in sketching remained untapped before the shorter version of this work \citep{sesia2022conformalized}, which appeared in the proceedings of the NeurIPS 2022 conference.
This extended manuscript contains novel methods and several original theoretical results, in Section~\ref{sec:coverage-unique}, studying the construction of confidence intervals with valid coverage for a large fraction of distinct queries.
This is stronger and more challenging guarantee compared to marginal coverage, and it is useful because it leads to more easily interpretable inferences when the data are discrete and may involve many repeated observations.
Further, we will show that the methodological extensions introduced in this paper improve the robustness to distribution shifts and other possible violations of the data exchangeability assumption \citep{tibshirani2019conformal,barber2023conformal}, which could be relevant for example when sketching streaming data \citep{cao2023meta}.
Finally, Sections~\ref{sec:experiments}--\ref{sec:app} of this manuscript contain several additional numerical results, and the whole paper has been re-organized to provide a more general description of the proposed methodology that better highlights its general applicability in combination with any sketching algorithm.

\section{Preliminaries on conformal prediction} \label{sec:conformal-review}

Consider {\it supervised learning}, with data pairs $(X_i, Y_i)$ where $X_i$ are a vector of {\it features} for the $i$-th observation and $Y_i$ are the corresponding {\it outcome} or {\it label}, which may be continuous- or discrete-valued.
The usual goal in supervised learning is to use $(X_1, Y_1), \ldots, (X_n, Y_n)$ to learn a predictor of an unseen label $Y_{n+1}$ using a new observation with features $X_{n+1}$. 
Related to this, conformal prediction
can be used to construct a prediction interval $[\hat{L}_{n,\alpha}$ $(X_{n+1})$, $\hat{U}_{n,\alpha}(X_{n+1})]$ with guaranteed marginal coverage,
\begin{align*}
\mathbb{P}[\hat{L}_{n,\alpha}(X_{n+1}) \leq Y_{n+1} \leq \hat{U}_{n,\alpha}(X_{n+1})] \geq 1-\alpha,
\end{align*}
for any fixed $\alpha \in (0,1)$,
assuming that $(X_1, Y_1), \ldots, (X_{n+1}, Y_{n+1})$ is an exchangeable random sample from some unknown distribution over $(X,Y)$.
Conformal prediction 
can leverage supervised learning methods 
to approximately reconstruct the relation between $X$ and $Y$,
capturing it in $\hat{L}_{n,\alpha}, \hat{U}_{n,\alpha}$, and it automatically calibrates such prediction interval to achieve marginal coverage.
%This can yield relatively short intervals satisfying marginal coverage.
While it is sufficient to focus on conformal intervals in this paper, similar techniques can also be used to construct more general prediction sets~\citep[e.g.,][etc]{vovk2005algorithmic, romano2020classification,angelopoulos2020uncertainty}.

A simple version of conformal prediction---known as {\em split} or {\em inductive} conformal prediction \citep{papadopoulos2002inductive,lei2018distribution}---begins by randomly splitting the observations into two disjoint subsets:
a \emph{training set} and a \emph{calibration set}. The first $n^{\mathrm{train}} \in \{1,\ldots,n\}$ data points are used as the training set, 
to fit a machine learning model for predicting $Y$ given $X$; e.g., a neural network or a random forest. 
The out-of-sample predictive accuracy of this model is then measured in terms of a  {\it conformity score} for each of the $n-n^{\mathrm{train}}$ held-out data points in the calibration set. In combination with the model learned from the training data, the quantiles of the empirical distribution of these scores are used to construct prediction intervals for future test points as a function of $X_{n+1}$.
As detailed shortly, these intervals are guaranteed to cover $Y_{n+1}$ with probability at least $1-\alpha$, treating all data as random.  Importantly, the coverage holds
in finite samples, regardless of the accuracy of the predictive model, as long as $X_{n+1}$ is exchangeable with the held-out data points.
It is unnecessary for the training data to be also exchangeable, as these may be viewed as fixed.

%The implementation of conformal inference depends on the choice of conformity scores.
One perspective on conformal prediction is to construct a {\it nested sequence}
\citep{vovk2005algorithmic,gupta2019nested} of prediction intervals $[\hat{L}_{n,\alpha}(x; t), \hat{U}_{n,\alpha}(x; t)]$, indexed by $t \in \mathcal{T} \subseteq \mathbb{R}$ for each $x$; based on the fitted machine learning model.
This sequence is nested,
in the sense that $\hat{L}_{n,\alpha}(x; t_2) \leq \hat{L}_{n,\alpha}(x; t_1)$ and $\hat{U}_{n,\alpha}(x; t_2) \geq \hat{U}_{n,\alpha}(x; t_1)$ for all $t_2 \geq t_1$. Further, assume there exists $ t_{\infty} \in \mathcal{T}$ such that $\hat{L}_{n,\alpha}(X; t_{\infty}) \leq Y \leq \hat{U}_{n,\alpha}(X; t_{\infty})$ almost surely.
For example, one may consider the sequences $\hat{\psi}_n(x) \pm t$, $t\ge 0$,
where $\hat{\psi}_n$ is a regression function for a bounded label $Y$ given $X$ output by machine learning model and $t$ plays the role of a  ``predictive standard error".
For one-sided (lower) confidence intervals $[\hat{L}_{n,\alpha}(x; t),  \infty)$, we may set $\hat{U}_{n,\alpha}(x; t) = \infty$.

Then, the conformity score for a point with $X=x$ and $Y=y$ is defined as the smallest---infimum---index $t$ such that $y$ is contained in the prediction interval $[\hat{L}_{n,\alpha}(x; t), \hat{U}_{n,\alpha}(x; t)]$:
\begin{align} \label{eq:conf-score}
  E(x,y) \defeq \inf \big\{ t \in \mathcal{T} : y \in [\hat{L}_{n,\alpha}(x; t), \hat{U}_{n,\alpha}(x; t)] \big\}.
\end{align}
Let $\mathcal{I}^{\mathrm{calib}} \subset \{1,\ldots,n\}$ be the subset of held-out data points, with cardinality $|\mathcal{I}^{\mathrm{calib}}|$. 
Let $\hat{Q}_{n, 1-\alpha}$ be the $\lceil (1-\alpha) (|\mathcal{I}^{\mathrm{calib}}|+1) \rceil$-th smallest conformity score $E(X_i,Y_i)$ among all $i \in \mathcal{I}^{\mathrm{calib}}$.
The conformal prediction interval for a new data point with features $X_{n+1}$ is:
\begin{align} \label{eq:conf-interval}
  \big[ \hat{L}_{n,\alpha}(X_{n+1}; \hat{Q}_{n, 1-\alpha}), \hat{U}_{n,\alpha}(X_{n+1}; \hat{Q}_{n, 1-\alpha}) \big].
\end{align}
Intuitively, this satisfies marginal coverage because $Y_{n+1}$ falls outside~\eqref{eq:conf-interval} if and only if $E(X_{n+1},Y_{n+1}) > \hat{Q}_{n, 1-\alpha}$.
The rest of the proof is a simple exchangeability argument; see~\cite{vovk2005algorithmic}, \citet{romano2019conformalized}, or the proof of Theorem~\ref{thm:coverage} in Appendix~\ref{sec:proofs}.

\section{Confidence intervals with marginal coverage} \label{sec:methods-marginal}

\subsection{Data exchangeability and conformal confidence intervals} \label{sec:sketch-problem}

As anticipated in Section~\ref{sec:problem-statement}, we study a sketching problem in which the query $Z_{m+1}$ and $m$ data points, $Z_1,\ldots,Z_m$, are an exchangeable random sample from some distribution $P_Z$ on $\mathscr{Z}$.
We assume that the full data set is too large to process directly.
Recall that our goal is to construct a confidence interval with guaranteed marginal coverage \eqref{eq:marginal-coverage} for the number of occurrences  $f_m(Z_{m+1})$---defined in \eqref{eq:f-true}---of the query $Z_{m+1}$ in the data set.
Since $Z_1,\ldots,Z_m$ cannot be observed, we rely on the information contained in the sketch $\phi(Z_1, \ldots, Z_m)$. Importantly, we would like to retain as much flexibility as possible with regard to the sketching function $\phi$.

To connect this problem with the conformal inference framework reviewed in Section~\ref{sec:conformal-review}, 
we need to define the appropriate features and outcomes.
Our approach is to store the true frequencies for all objects in the first $n$ observations in a {\it warm-up} stage, for some fixed $n \ll m$ that is sufficiently large subject to memory constraints\footnote{Note that the index $n+1$ of the test point from Section~\ref{sec:conformal-review} is now replaced by $m+1$.}.
An extension of this method allowing $n$ to be data-dependent will be discussed later in Section~\ref{sec:adaptive-warm-up}.
%Without loss of generality, assume $n \ll m$; otherwise, the problem becomes trivial.
Let $n_0 \leq n$ indicate the number of distinct objects among the first $n$ observations.
The memory required to store these frequencies is $O(n_0)$, which is typically negligible if $n$ is small compared to the size of the sketch.
We use these stored frequencies to define features and outcomes, transforming our task into supervised prediction, as detailed below.

During the warm-up phase, 
we store the frequencies of the distinct objects among the first $n$ observations $Z_1,\ldots,Z_{n}$ from the data stream. 
We denote these counts as $f^{\mathrm{wu}}_{n}(z)$, defined for all $z\in \mathcal{Z}$ as
\begin{align} \label{eq:warm-up-freq}
  f_{n}^{\mathrm{wu}}(z) \defeq \sum_{i=0}^{n} \I{Z_i = z}.
\end{align}
Next, the remaining $m-n$ data points are streamed and compressed using any black-box sketching function $\phi$ of choice. 
At the same time, however, we also keep track of the true frequencies for all instances of objects already seen during the warm-up phase.
In other words, the following counters are computed and stored along with the sketch\footnote{Compared to the setup from Section \ref{sec:problem-statement}, here the sketch is only applied to the observations $Z_{n+1}, \ldots, Z_{m}$ instead of $Z_{1}, \ldots, Z_{m}$, because the frequencies of the first $n$ observations are already stored exactly.} 
$\phi(Z_{n+1}, \ldots, Z_{m})$:
\begin{align}
  f_{m-n}^{\mathrm{sv}}(z) \defeq
  \begin{cases}
    \sum_{i=n+1}^{m} \I{Z_i = z}, & \text{if } f_{n}^{\mathrm{wu}}(z) > 0, \\
    0, & \text{otherwise.}
  \end{cases}
\end{align}
Again, this requires only $O(n_0)$ memory.
Next, we define the variables $Y_i$ for all $i \in \{1,\ldots,n\}$ $ \cup \{m+1\}$ as the true frequencies of $Z_i$ among $Z_{n+1}, \ldots, Z_{m}$:
\begin{align} \label{eq:Y-def}
  Y_i \defeq \sum_{i'=n+1}^{m} \I{Z_{i'} = Z_i}.
\end{align}
For all $i \in \{1,\ldots,n\}$ $ \cup \{m+1\}$,
the frequencies of $Z_i$ can be written as $f_{m}(Z_{i})  = Y_{i} + f_{n}^{\mathrm{wu}}(Z_{i})$. 
%Note that $Y_i$ and $f_{n}^{\mathrm{wu}}(Z_{i})$ together are perfectly predictive of $f_{m}(Z_i)$, as . 
Thus, $f_{n}^{\mathrm{wu}}(Z_{i})$ and $Y_i$  together determine the outcome $f_{m}(Z_{i})$ of interest.
For  $i \in \{1,\ldots,n\}$, 
$Y_i$ are observed 
and equal $Y_i = f_{m-n}^{\mathrm{sv}}(Z_i)$.
In contrast, for the  query
$Z_{m+1}$, we have $Y_{m+1} = f_{m-n}^{\mathrm{sv}}(Z_{m+1})$ only if the value of $Z_{m+1}$ has occurred among $Z_1,\ldots, Z_{n}$ and thus the frequency of $Z_{m+1}$ has been stored.
Otherwise, the value of $Y_{m+1}$ is not known.
Since $f_{n}^{\mathrm{wu}}(Z_{m+1})$ and $Y_{m+1}$ determine $f_{m}(Z_{m+1})$, it is reasonable to aim to build a predictive model or conformal interval for $Y_{m+1}$ based on the observed data $Z_{m+1}$ and the sketch.

%Thus, if we can predict $Y_i$ based on $Z_i$, then, after observing a new query $Z_{m+1}$, we can use that to predict $Y_{m+1}$, and hence $f_{m}(Z_{m+1})$. Below, we will show how we can leverage existing sketching algorithms for this prediction step.

%For a new query $Z_{m+1}$, 
%$Y_{m+1}$ is not known---in truth, the ultimate target is $f_{m}(Z_{m+1})  = Y_{m+1} + f_{n}^{\mathrm{wu}}(Z_{m+1})$, but the second term is already known exactly, as $f_{n}^{\mathrm{wu}}(z)$ is recorded for all $z$.

To formalize this, for each $i \in \{1,\ldots,n\} \cup \{m+1\}$, define the features $X_i$ as the vectors containing the data point $Z_i$
and
the information in the sketch:
\begin{align} \label{eq:X-def}
  X_i \defeq \left(Z_i, \phi(Z_{n+1}, \ldots, Z_{m}) \right).
\end{align}

To obtain a conformal guarantee, we 
will rely on result that the pairs $(X_1,Y_1),\ldots$, $(X_{n},Y_{n})$, $(X_{m+1},Y_{m+1})$ are exchangeable with one another---where, as discussed, $Y_{m+1}$ is possibly unobserved.
All mathematical proofs are in Appendix~\ref{sec:proofs}.
\begin{prop} \label{eq:exchangeability-XY}
  If the unsketched data points $Z_1,\ldots,Z_{m+1}$ are exchangeable, then the pairs of random variables $(X_1,Y_1),\ldots,(X_{n},Y_{n}),(X_{m+1},Y_{m+1})$  in~\eqref{eq:Y-def}--\eqref{eq:X-def} are also exchangeable with one another.
\end{prop}

Proposition~\ref{eq:exchangeability-XY} opens the door to applying
conformal inference to the supervised observations $(X_1,Y_1),\ldots,(X_{n},Y_{n})$ in order to predict $Y_{m+1}$ given $X_{m+1}$, guaranteeing marginal coverage.
In particular, using the inductive/split conformal prediction methodology reviewed in Section \ref{sec:conformal-review}, 
one can randomly split the observations indexed by $\{1,\ldots,n\}$ into a training subset indexed by $\{1,\ldots,n^{\text{train}}\}$ for some fixed $n^{\text{train}} < n$, 
and a disjoint calibration subset indexed by $\{n^{\text{train}}+1,\ldots,n\}$.
The training set is used for fitting a predictor
for computing nested confidence intervals, $[\hat{L}_{m,\alpha}(\cdot; t), \hat{U}_{m,\alpha}(\cdot; t)]$, $t\in\mathcal{T}$;
see the next sections for further implementation details.
The aim is that
$Y\in [\hat{L}_{m,\alpha}(X; t), \hat{U}_{m,\alpha}(X; t)]$ holds with large probability; and thus this interval can be used to predict $Y$ and hence also $f_m(Z) = Y + f_{n}^{\mathrm{wu}}(Z)$.
In certain cases, this predictor will leverage a classical deterministic sketching method, making the training step unnecessary.

To choose a suitable value for the parameter $t$,
following the general approach reviewed in Section~\ref{sec:conformal-review},
the calibration set of observations indexed by $\{n^{\text{train}}+1,\ldots,n\}$ is used to compute conformity scores  $E(X_i,Y_i)$ for $i \in \{n^{\mathrm{train}}+1,\ldots,n\}$.
Then, with $n^{\mathrm{cal}} = n- n^{\mathrm{train}}$, 
the conformal interval is constructed as in \eqref{eq:conf-interval}, by setting $t$ as the $\lceil (1-\alpha) (n^{\mathrm{cal}}+1) \rceil$-th smallest score of $E(X_i,Y_i)$ for $i \in \{n^{\mathrm{train}}+1,\ldots,n\}$.
The resulting interval from \eqref{eq:conf-interval} guarantees valid marginal coverage for a new test query in finite samples.

\begin{figure}[!htb]
\begin{center}
\includegraphics[scale=1]{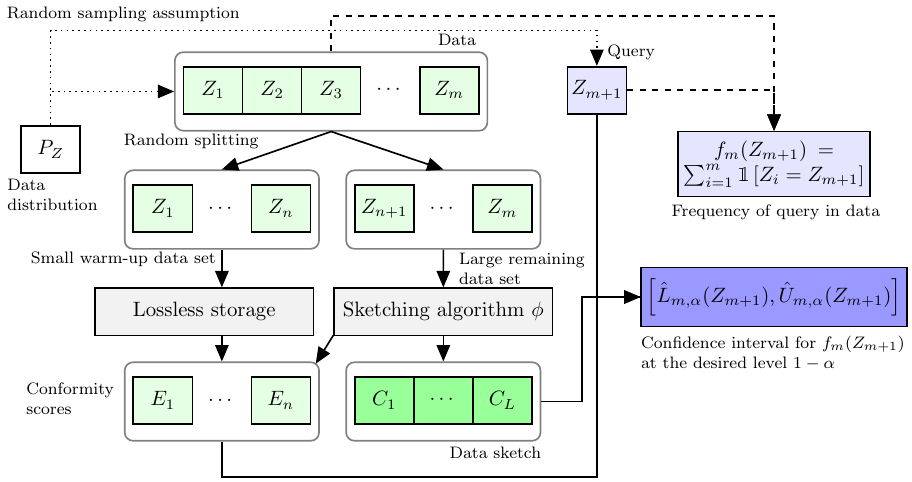}
\caption{A diagram of the conformalized sketching method in Algorithms~\ref{alg:conformal-sketch-train}--\ref{alg:conformal-sketch}.}
\label{fig:sketch-diagram-method}
\end{center}
\end{figure}

This solution is outlined by Algorithms~\ref{alg:conformal-sketch-train}--\ref{alg:conformal-sketch} and visualized schematically in Figure~\ref{fig:sketch-diagram-method}.
Algorithm~\ref{alg:conformal-sketch} outputs the final confidence interval after Algorithm~\ref{alg:conformal-sketch-train} sketches and pre-processes the data. This modular organization will prove useful in the following sections to simplify the exposition of extensions of our methodology.
The following result states that the confidence interval output by Algorithm~\ref{alg:conformal-sketch} has the desired marginal coverage.

\begin{algorithm}[!htb]
   \caption{Conformalized sketching (data sketching, training, and calibration)}
   \label{alg:conformal-sketch-train}
\begin{algorithmic}
  \STATE {\bfseries Input:} Data set $Z_1, \ldots, Z_m$. Sketching function $\phi$. Warm-up period $n \ll m$.
  \STATE {\bfseries \textcolor{white}{Input:}} A trainable predictor to compute nested intervals $[\hat{L}_{m,\alpha}(\cdot; t), \hat{U}_{m,\alpha}(\cdot; t)]_{t \in \mathcal{T}}$.
  \STATE {\bfseries \textcolor{white}{Input:}} Number of data points $n^{\mathrm{train}} < n$ used for training $[\hat{L}_{m,\alpha}(\cdot; t), \hat{U}_{m,\alpha}(\cdot; t)]$.
%  \STATE {\bfseries \textcolor{white}{Input:}} A partition $\mathcal{B} = (B_1,\ldots,B_L)$ of $\{0,\ldots,m\}$ into $L$ intervals.
%  \STATE {\bfseries \textcolor{white}{Input:}} Random query $Z_{m+1}$. Desired coverage level $1-\alpha \in (0,1)$.
  \STATE{\bfseries Initialize} a sparse counter $f_{n}^{\mathrm{wu}}(z) = 0, \forall z \in \mathcal{Z}$.
  \FOR{$i = 1,\ldots,n$}
  \STATE{\bfseries Increment} $f_{n}^{\mathrm{wu}}(Z_i) \leftarrow f_{n}^{\mathrm{wu}}(Z_i)  +1$.
  \ENDFOR
  \STATE{\bfseries Initialize} a sparse counter $f_{m-n}^{\mathrm{sv}}(z) = 0, \forall z \in \mathcal{Z}$.
  \STATE{\bfseries Initialize} an empty sketch $\phi(\emptyset)$.
    \FOR{$i = n+1,\ldots,m$}
    \STATE{\bfseries Update} the sketch $\phi$ with the new observation $Z_i$.
    \IF{$f_{n}^{\mathrm{wu}}(Z_i) > 0$}
    \STATE{\bfseries Increment} $f_{m-n}^{\mathrm{sv}}(Z_i) \leftarrow f_{m-n}^{\mathrm{sv}}(Z_i)  +1$.
    \ENDIF
    \ENDFOR

  \FOR{$i = 1,\ldots,n$}
  \STATE{\bfseries Set}  $X_i = \left(Z_i, \phi(Z_{n+1}, \ldots, Z_{m}) \right)$ as in~\eqref{eq:X-def}.
  \STATE{\bfseries Set}  $Y_i = f_{m-n}^{\mathrm{sv}}(Z_i)$.
  \ENDFOR
  \STATE{\bfseries Train} $[\hat{L}_{m,\alpha}(\cdot; t), \hat{U}_{m,\alpha}(\cdot; t)]$, $t\in \mathcal{T}$, using the data in $\{(X_i,Y_i)\}_{i=1}^{n^{\mathrm{train}}}$.
  \FOR{$i = n^{\mathrm{train}}+1,\ldots,n$}
  \STATE{\bfseries Compute} the conformity score $E(X_i,Y_i)$ with~\eqref{eq:conf-score}, using $[\hat{L}_{m,\alpha}(\cdot; t), \hat{U}_{m,\alpha}(\cdot; t)]$.
  \ENDFOR
  \STATE {\bfseries Output:} Data sketch $\phi$;
\STATE {\color{white} \bfseries Output:} Sparse counter $f_{n}^{\mathrm{wu}}(z), \forall z \in \mathcal{Z}$;
  \STATE {\color{white} \bfseries Output:} Trained predictor $[\hat{L}_{m,\alpha}(\cdot; t), \hat{U}_{m,\alpha}(\cdot; t)]$;
  \STATE {\color{white} \bfseries Output:} Conformity scores $E(X_i,Y_i)$ for all $i \in \{n^{\mathrm{train}}+1,\ldots,n\}$.
\end{algorithmic}
\end{algorithm}

\begin{algorithm}[!htb]
   \caption{Conformalized sketching with marginal coverage}
   \label{alg:conformal-sketch}
\begin{algorithmic}
  \STATE {\bfseries Input:} Same as for Algorithm~\ref{alg:conformal-sketch-train}.
%  \STATE {\bfseries \textcolor{white}{Input:}} A partition $\mathcal{B} = (B_1,\ldots,B_L)$ of $\{0,\ldots,m\}$ into $L$ intervals.
  \STATE {\bfseries \textcolor{white}{Input:}} Random query $Z_{m+1}$. Desired coverage level $1-\alpha \in (0,1)$.
  \STATE{\bfseries Compute} using Algorithm~\ref{alg:conformal-sketch-train}: \\
  \STATE{\color{white} \bfseries Input:} Data sketch $\phi$; a sparse counter $f_{n}^{\mathrm{wu}}(z), \forall z \in \mathcal{Z}$;
  \STATE{\color{white} \bfseries Input:} Variables $X_i = \left(Z_i, \phi(Z_{n+1}, \ldots, Z_{m}) \right)$ and $Y_i = f_{m-n}^{\mathrm{sv}}(Z_i)$ for $i \in \{1,\ldots,n\}$.
  \STATE{\color{white} \bfseries Input:} Trained predictor for computing nested intervals $[\hat{L}_{m,\alpha}(\cdot; t), \hat{U}_{m,\alpha}(\cdot; t)]_{t \in \mathcal{T}}$;
  \STATE{\color{white} \bfseries Input:} Conformity scores $E(X_i,Y_i)$ for all $i \in \{n^{\mathrm{train}}+1,\ldots,n\}$.
  \STATE{\bfseries Compute} $\hat{Q}_{n^{\mathrm{cal}}, 1-\alpha}$ as the $\lceil (1-\alpha) (n^{\mathrm{cal}}+1) \rceil$-th smallest score, with $n^{\mathrm{cal}} = n- n^{\mathrm{train}}$.
  \STATE{\bfseries Set}  $X_{m+1} = \left(Z_{m+1}, \phi(Z_{n+1}, \ldots, Z_{m}) \right)$ as in~\eqref{eq:X-def}.
  \STATE {\bfseries Output:} a $(1-\alpha)$-level confidence interval $$\left[ f_{n}^{\mathrm{wu}}(Z_{m+1}) + \hat{L}_{m,\alpha}(X_{m+1}; \hat{Q}_{n^{\mathrm{cal}}, 1-\alpha}), f_{n}^{\mathrm{wu}}(Z_{m+1}) +  \hat{U}_{m,\alpha}(X_{m+1}; \hat{Q}_{n^{\mathrm{cal}}, 1-\alpha}) \right]$$ for the unobserved frequency $f_m(Z_{m+1})$ of $Z_{m+1}$ defined in~\eqref{eq:f-true}.
\end{algorithmic}
\end{algorithm}

\begin{theorem} \label{thm:coverage}
  If the data $Z_1,\ldots,Z_{m+1}$ are exchangeable, the confidence interval output by Algorithm~\ref{alg:conformal-sketch} satisfies the marginal coverage property defined in~\eqref{eq:marginal-coverage}.
\end{theorem}

\textbf{Remark.} Algorithm~\ref{alg:conformal-sketch} could be trivially modified to output perfect ``singleton'' confidence intervals for any new queries that happen to be identical to an object previously observed during the warm-up phase. We will not take advantage of this option in the experiments presented in this paper in order to provide a fairer comparison with alternative methods which do not involve a similar warm-up phase.

\subsection{Conformity scores for confidence intervals with fixed width} \label{sec:conf-scores-fixed}

The method described by Algorithms~\ref{alg:conformal-sketch-train}--\ref{alg:conformal-sketch} can accommodate any data-adaptive intervals $[\hat{L}_{m,\alpha}(x; t), \hat{U}_{m,\alpha}(x; t)]$---for computing nested confidence intervals, which may depend on the sketch $\phi \defeq \phi(Z_{n+1}, \ldots, Z_{m})$.
A simple one-sided construction of these confidence intervals is possible if the sketching algorithm provides us with a non-trivial deterministic upper bound 
$\hat{f}_{\mathrm{up}}(X_{m+1})=\hat{f}_{\mathrm{up}}(Z_{m+1},\phi)$ for the query frequency---such that $f(X_{m+1})\le \hat{f}_{\mathrm{up}}(Z_{m+1},\phi)$ for all $Z_{m+1}$---as it is the case with the CMS \citep{cormode2005improved}.
In those cases, we suggest to calibrate the parameter $t$ of the following sequence of potential lower bounds on the query frequency:
\begin{align} \label{eq:rule-fixed}
  \hat{L}^{\text{fixed}}_{m,\alpha}((Z_{m+1},\phi); t) \defeq \max\{0, \hat{f}_{\mathrm{up}}(Z_{m+1},\phi) - t\}, \qquad t \in \{0, 1, \ldots, m\}.
\end{align}
In words, a potential lower bound for $f_m(Z_{m+1})$ in~\eqref{eq:rule-fixed} is defined by shifting the deterministic upper bound down by a constant $t$.
The appropriate value of $t$ guaranteeing marginal coverage is chosen by applying Algorithm~\ref{alg:conformal-sketch} at the nominal level $\alpha$.
If $Y_i \le \hat{f}_{\mathrm{up}}(X_{i})$ for all $i \in \{n^{\mathrm{train}}+1,\ldots,n\}$, then the chosen $t$ can also be written as the  $\lceil (1-\alpha) (n^{\mathrm{cal}}+1) \rceil$-th smallest value of $\hat{f}_{\mathrm{up}}(X_{i})-Y_i$ among $i \in \{n^{\mathrm{train}}+1,\ldots,n\}$.

This approach does not require training data, in the sense that it allows one to use $n^{\text{train}}=0$ and use all $n$ observations with tracked frequencies for calibration.
Further, two-sided conformal confidence intervals can be constructed as explained in Appendix~\ref{sec:conf-scores-twosided}.

\subsection{Conformity scores for confidence intervals with adaptive width} \label{sec:conf-scores-adaptive}

A more flexible confidence interval construction with query-dependent width can sometimes lead to more informative predictions compared to the simpler method described in Section~\ref{sec:conf-scores-fixed}.
This approach, which we call ``adaptive", involves training a machine learning model to approximate the optimal width of the confidence intervals, and it is inspired by the methods of~\cite{chernozhukov2021distributional} and~\citet{sesia2021conformal}.
For simplicity, we focus here on the construction of one-sided intervals, assuming that a deterministic upper bound $\hat{f}_{\mathrm{up}}$ for the desired query frequency is already available (e.g., as in the case of the CMS). However, the same idea can be generalized easily to construct instead two-sided confidence intervals; see Appendix~\ref{sec:conf-scores-twosided}.

Consider a machine learning model taking as input the deterministic upper bound $\hat{f}_{\mathrm{up}}(Z_i)$ and estimating the conditional distribution of $\hat{f}_{\mathrm{up}}(Z_i)-f_m(Z_i)$ given $\hat{f}_{\mathrm{up}}(Z_i)$. For example, think of a quantile neural network~\citep{taylor2000quantile} or a quantile random forest~\citep{meinshausen2006quantile}.
After fitting this model on the training data set of size  $n^{\mathrm{train}}$,
let $\hat{q}_t$ be the estimated $\alpha_t$-th lower quantile of $\hat{f}_{\mathrm{up}}(Z_i) - f_m(Z_i) \mid \hat{f}_{\mathrm{up}}(Z_i)$, for all $t \in \{1,\ldots,T\}$ and some fixed sequence $0=\alpha_1 < \ldots < \alpha_T = 1$. Without loss of generality, assume that quantile crossings do not occur \citep{he1997quantile} and let
$\hat{q}_0 = 0$, $\hat{q}_T = m-n$. Then, define the following monotone sequence of conformal lower bounds, recalling that $X_{m+1}=(Z_{m+1},\phi)$:
\begin{align*} %\label{eq:rule-fixed}
  \hat{L}^{\mathrm{adaptive}}_{m,\alpha}((Z_{m+1},\phi); t) \defeq \max \left\{0, \hat{f}_{\mathrm{up}}(X_{m+1}) - \hat{q}_t\left( \hat{f}_{\mathrm{up}}(X_{m+1}) \right) \right\}, \qquad t \in \{0, 1, \ldots, m\}.
\end{align*}
Finally, the calibrated value of $t$ guaranteeing marginal coverage is obtained by applying Algorithm~\ref{alg:conformal-sketch} at the nominal level $\alpha$.
This approach can lead to a lower confidence bound whose distance from the upper bound is adaptive to the test instance $X_{m+1}$.
This can be an advantage because the sketching algorithms may introduce higher uncertainty about common queries compared to rarer ones, or vice versa, depending on the data distribution, and such patterns may be learnt given a sufficient number of observations.

\subsection{Data-adaptive warm-up} \label{sec:adaptive-warm-up}

Algorithm~\ref{alg:conformal-sketch-train} requires pre-specifying the total number of data points $n$ processed during the warm-up phase.
A possible limitation of this approach is that the number of distinct objects $n_0 \leq n$ among the first $n$ observations depends on the data distribution $P_Z$ and cannot be known in advance. In particular, if the data follow a distribution with a power-law tail behaviour, as it is often the case in many practical applications \citep{clauset2009power}, some types of objects may be much more likely than others to be observed, resulting in $n_0 \ll n$.
Given that the memory cost of Algorithm~\ref{alg:conformal-sketch-train} depends on the number of distinct objects observed during the warm-up phase, it would be more intuitive to allow the user to control the duration of the warm-up phase by directly specifying the desired value of $n_0$ instead of $n$.
In other words, one may want to run the warm-up phase of Algorithm~\ref{alg:conformal-sketch-train} for a flexible number of steps $n$, until exactly $n_0$ distinct objects are observed.
Unfortunately, a straightforward implementation of this alternative strategy, which is outlined by Algorithm~\ref{alg:conformal-sketch-train-heuristic} in Appendix~\ref{app:adaptive-warmup}, does not lead to theoretically valid conformal inferences because the randomness in $n$ breaks the desired exchangeability of the calibration data with the test query $Z_{m+1}$.
Nonetheless, Algorithm~\ref{alg:conformal-sketch-train-heuristic} does provide a reasonable heuristic that often works well in practice, as we shall see empirically in Section~\ref{sec:experiments}.

Alternatively, a rigorous solution can be obtained by modifying Algorithm~\ref{alg:conformal-sketch-train-heuristic} so that the conformal inferences are calibrated using only the observations collected during a second distinct warm-up phase, whose duration is fixed conditional on the first warm-up phase.
In particular, the duration of the second warm-up phase is set equal to $n$, namely the (random) number of data points collected until $n_0$ distinct objects are observed during the first warm-up phase. By the exchangeability of the data, one thus expects to observe approximately $n_0$ distinct objects in the second warm-up phase.
Further, this two-step preserves the exchangeability of the calibration data with the test query $Z_{m+1}$ conditional on the value of $n$ and on all the data observed during the first warm-up phase, thus enabling theoretically valid conformal inferences. See Algorithm~\ref{alg:conformal-sketch-train-adaptive} for an outline of this procedure.

\section{Confidence intervals with frequency-conditional coverage} \label{sec:frequency-coverage}

As explained in Section~\ref{sec:problem-statement}, marginal coverage is not fully satisfactory because our data are discrete and more common queries should not be over-counted.
Therefore, we begin to address the limitations of marginal coverage by extending the method presented in Section~\ref{sec:methods-marginal} to obtain confidence intervals valid simultaneously for both rarer and more common queries.
Our approach is inspired by Mondrian conformal inference~\citep{vovk2003mondrian}, which has been previously used---for instance---to construct prediction sets with label-conditional coverage for classification problems~\citep{vovk2005algorithmic,sadinle2019least,romano2020classification}.
However, departing from multi-class classification, we will not seek perfect coverage conditional on the exact frequency of the queried object. In fact, that problem is likely to be impossible to solve without stronger assumptions~\citep{foygel2021limits}, as $f_m(Z_{m+1})$ can take a very large number of values when the sketched data set is big.
Instead, we will focus on achieving a relaxed version of frequency-conditional coverage which groups together queries with similar frequencies.

Fix any partition $\mathcal{B} = (B_1,\ldots,B_L)$ of $\{1,\ldots,m\}$ into $L$ sub-intervals, for some relatively small integer $L$.
The choice of $\mathcal{B}$ and $L$ will be discussed below.
For the time being, it suffices to emphasize that this partition may be arbitrary, as long it is fixed prior to seeing the data $Z_1,\ldots, Z_{m+1}$.
Our goal is to construct a confidence interval $[\hat{L}_{m,\alpha}(Z_{m+1}), \hat{U}_{m,\alpha}(Z_{m+1})]$ depending on $\phi(Z_1,\ldots,Z_m)$ and $\mathcal{B}$ that is reasonably short in practice and guarantees {\em frequency-range conditional coverage}:
\begin{align} \label{eq:conf-int-cond}
  \Pbig{\hat{L}_{m,\alpha}(Z_{m+1}) \leq f_m(Z_{m+1}) \leq \hat{U}_{m,\alpha}(Z_{m+1}) \mid f_m(Z_{m+1}) \in B } \geq 1-\alpha, \qquad \forall B \in \mathcal{B}.
\end{align}
Thus, coverage is guaranteed for observations $Z_{m+1}$ with $f_m(Z_{m+1}) \in B$, for each $B$.
Confidence intervals satisfying~\eqref{eq:conf-int-cond} can be obtained by modifying Algorithm~\ref{alg:conformal-sketch} as outlined in Algorithm~\ref{alg:conformal-sketch-frequency};
by computing empirical quantiles for the conformity scores corresponding to the calibration data points in each frequency bin separately. Then, the final confidence interval for the random query is computed based on the largest quantile across all bins.
The theoretical validity of this solution is established below in Theorem~\ref{thm:coverage-cond}.

\begin{algorithm}[!htb]
   \caption{Conformalized sketching with frequency-conditional coverage}
   \label{alg:conformal-sketch-frequency}
\begin{algorithmic}
  \STATE {\bfseries Input:} Data set $Z_1, \ldots, Z_m$. Sketching function $\phi$. Warm-up period $n \ll m$.
  \STATE {\bfseries \textcolor{white}{Input:}} A (trainable) predictor to compute nested intervals $[\hat{L}_{m,\alpha}(\cdot; t), \hat{U}_{m,\alpha}(\cdot; t)]_{t \in \mathcal{T}}$.
  \STATE {\bfseries \textcolor{white}{Input:}} Number of data points $n^{\mathrm{train}} < n$ used for training $[\hat{L}_{m,\alpha}(\cdot; t), \hat{U}_{m,\alpha}(\cdot; t)]$.
  \STATE {\bfseries \textcolor{white}{Input:}} A partition $\mathcal{B} = (B_1,\ldots,B_L)$ of $\{0,\ldots,m\}$ into $L$ intervals.
  \STATE {\bfseries \textcolor{white}{Input:}} Random query $Z_{m+1}$. Desired coverage level $1-\alpha \in (0,1)$.
  \STATE{\bfseries Compute} using Algorithm~\ref{alg:conformal-sketch-train}: \\
  \STATE{\color{white} \bfseries Input:} Data sketch $\phi$; a sparse counter $f_{n}^{\mathrm{wu}}(z), \forall z \in \mathcal{Z}$;
  \STATE{\color{white} \bfseries Input:} Variables $X_i = \left(Z_i, \phi(Z_{n+1}, \ldots, Z_{m}) \right)$ and $Y_i = f_{m-n}^{\mathrm{sv}}(Z_i)$ for $i \in \{1,\ldots,n\}$.
  \STATE{\color{white} \bfseries Input:} Trained predictor $[\hat{L}_{m,\alpha}(\cdot; t), \hat{U}_{m,\alpha}(\cdot; t)]$;
  \STATE{\color{white} \bfseries Input:} Conformity scores $E(X_i,Y_i)$ for all $i \in \{n^{\mathrm{train}}+1,\ldots,n\}$.
  \FOR{$i = n^{\mathrm{train}}+1,\ldots,n$}
  \STATE{\bfseries Assign} each score $E(X_i,Y_i)$ to an appropriate frequency bin $B \in \mathcal{B}$ based on $Y_i$.
  \ENDFOR
  \FOR{$l = 1,\ldots,L$}
  \STATE{\bfseries Compute} the number $n_l$ of scores assigned to bin $B_l$.
  \STATE{\bfseries Compute} $\hat{Q}_{n_l, 1-\alpha}(B_l)$ as the $\lceil (1-\alpha) (n_l+1) \rceil$-th smallest score in bin $B_l$.
  \ENDFOR
  \STATE{\bfseries Set} $\hat{Q}^*_{n, 1-\alpha} = \max_{l}\hat{Q}_{n_l, 1-\alpha}(B_l)$.
  \STATE{\bfseries Set}  $X_{m+1} = \left(Z_{m+1}, \phi(Z_{n+1}, \ldots, Z_{m}) \right)$ as in~\eqref{eq:X-def}.
  \STATE {\bfseries Output:} a $(1-\alpha)$-level confidence interval $$\left[ f_{n}^{\mathrm{wu}}(Z_{m+1}) + \hat{L}_{m,\alpha}(X_{m+1}; \hat{Q}^*_{n, 1-\alpha}), f_{n}^{\mathrm{wu}}(Z_{m+1}) +  \hat{U}_{m,\alpha}(X_{m+1}; \hat{Q}^*_{n, 1-\alpha}) \right]$$ for the unobserved frequency $f_m(Z_{m+1})$ of $Z_{m+1}$ defined in~\eqref{eq:f-true}.
\end{algorithmic}
\end{algorithm}

\begin{theorem} \label{thm:coverage-cond}
  If the data $Z_1,\ldots,Z_{m+1}$ are exchangeable, the confidence interval output by Algorithm~\ref{alg:conformal-sketch-frequency} satisfies the frequency-conditional  property defined in~\eqref{eq:conf-int-cond}.
\end{theorem}

\textbf{Remark.}
The choice of the partition $\mathcal{B}$ involves an important trade-off. On the one side, frequency-conditional coverage~\eqref{eq:conf-int-cond} becomes stronger with finer partitions; a larger value of $|\mathcal{B}|$ tends to yield more reliable intervals.
On the other side, coarser partitions (smaller $|\mathcal{B}|$) enable a larger calibration sample within each bin, leading to tighter and more stable intervals.
Concretely, the illustrations described in this paper will adopt $|\mathcal{B}| = 5$, although finer partitions may be used when working with very large data sets.

As $|\mathcal{B}|$ should be small relative to the number $n$ of calibration data points to have short intervals, frequency-conditional coverage can only be guaranteed conditional on a relatively rough approximation of the true empirical frequency of a new query. Therefore, rarer queries may still suffer from lower coverage compared to more common queries within the same frequency bin, as we shall see empirically in Section~\ref{sec:experiments}.
This remaining limitation motivates the more sophisticated approach presented below, which is designed to guarantee valid coverage for a sufficiently large fraction of all distinct queries occurring in random test set with repetitions, regardless of their relative frequencies.

\section{Confidence intervals with valid coverage for distinct  queries} \label{sec:coverage-unique}

Section \ref{51} describes our methodology for constructing confidence intervals with valid coverage for distinct queries.
Then, Sections \ref{52} and \ref{53}  study some of its robustness properties.

\subsection{Construction of confidence intervals with coverage for distinct queries}
\label{51}

First, we introduce some notations.
Recall that a \emph{multiset} $V$ of objects $\{v_1,\ldots,v_m\}$ is simply the set of $v_1,\ldots,v_m$
 with repetitions.
Since we are dealing with settings where there are potentially a lot of repeated values, 
it is helpful to refer to the {multiset} 
 $\mathcal{Z}^{\text{cal}}$  of calibration data points $Z_i$ for all $i \in \mathcal{I}^{\text{cal}} = \{n^{\text{train}}+1,\ldots,n\}$, 
 for an appropriate $n < m$. 
 As above, we define $n^{\mathrm{cal}}$ as the cardinality of $\mathcal{I}^{\text{cal}}$.

Next, for some $M>0$, we aim for coverage for distinct queries among $M$  new queries.
Therefore, we consider a multiset $\mathcal{Z}^{\text{test}}$ of $M$ queries, indexed by $\mathcal{I}^{\text{test}} = \{m+1,\ldots,m+M\}$, which we assume to be sampled from $P_Z$ exchangeably with one another as well as with the $m$ sketched data points. This generalizes the setting considered so far, where we had considered $M=1$.

Define also $\textsc{Unique}(\mathcal{Z}^{\text{test}}) \subseteq \mathcal{Z}^{\text{test}}$ as the subset of unique objects in $\mathcal{Z}^{\text{test}}$.
Then, we formalize  ``coverage over uniques" by first sampling  $Z^*$ from the uniform distribution over $\textsc{Unique}(\mathcal{Z}^{\text{test}})$:
\begin{align}  \label{eq:model-unique}
\begin{split}
  Z_1, \ldots, Z_m, Z_{m+1}, \ldots, Z_{m+M} & \overset{\text{exch.}}{ \sim} P_Z, \\
  Z^* & \;\;\sim  \text{Uniform}\left[ \textsc{Unique}(Z_{m+1}, \ldots, Z_{m+M} ) \right].
\end{split}
\end{align}
Then, the goal is to construct a confidence interval $[\hat{L}_{m,\alpha}(Z^*), \hat{U}_{m,\alpha}(Z^*)]$ achieving coverage of $f_m(Z^*)$ over the random draw of $Z^*$, i.e., on average over the uniques in the test set:
\begin{align} \label{eq:conf-int-unique}
  \mathbb{P}^{*}\left[\hat{L}_{m,\alpha}(Z^*) \leq f_m(Z^*) \leq \hat{U}_{m,\alpha}(Z^*) \right] \geq 1-\alpha,
\end{align}
for any desired $\alpha \in (0,1)$. Above, the probability $\mathbb{P}^{*}$ is taken with respect to $Z_1,\ldots,Z_{m+M}$ as well as to the randomness in $Z^*$, according to the model defined in~\eqref{eq:model-unique}.
Equations~\eqref{eq:model-unique}--\eqref{eq:conf-int-unique} say that our goal is to cover at least a fraction $1-\alpha$ of the distinct queries in the test set; on average over the distribution of the test and calibration data.
In the special case of a test set with cardinality $M=1$, the property in~\eqref{eq:conf-int-unique} reduces to marginal coverage.

To achieve~\eqref{eq:conf-int-unique} with any value of $M$, we follow an approach inspired by~\cite{dunn2018distribution,park2022pac}. 
We randomly partition the calibration data into $G = \lfloor n^{\mathrm{cal}} / M \rfloor$ multisets $\mathcal{Z}^{\text{cal}}_g$, for $g \in [G]:=\{1,\ldots,G\}$, called \emph{calibration shards}. Without loss of generality, assume the cardinality of each $\mathcal{Z}^{\text{cal}}_g$ is $M$.
For our method to be powerful, we will need $n^{\mathrm{cal}} > M$, and ideally we would like $n^{\mathrm{cal}} \gg M$; or equivalently a large $G$.

Following the same notation as above, let $\textsc{Unique}(\mathcal{Z}^{\text{cal}}_g) \subseteq \mathcal{Z}^{\text{cal}}_g$ denote the subset of unique objects in the calibration shard $\mathcal{Z}^{\text{cal}}_g$, for all $g \in [G]$.
Then, for each $g \in [G]$, pick an element from each calibration shard $\mathcal{Z}^{\text{cal}}_g$ uniformly at random and call it $\tilde{Z}_g \in \mathcal{Z}$.
By construction, the shard-unique element pairs $(\mathcal{Z}^{\text{cal}}_g, \tilde{Z}_g)$, $g\in [G]$, are exchangeable with one another as well as with $(\mathcal{Z}^{\text{test}}, Z^*)$, for all $g \in [G]$. Therefore, a confidence interval $[\hat{L}_{m,\alpha}(Z^*), \hat{U}_{m,\alpha}(Z^*)]$ satisfying~\eqref{eq:conf-int-unique} can be obtained by applying the method from Section~\ref{sec:methods-marginal} with the calibration set $\mathcal{Z}^{\text{cal}}$ replaced by $(\tilde{Z}_1, \ldots, \tilde{Z}_G)$.

This solution is outlined in Algorithm~\ref{alg:conformal-sketch-unique} and its theoretical validity is established by Theorem~\ref{eq:coverage-unique}.
Algorithm~\ref{alg:conformal-sketch-unique} is written as to potentially allow the size $M'$ of each of the $G$ calibration shards to be different from the size $M$ of the test set. This generalization of Algorithm~\ref{alg:conformal-sketch-unique} with $M' \neq M$ will be studied theoretically in the next section, and it is worth considering because one may sometimes be tempted to apply Algorithm~\ref{alg:conformal-sketch-unique} with $M' < M$ in practical applications with limited amounts of data. However, the remainder of this section will continue to focus on the standard choice of $M'=M$.

\begin{algorithm}[!htb]
   \caption{Conformalized sketching with valid coverage for distinct queries}
   \label{alg:conformal-sketch-unique}
\begin{algorithmic}
%  \STATE {\bfseries Input:} Same as for Algorithm~\ref{alg:conformal-sketch}, with $M$ random queries $Z_{m+1}, \ldots, Z_{m+M}$.
  \STATE {\bfseries Input:} Same as for Algorithm~\ref{alg:conformal-sketch}, with query $Z^*$.
  \STATE {\color{white} \bfseries Input:} Calibration set size $M'$.
  \STATE{\bfseries Compute} using Algorithm~\ref{alg:conformal-sketch-train}: \\
  \STATE{\color{white} \bfseries Input:} Data sketch $\phi$; a sparse counter $f_{n}^{\mathrm{wu}}(z), \forall z \in \mathcal{Z}$;
  \STATE{\color{white} \bfseries Input:} Variables $X_i = \left(Z_i, \phi(Z_{n+1}, \ldots, Z_{m}) \right)$ and $Y_i = f_{m-n}^{\mathrm{sv}}(Z_i)$ for $i \in \{1,\ldots,n\}$.
  \STATE{\color{white} \bfseries Input:} Trained predictor for computing nested intervals $[\hat{L}_{m,\alpha}(\cdot; t), \hat{U}_{m,\alpha}(\cdot; t)]$;
  \STATE{\color{white} \bfseries Input:} Conformity scores $E(X_i,Y_i)$ for all $i \in \{n^{\mathrm{train}}+1,\ldots,n\}$.
  \STATE{\bfseries Define} $G = \lfloor n^{\mathrm{cal}} / M' \rfloor$, where $n^{\mathrm{cal}} = n- n^{\mathrm{train}}$.
  \STATE{\bfseries Partition} at random $\{n^{\mathrm{train}}+1,\ldots,n\}$ into $G$ subsets $\mathcal{I}^{\text{cal}}_g$.
  \FOR{$g = 1,\ldots,G$}
  \STATE{\bfseries Pick} uniformly at random one value $Z_g^*$ from the set $\textsc{Unique}(\{Z_i\}_{i \in \mathcal{I}^{\text{cal}}_g})$.
  \STATE{\bfseries Set}  $X_g^* = \left(Z^*_g, \phi(Z_{n+1}, \ldots, Z_{m}) \right)$ as in~\eqref{eq:X-def}, and  $Y_g^* = f_{m-n}^{\mathrm{sv}}(Z^*)$.
  \STATE{\bfseries Set} $E^*_g = E(X_g^*,Y_g^*)$.
  \ENDFOR
  \STATE{\bfseries Compute} $\hat{Q}_{G, 1-\alpha}$ as the $\lceil (1-\alpha) (G+1) \rceil$-th smallest score in $\{E^*_g\}_{g=1}^{G}$.
%  \STATE{\bfseries Pick} $Z^*$ uniformly at random from $\textsc{Unique}(\{Z_{m+1}, \ldots, Z_{m+M}\})$.
  \STATE{\bfseries Set}  $X^* = \left(Z^*, \phi(Z_{n+1}, \ldots, Z_{m}) \right)$ as in~\eqref{eq:X-def}.
  \STATE {\bfseries Output:} a $(1-\alpha)$-level confidence interval for the frequency $f_m(Z^*)$ of $Z^*$ defined in~\eqref{eq:f-true}: $$\left[ f_{n}^{\mathrm{wu}}(Z^*) + \hat{L}_{m,\alpha}(X^*; \hat{Q}^*_{n^{\mathrm{cal}}, 1-\alpha}), f_{n}^{\mathrm{wu}}(Z^*) +  \hat{U}_{m,\alpha}(X^*; \hat{Q}^*_{n^{\mathrm{cal}}, 1-\alpha}) \right].$$
\end{algorithmic}
\end{algorithm}

\begin{theorem} \label{eq:coverage-unique}
  Assume the data $Z_1,\ldots,Z_{m+M}$ are exchangeable and the query $Z^*$ is sampled according to \eqref{eq:model-unique}. If Algorithm~\ref{alg:conformal-sketch-unique} is applied with parameter $M'$ equal to the test set size $M$, the output confidence interval satisfies the distinct-query coverage property defined in~\eqref{eq:conf-int-unique}.
\end{theorem}

\textbf{Remark.}
The cardinality $M$ of the query set controls the trade-off between the power and reliability of the confidence intervals output by Algorithm~\ref{alg:conformal-sketch-unique}, assuming the latter is applied with parameter $M'=M$ as prescribed by Theorem~\ref{eq:coverage-unique}.
On the one hand, smaller values of $M$ lead to tighter and more stable intervals due to a larger number $G$ of data points available for calibration. 
On the other hand, larger values of $M$ lead to stronger theoretical guarantees, as they reduce the dependence between the expected conditional coverage for a particular query and the population frequency of that query.
In general, we recommend that Algorithm~\ref{alg:conformal-sketch-unique} should be applied with values of $M$ so large as to result in a number $G$ of final calibration data points in the hundreds. Concretely, the numerical experiments presented in this paper will apply Algorithm~\ref{alg:conformal-sketch-unique} with values of $M$ allowing $G\ge 100$.

We conclude this section by emphasizing that Algorithm~\ref{alg:conformal-sketch-unique} and Algorithm~\ref{alg:conformal-sketch-frequency} differ in their formally stated goals (achieving distinct-query coverage and frequency-conditional coverage, respectively), but they are designed to mitigate the same limitation of confidence intervals with marginal coverage.
On the one hand, distinct-query coverage is intuitively more appealing and easier to explain compared to frequency-conditional coverage, as anticipated in Section~\ref{sec:frequency-coverage}. On the other hand, Algorithms~\ref{alg:conformal-sketch-unique} and Algorithm~\ref{alg:conformal-sketch-frequency} require a calibration set that is large relative to the size of the query set.
Therefore, the relative advantages of Algorithms~\ref{alg:conformal-sketch-frequency} and~\ref{alg:conformal-sketch-unique} in finite samples may not necessarily be straightforward to see, suggesting the need for a deeper theoretical study of Algorithm~\ref{alg:conformal-sketch-unique} (in the remainder of this section) as well as careful simulations (in Section~\ref{sec:app}).

\subsection{Robustness to sample inflation}
\label{52}

To better understand the benefits of Algorithm~\ref{alg:conformal-sketch-unique}, we study the robustness of its distinct-query coverage guarantee in situations where it is not applied with the default settings, due to a limited sample size.
In particular, we are interested in understanding what happens if the size $M'$ of the calibration shards
$\mathcal{Z}^{\text{cal}}_g$, for all $g \in [G]$, is smaller than
the test sample size $M$.
As mentioned, this scenario is motivated when we aim to reach a strong unique-coverage guarantee with a large $M$ despite having only a relatively small calibration sample size.

Let us begin the analysis by recalling the key modelling assumption used throughout this paper: all data points are sampled exchangeably from a discrete distribution $P_Z$ with support on some countable dictionary $\mathcal{Z}$.
To facilitate the analysis hereafter, we further assume the data are independent; that is, $Z_i \overset{\text{i.i.d.}}{\sim} P_Z$, for all $i \in [m+M]$.
Moreover, we denote
 $P_Z = \sum_{j \in \mathbb{N}} p_j \delta_{a_j} $, where the $a_j \in \mathcal{Z}$ are the distinct symbols in the dictionary $\mathcal{Z}$, while $p_j \ge 0$ are their respective probabilities for all $j \in \NN$, such that $\sum_{j \in \NN} p_j=1$.
%A discrete data distribution is a natural assumption in the context of sketching---practical hashing computations can only operate with discrete values---and it is the only interesting case to consider in a theoretical analysis of Algorithm~\ref{alg:conformal-sketch-unique}. In fact, the set of distinct values in any finite sample from a continuous distribution is almost surely the same as the original sample.

Let $V = \textsc{Unique}(\mathcal{Z}^{\text{test}})$ denote the set of unique values in the test set $\mathcal{Z}^{\text{test}}$, which contains all $Z_i$ indexed by the test index set $i\in \mathcal{I}^{\text{test}}$.
For any positive integers $k$ and $M$ such that $M \ge k$, let $C_{M,k}$ be the set of \emph{$k$-compositions} of $M$: these are the sequences $c=(c_1,\ldots,c_k)$ of positive integers $c_j \ge 1$ such that $\sum_{j=1}^k c_j = M$.
For instance, $(1,1,2)$ is a $k=3$-composition of $M=4$.
It is known that the number of such sequences is $|C_{M,k}| = \binom{M-1}{k-1}$; see e.g., \citet{riordan2012introduction}.
For instance, $(1,1,2)$, $(1,2,1)$, and $(2,1,1)$ are all $k=3$-compositions of $M=4$, and their number is $\binom{3}{1}=3$.

With this notation, we can characterize the probability distribution of the set $V$ of unique values among a random sample from $P_Z$; see Proposition~\ref{unipro} in Appendix~\ref{sec:theory-supp}. From there, we obtain the following result characterizing the distribution $U_Z^{[M]}$ of a uniformly sampled element $\zeta$ over the set of uniques $V$, when $V \sim P_Z^{[M]}$.
This result will be useful in our analysis of the robustness of Algorithm~\ref{alg:conformal-sketch-unique} to situations in which $M' \neq M$.
We are not aware of Propositions~\ref{unipro} and \ref{uni2} being known in the literature; we believe they may be of independent interest and could find uses in future analyses of coverage over unique/distinct elements.

\begin{proposition}[Uniform distribution over unique elements]\label{uni2}
Let $\mathcal{Z}^{\text{test}}$ be an i.i.d. sample of size $M$ from a discrete distribution
$P_Z = \sum_{i \in \mathbb{N}} p_j \delta_{a_j} $, where $a_j \in \mathcal{Z}$ are distinct, and $p_j\ge 0$ for all $j\in \NN$.
%Let $P_Z^{[M]}$ be the probability distribution of
%$\mathcal{Z}^{\text{test}}$
%Under the conditions of Proposition \ref{unipro},
Let $U_Z^{[M]}$  be the distribution of a uniformly sampled element $\zeta$ of $V = \textsc{Unique}(\mathcal{Z}^{\text{test}})$.
Then, for all  $j_1\in \NN$,
the probability mass function of $\zeta$ at
$a_{j_1}$ is
\begin{equation}\label{uz}
    U_Z^{[M]}(\zeta = a_{j_1})
    =
    \sum_{k=1}^M
    \frac1k
    \sum_{J=\{j_1,\ldots, j_k\}\subset \NN^k,\, |J|  = k}\,
    \sum_{c\in C_{M,k}} \binom{M}{c_1\,c_2\,\ldots c_k}
    p_{j_1}^{c_1}\cdots p_{j_k}^{c_k}.
\end{equation}
In particular, $U_Z^{[1]}=U_Z^{[2]} = P_Z$, and for all  $j_1\in \NN$,
\begin{align*}
    &U_Z^{[3]}(\zeta = a_{j_1})
    =\frac
    {p_{j_1}(2p_{j_1}^2-3p_{j_1}+3)}
    {2}
    +
    \frac{p_{j_1}}{2}
    \sum_{\{j_2, j_3\}\subset (\NN\setminus\{j_1\})^2,\, |J|  = 2}\,
     p_{j_2} p_{j_3}.
\end{align*}
\end{proposition}

    Proposition \ref{uni2} suggests that one should generally expect to lose coverage over distinct queries when applying Algorithm~\ref{alg:conformal-sketch-unique} with a calibration set size $M'$ that is different from the size $M$ of the test set.
    Indeed, the $U_Z^{[M]}$-probability of
    the event $\zeta = a_{j}$ can either increase or decrease as a function of $M$, depending on
    the probability $p_{j}$ of $a_j$ under $P_Z$.
    %Indeed, for all $i$,  $U_Z^{[1]}(a_{i}) =p_{i}$.
    To see this, define the function $\tau:[0,1]\to [0,1]$, such that for all $p\in[0,1]$,
    \begin{align}\label{t}
    &\tau(p)
    =\frac
    {p(2p^2-3p+3)}
    {2}.
\end{align}
A plot of $\tau$ is in Figure \ref{fig:t-nu}~(a), Appendix~\ref{app:figures}.
    Then, for $P_Z$ taking only two possible distinct
    values $a_1$ and $a_2$, with probabilities $p_1$ and $p_2$, respectively, Proposition \ref{uni2} implies that
    for $j=1,2$, $U_Z^{[3]}(\zeta = a_{j}) =\tau(p_{j})$.
Now, for $p\in[0,1/2)$, $\tau(p)<p$,
while for $p\in(1/2,1]$, $\tau(p)>p$.
Assuming $p_1<p_2$,
we have $U_Z^{[3]}(\zeta = a_{1})<U_Z^{[2]}(\zeta = a_{1})$,
while
$U_Z^{[3]}(\zeta = a_{2})>U_Z^{[2]}(\zeta = a_{2})$.
Thus, the probability of $\zeta = a_{i}$
can either increase or decrease as a function of $M$, depending on $p_i$.
    Hence, we expect that the probability of the coverage event using calibration data points of size $M'$, which is a union of such elementary events,
    can also increase or decrease as a function of $M$.

More specifically, let $\mathcal{E} = \{\hat{L}_{m,\alpha}(Z^*) \leq f_m(Z^*) \leq \hat{U}_{m,\alpha}(Z^*)\}$
be the coverage event from \eqref{eq:conf-int-unique},
whose probability is lower bounded in
Theorem \ref{eq:coverage-unique}.
Let the random variables $Z_i$, $i \in \mathcal{I}^{\text{cal}}$
that constitute the calibration set of size $n^{\mathrm{cal}}$
and
$i \in \mathcal{I}^{\text{test}}$ that are test set of size $M$
be i.i.d.~according to $P_Z$.
The probability of coverage can be written in terms of the variables
$\tilde Z_g$, for $g\in [G]$, chosen from the calibration shards,
which are i.i.d.~following the distribution $U_Z^{[M']}$---abbreviated as
$\tilde Z_{1:G}\sim (U_Z^{[M']})^{|G|}$---and an independent random variable
$Z^*$ chosen uniformly over the test set,
which follows the distribution
$U_Z^{[M]}$, as
\begin{equation}\label{ee}
 \mathbb{P}_{Z^*\sim U_Z^{[M]},\,\tilde Z_{1:G}\sim (U_Z^{[M']})^{|G|} }[\mathcal{E}]
 =
 \mathbb{E}_{Z^*\sim U_Z^{[M]}}
 \mathbb{P}_{\tilde Z_{1:G}\sim (U_Z^{[M']})^{|G|} }
 [\mathcal{E}]
 =
 \mathbb{E}_{Z^*\sim U_Z^{[M]}}\, e(Z^*).
\end{equation}
Above, we defined $e(Z^*) = \mathbb{P}_{\tilde Z_{1:G}\sim (U_Z^{[M']})^{|G|} }
 [\mathcal{E}]$
 to be the conditional probability of the coverage event $\mathcal{E}$, given $Z^*$.
Theorem \ref{eq:coverage-unique} says the expectation in~\eqref{ee}
is at least $1-\alpha$ if $M' = M$.
However, when $M'\neq M$,
we have showed that $U_Z^{[M]}$ can be different from $U_Z^{[M']}$.
Thus the above expectation of $e(Z^*)$ may decrease, and the method may lose coverage if $M'\neq M$.

    Aiming to understand the extent by which the coverage can be affected, we
    let $\mathcal{P}_{\NN}(\mathcal{Z};K)$ be the set of discrete probability distributions
    over $\mathcal{Z}$
    supported on at most $K$ distinct values.
    This is of interest especially because smaller $K$ leads to a more analytically tractable theory, as described below.
    Then, we introduce the quantity
    $$\Delta(M,M';K) =
    \sup_{P_Z\in \mathcal{P}_{\NN}(\mathcal{Z};K)
    }\, \sup_{j \in \NN}
    \left|U_Z^{[M]}(\{a_j\})
    -U_Z^{[M']}(\{a_j\})\right|,
    $$
    which measures the worst-case
    difference between
    the probabilities of observing
    a value
    $a_{j}$
    according to the distributions
    $U_Z^{[M]}$ and $U_Z^{[M']}$.
    Here, we are thinking of $U_Z^{[M']}$ as the calibration distribution and $U_Z^{[M]}$ as the test distribution.
    Thus, if our conformal prediction algorithm outputs sets of size at most $s\ge 0$, then the probability of those sets
     differs by at most
    $s \cdot \Delta(M,M';K)$
    between the training at  test distributions.

    Studying $\Delta(M,M';K)$
    seems challenging in general, as it involves
    maximizing differences of probabilities given in \eqref{uz}.
    These are nontrivial quantities to deal with, because (a) large values of $M$ lead to large-degree polynomials in the expressions for the $p_{j}$s, and (b) large values of $K$ lead to large numbers of degrees of freedom (i.e., many different $p_{j}$s).
    
    To illustrate some of the difficulties, 
    consider for instance the case $K=3$. 
    Denoting the three objects in $P_Z$ by $a_1$, $a_2$, $a_3$, one can verify using \eqref{q2} in Proposition~\ref{unipro} and \eqref{uz} in Proposition~\ref{uni2} that, for $j=1,2,3$,
    \begin{equation*}
    U_Z^{[M]}(\zeta = a_{j})
    =
    \frac{1+p_{j}^M -(1-p_{j})^M+ 1/2 \sum_{l\neq j} (p_j+p_l)^M }3.
\end{equation*}
Therefore,
    \begin{align*}
    \Delta(M,M';3)  =&
    \frac13
    \sup_{p,q,r\in[0,1]: p+q+r=1}
    \left|
    p^M-(1-p)^M + 1/2[(p+q)^M+(p+r)^M]\right.\\
    &\qquad\left.- \left[p^{M'}-(1-p)^{M'} 1/2[(p+q)^{M'}+(p+r)^{M'}] \right]\right|.
    \end{align*}
Denoting $a = p+q$, $b=p+r$, and noting $p = a+b-1$, 
with the function 
$$
\Lambda_M(a,b) = (a+b-1)^M-(2-a-b)^M + 1/2(a^M+b^M),
$$
we find
$$\Delta(M,M';3)  =
    \frac13
    \sup_{a,b\in[0,1]: a+b\ge 1}
    \left|
    \Lambda_M(a,b)- \Lambda_{M'}(a,b)\right|.
$$
This expression does not appear to be straightforward to analyze using standard tools. In particular, setting the gradients of the objective to zero in order to understand the maximizing $a,b$ does not seem to lead to a tractable answers, due to the high order polynomials involved.
Moreover the problem seems to get even more complicated for larger $K$, with more complicated polynomials to analyze.

The above results have illustrated some of the theoretical challenges that arise when analyzing $\Delta(M,M';K)$.
  Therefore, in order to provide some theoretical results, we focus on the
    simpler but still non-trivial case of $K=2$; i.e., we imagine there are only two distinct objects in the population $P_Z$.
However in our experiments we will continue to use general $K$, and will see experimental results that broadly agree with the message of the theory.

   To do this,
we can assume without loss of generality that 
the size $M'$ of the available calibration shards
$\mathcal{Z}^{\text{cal}}_g$, for all $g \in [G]$, is smaller than
the test sample size $M$, i.e.,
$M > M'$,
as $\Delta$ is symmetric in $M,M'$.
Moreover, we can also
assume without loss of generality that $M'\ge 2$,
    since
    $U_Z^{[1]}=U_Z^{[2]}$ and thus the cases $M'=1$ and $M'=2$ are equivalent.
    For fixed $M>M'\ge 2$, our theoretical results are presented in terms of  the function $h:[0,\infty)\to \mathbb{R}$ defined, for all $\delta\in[0,\infty)$, as
    \begin{equation}\label{h}
    h(\delta)=\ln\frac{1+\delta^{M-1}}{1+\delta^{M'-1}}
    - (M-M')\ln(1+\delta).
    \end{equation}
    This function comes up after suitable calculations when maximizing $\Delta$.
    Our next result characterizes $\Delta(M,M';2)$
    based on the function $h$.
    The proof relies on carefully studying the monotonicity properties of $\Delta$ using calculus; see Section \ref{pf:d2}.

    \begin{proposition}[Characterizing $\Delta(M,M';2)$]\label{d2}
    Fix $M>M'\ge 2$ and take the function $h$ as defined in~\eqref{h}.
    There is a unique solution $\delta_* \in [0,1]$ to $h(\delta_*)=\ln(M'/M)$,
    and
    $$\Delta(M,M';2) =
    \frac12
    \left|
    \frac{1-\delta_*^M}{(1+\delta_*)^M} -
    \frac{1-\delta_*^{M'}}{(1+\delta_*)^{M'}}\right|.
    $$
    \end{proposition}

As an illustration,
we consider the setting where
%$M = M'+1$.
$M = aM'$, for some $a>1$.
%\ed{M' = aM}
This corresponds to applying Algorithm~\ref{alg:conformal-sketch-unique} using calibration shards of size $M'$, with $M'$ being smaller than the target test set size $M$ by a factor $1/a$.
Naturally, one would like to know how low the coverage can be in this case  compared to the ideal situation in which $M'=M$.
Our next result shows that the loss in coverage may remain relatively bounded, as long as $a$ is moderate and $M$ is large.
The proof leverages Proposition~\ref{d2} and relies on a detailed analysis of the polynomial equation satisfied by  $\delta_*$; see Section \ref{pf:d3}.

\begin{corollary}[Asymptotics of $\Delta(M,M/a;2)$]\label{d3}
For $M\ge a\max \{2/(a-1),2+ \log_2[a/(a-1)]\}$,
with $\nu(a):=a^{-\frac{1}{a-1}}(1-\frac{1}{a})$
and 
\begin{align}\label{b}
&\beta(M,a):=
    2^{3-M/a} a^{-1/(a-1)}/(a-1)
    +2 [M(1-1/a)]^{-M/a},
\end{align} 
we have  $|\Delta(M,M/a;2)-\nu(a)| \le \beta(M,a)$.
\end{corollary}
The error term $\beta(M,a)$ is exponentially small in $M$ for a fixed $a>1$, and can be viewed as negligible. 
Moreover, the main term $\nu(a)$ is also quite small; for instance, if $a=1.1$, we have $\nu(a) \approx 0.035$.
 Combined with \eqref{ee} and Theorem \ref{eq:coverage-unique}, Corollary~\ref{d3} implies that the coverage over unique values for a test set of size $M$ and calibration sets of size $M'=M/a$ satisfies, for $M$ large enough as specified in Corollary \ref{d3},
$$
\mathbb{P}_{Z^*\sim U_Z^{[M]},\,\tilde Z_{1:G}\sim (U_Z^{[Ma]})^{|G|} }[\mathcal{E}]
\ge
1-\alpha
-2\cdot \nu(a)
-\beta(M,a).
$$
%\note{Edgar: Can we use 'inf' and equality to make this result look sharper?}
This immediately gives the following result, which guarantees that the coverage of Algorithm~\ref{alg:conformal-sketch-unique} when applied with $M' \neq M$ is correct up to a small error term $2\nu(a)$.

\begin{theorem} \label{eq:coverage-unique-robust}
  Assume that the data $Z_1,\ldots,Z_{m+M}$ are exchangeable and let  Algorithm~\ref{alg:conformal-sketch-unique} be applied at the nominal coverage level $\alpha \in (0,1)$ with parameter $M'=M/a$ for some $a > 1$, where $M$ is the size of the test set. Then,
  the output confidence interval satisfies the distinct-query coverage property defined in~\eqref{eq:conf-int-unique} at level
$ \alpha +  2\cdot \nu(a) + \beta(M,a)$, where $\nu(a)=a^{-1/(a-1)}(1-1/a)$ and $\beta$ is defined in \eqref{b}.
\end{theorem}

To better understand this result, it helps to look at the plot of the function $\nu$ shown in Figure \ref{fig:t-nu}~(b).
For instance, if $a=1.2$, we have $\nu(a) \approx 0.067$; therefore, a 95\% nominal coverage level may result empirical coverage over distinct queries that is as low as 80.6\% when $M=100$. If $a=1.1$,  we have, as already mentioned, $\nu(a) \approx 0.035$; therefore, a 95\% nominal coverage level may result empirical coverage over distinct queries that is as low as 87.0\% when $M=100$.
Of course, Theorem~\ref{eq:coverage-unique-robust} gives a conservative lower bound for the coverage over distinct queries which refers to the worst-case scenario over all data distributions $P_Z$. In practice, Algorithm~\ref{alg:conformal-sketch-unique} applied with $M' < M$ may sometimes result in higher coverage than anticipated by Theorem~\ref{eq:coverage-unique-robust}, as we will see empirically in Sections~\ref{sec:experiments}--\ref{sec:app}.

% \begin{figure}[!htb]
% \begin{center}
% \includegraphics[scale=0.5]{figures/nu.png}
% \caption{}
% \label{fig:nu}
% \end{center}
% \end{figure}

\subsection{Robustness to distribution shift}
\label{53}

An additional advantage of the distinct-query coverage property defined in~\eqref{eq:model-unique} is that it tends to be more ``robust'' to
certain types of distribution shift compared to the standard notion of marginal coverage.
In other words, if Algorithm~\ref{alg:conformal-sketch-unique} is applied in a situation where the queried objects are not sampled from the same distribution as the sketched data, its effective coverage over distinct queries may be lower than the ideal $1-\alpha$ expected under perfect exchangeability, but this loss may not be as large as that of Algorithm~\ref{alg:conformal-sketch}.

Recall that $U_Z^{[M]}$ is the distribution of unique values in a sample $Z_1,\ldots,Z_M$ of size $M$ from $P_Z$; and that the coverage over uniques from \eqref{eq:conf-int-unique} refers to a test data point from $U_Z^{[M]}$.
The next result establishes that,
in the special case of a support of size $K=2$ studied above,
the probabilities
shift less
in the worst case
under the distribution  $U_Z^{[M]}$ of unique values
than under the original distribution $P_Z$, for a large range of probability values $p_i$ of $P_Z$.
Experiments presented in Sections~\ref{sec:experiments}--\ref{sec:app} 
show similar results for larger $K$ as well.
The proof relies on the mean value theorem and can be found in Section \ref{pf:uni3}.

\begin{theorem}[Bounding the effect of distribution shift]\label{thm:shift}
Let $Z$ and $Z'$ take two values
with probabilities $p_1,p_2$, and  $p_1',p_2'$,  respectively.
For $M\ge 3$,
let $U_Z^{[M]}$ be the distribution of a uniformly sampled element over $\textsc{Unique}(V)$, when $V \sim P_Z^{[M]}$; and define $U_{Z'}^{[M]}$ similarly.
Define $c\in (0,1/2)$ as the unique solution of
\begin{equation}\label{c}
    c^{M-1}+(1-c)^{M-1} = \frac2M.
\end{equation}
Let
$$
S_c
=
\left\{
P_Z=(p_1,p_2):\,
p_j\in (c, 1-c), \, j=1,2
\right\}.
$$
Then, for all $P_Z,P_{Z'}\in S_c$, with $P_Z\neq P_{Z'}$,
%Following \eqref{ee}
\begin{align*}
 &\sup_{
 \mathcal{E}\subset \{a_1,a_2\}^{|G|+1}}
 \left|\mathbb{P}_{Z^*\sim U_Z^{[M]},\,\tilde Z_{1:G}\sim \left(U_Z^{[M]}\right)^{|G|} }[\mathcal{E}]
 - \mathbb{P}_{Z^*\sim U_{Z'}^{[M]},\,\tilde Z_{1:G}\sim \left(U_Z^{[M]}\right)^{|G|} }[\mathcal{E}]\right| <\\
 &
 \qquad\qquad
 \sup_{
 \mathcal{E}\subset \{a_1,a_2\}^{|G|+1}}
 \left|\mathbb{P}_{Z^*\sim P_Z,\,\tilde Z_{1:G}\sim P_Z^{|G|} }[\mathcal{E}]
 - \mathbb{P}_{Z^*\sim P_{Z'},\,\tilde Z_{1:G}\sim P_Z^{|G|} }[\mathcal{E}]\right|.
\end{align*}
\end{theorem}

In other words,
since  the coverage event $\mathcal{E} = \{\hat{L}_{m,\alpha}(Z^*) \leq f_m(Z^*) \leq \hat{U}_{m,\alpha}(Z^*)\}$ from \eqref{eq:conf-int-unique} is included among the sets where the supremum is evaluated,
Theorem~\ref{thm:shift} tells us that the
coverage of the sets output by Algorithm~\ref{alg:conformal-sketch-unique}
tends to be relatively stable for certain classes of data distributions $P_Z$.
Specifically, for a given $P_Z$, the change in coverage when shifting from the distribution of uniques $U_{Z}^{[M]}$ to the distribution of uniques $U_{Z'}^{[M]}$ is strictly smaller, in the worst case, than the corresponding change in coverage when shifting from $P_Z$ to $P_{Z'}$.
This suggests that Algorithm~\ref{alg:conformal-sketch-unique} may be relatively robust to distribution shifts in the query set.

Now, we can try to better understand the family of $P_Z$ over which the distribution of unique values is more stable. Since $c< 1/2$, we have $c< 1-c$; thus, it follows from \eqref{c} that $(1-c)^{M-1} \le 2/M\le 2(1-c)^{M-1}$, which can be rearranged to obtain:
$$1-(2/M)^{1/(M-1)} \le c\le 1-(1/M)^{1/(M-1)}.$$
Therefore, $c = O(M^{-1} \ln M)$ for large $M$.
This implies that the distribution $U_Z^{[M]}$ of the unique values is less affected by changes in the distribution of probabilities in $P_Z$ than $P_Z$ itself, for a large range of possible values of $p_j$ from $O(M^{-1} \ln M)$ to $1-O(M^{-1} \ln M)$.

While Theorem~\ref{thm:shift} focuses on a special case in which the data distribution $P_Z$ has support on only two possible objects in order to simplify the theoretical analysis, the relative robustness of Algorithm~\ref{alg:conformal-sketch-unique} to distribution shift in more general settings is supported by empirical experiments, as shown in Sections~\ref{sec:experiments}--\ref{sec:app}.

\section{Experiments with synthetic data} \label{sec:experiments}

Section~\ref{sec:app-synthetic} describes experiments in which we seek marginal or frequency-conditional coverage using the CMS sketch.
Section~\ref{sec:app-synthetic-cu} presents similar experiments based on the CMS-CU.
Section~\ref{sec:app-synthetic-distinct} focuses on coverage for distinct queries. 
Section~\ref{sec:app-synthetic-shift} studies robustness to distribution shifts.
Section~\ref{sec:ml-sketching} applies our methods in combination with a learning-based sketch \citep{bertsimas2021frequency} and with the CS sketch \citep{charikar2002finding}.
Section~\ref{sec:experiments-extra} summarizes additional results presented in the appendix.

%Section \ref{51} describes our methodology for constructing confidence intervals with valid coverage for distinct queries.
%Then, Sections \ref{52} and \ref{53}  study some of its robustness properties.

\subsection{Marginal and frequency-conditional coverage with the CMS} \label{sec:app-synthetic}

We apply Algorithm~\ref{alg:conformal-sketch-frequency} in combination with the CMS \citep{cormode2005improved} on synthetic data.
The CMS is implemented using $d=3$ random hash functions of width $w=1000$.
As this sketch already gives a deterministic upper bound for any frequency query, the goal of our experiments is to compute corresponding lower bounds for 95\% coverage.

The data are generated i.i.d.~from a Zipf distribution---a standard option to describe power-law tail behavior~\citep{zipf2016human}.
Power-law distributions are observed in many scientific applications, and they are useful to understand many natural and social phenomena \citep{ferrer2001two,adamic2002zipf,clauset2009power,muchnik2013origins}.
To be precise, we sample a random query $Z_{m+1}$ and $m = 100,000$ data points according to the law $\P{Z_i = z} = z^{-a}/\zeta(a)$ for all $z \in \{1,2\ldots,\}$, where $\zeta$ is the Riemann Zeta function and $a>1$ controls the power-law tail behavior.

Prior literature has already studied the problem of uncertainty estimation for frequency queries based on the CMS \citep{cormode2020small,ting2018count,cai2018bayesian, dolera2021bayesian}, which provides us with three informative benchmarks.
The first one is the {\it classical} 95\% lower bound \citep{cormode2005improved} obtained by treating the data as fixed and modeling only the randomness in hash functions, as explained in Appendix~\ref{sec:cms-classical-lower}. This approach is often too conservative when applied to non-adversarial data \citep{ting2018count}.

The second benchmark is the {\it Bayesian} method of~\citet{cai2018bayesian}, which assumes a non-parametric Dirichlet process prior for the distribution of the data, estimates its scaling parameter by maximizing the marginal likelihood of the observed sketch, and then computes the posterior distribution of the queried frequency. The lower 5\% quantile of this posterior distribution is taken as the lower confidence bound for a frequency query.
The third benchmark is the bootstrap method of~\citet{ting2018count}, which is also designed for the CMS and does not extend to other non-linear sketches (which we will study later).

Algorithm~\ref{alg:conformal-sketch-frequency} is applied using the first $n=5000$ data points for warm-up and then sketching the remaining $95,000$ data points with the CMS, as explained in Section~\ref{sec:sketch-problem}.
Two versions of our method are compared: one based on fixed-width conformity scores described in Section~\ref{sec:conf-scores-fixed}, and one based on the adaptive-width scores from Section~\ref{sec:conf-scores-adaptive}. The latter are implemented using an isotonic distributional regression model \citep{henzi2019isotonic}.
In each case, the conformity scores are evaluated separately within $L=5$ frequency bins, seeking frequency-range conditional coverage~\eqref{eq:conf-int-cond}. The bins are determined so that each contains approximately the same probability mass,
by partitioning the range of frequencies for the objects tracked exactly according to the observed empirical quantiles.

Each method is applied to construct one-sided confidence intervals for the frequencies of $10,000$ random queries, sampled i.i.d.~from the same distribution as the sketched data. The confidence intervals are evaluated based with two metrics: their average {\it length} and their {\em empirical coverage}---the latter is the proportion of queries for which the true frequency is covered. These performance metrics are averaged over 10 independent experiments.

\begin{figure}[!htb]
\begin{center}
\includegraphics[width=0.8\linewidth]{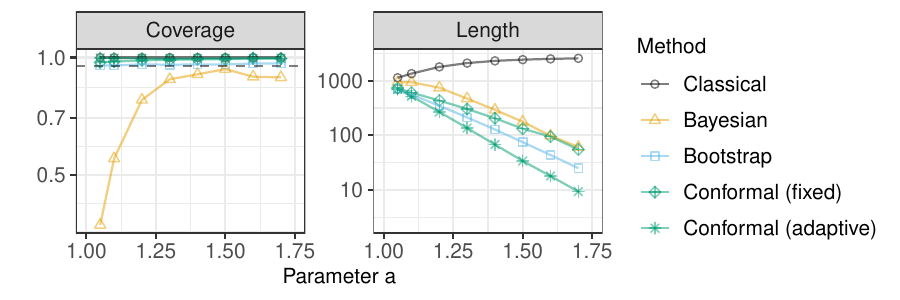}
\caption{Performance of 95\% confidence intervals with simulated Zipf data sketched with the CMS. The results are shown as a function of the Zipf tail parameter $a$. }
\label{fig:exp-zipf-marginal}
\end{center}
\end{figure}

Figure~\ref{fig:exp-zipf-marginal} compares our method to three benchmarks on the Zipf data. All methods achieve marginal coverage~\eqref{eq:marginal-coverage}, with the exception of the Bayesian approach whose prior does not match the true data distribution in this case, especially when the tail parameter $a$ is small.
As expected, the classical approach turns out to be very conservative, while the bootstrap and conformal methods provide relatively informative confidence intervals, particularly when the tail parameter $a$ is larger and hash collisions become rarer.

\begin{figure}[!htb]
\begin{center}
\includegraphics[width=\linewidth]{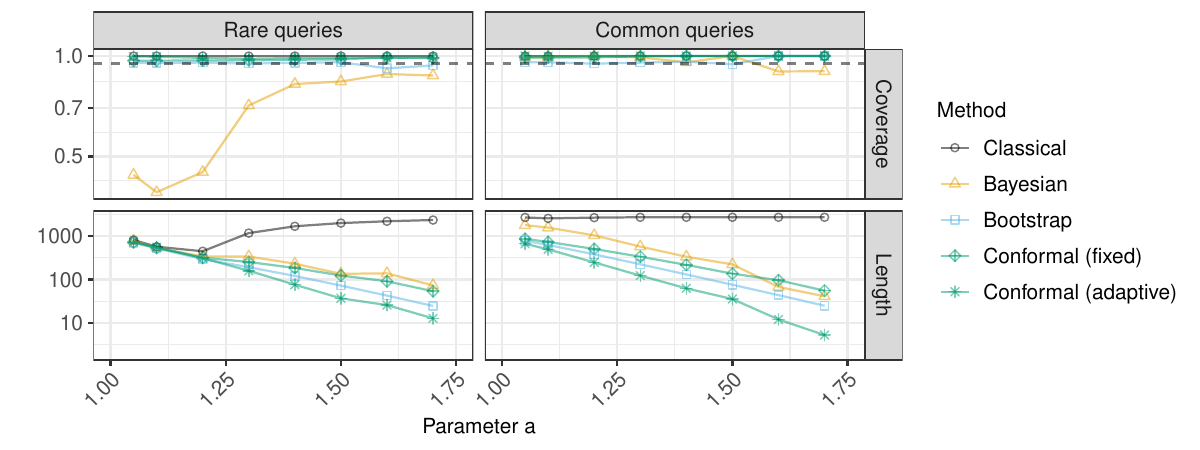}
\caption{Performance of confidence intervals stratified by the true query frequency. Left: frequency below median; right: above median. Other details are as in Figure~\ref{fig:exp-zipf-marginal}.}
\label{fig:exp-zipf-conditional}
\end{center}
\end{figure}

The conformal intervals produced by Algorithm~\ref{alg:conformal-sketch-frequency} are the shortest among all alternatives, especially if they use adaptive conformity scores.
The standard errors are omitted because they are relatively small but would clutter the display.
Further, Figure~\ref{fig:exp-zipf-conditional} stratifies the results of Figure~\ref{fig:exp-zipf-marginal} based on the true frequency of each random query.
This shows that Algorithm~\ref{alg:conformal-sketch-frequency} produces valid inferences for both rarer and more common queries, at least within the resolution level considered here. This should not be surprising given that our method controls the frequency-conditional coverage defined in~\eqref{eq:conf-int-cond}.
Note that this notion of frequency-conditional coverage would not necessarily be satisfied if the conformal confidence intervals were constructed using Algorithm~\ref{alg:conformal-sketch}, which guarantees only marginal coverage~\eqref{eq:marginal-coverage}, instead of Algorithm~\ref{alg:conformal-sketch-frequency}; see Figure~\ref{fig:exp-zipf-conditional-1bin} in Appendix~\ref{app:figures}.

As explained in Sections~\ref{sec:problem-statement} and~\ref{sec:frequency-coverage}, frequency-range conditional coverage~\eqref{eq:conf-int-cond} is not always fully satisfactory. In practice, one may ask instead what is the average proportion of unique queries in the random test set for which the conformal confidence intervals constructed above indeed provide coverage---this is a more challenging notion of coverage that could be guaranteed rigorously by applying the alternative methods from Section~\ref{sec:coverage-unique}.
Figure~\ref{fig:exp-zipf-unique-cms-bins5} addresses this question by reporting performance metrics for the intervals output by Algorithm~\ref{alg:conformal-sketch-frequency} that are analogous to those in Figure~\ref{fig:exp-zipf-marginal} but evaluated only on distinct queries. The results are encouraging in this case: Algorithm~\ref{alg:conformal-sketch-frequency} produces intervals that are empirically valid for more than 95\% of unique queries across all values of the Zipf tail parameter considered here.
Unsurprisingly, though, the same is not true of weaker conformal confidence intervals produced by Algorithm~\ref{alg:conformal-sketch}, which target the weaker notion of marginal coverage \eqref{eq:marginal-coverage}; see Figure~\ref{fig:exp-zipf-unique-cms-bins1}.
Further, we shall see in the next section that even Algorithm~\ref{alg:conformal-sketch-frequency} can practically fail to produce intervals that are valid for a high proportion of unique queries if the data are compressed by a more powerful sketching algorithm; this is what motivates the methods presented in Section~\ref{sec:coverage-unique}, which we will apply later in Section~\ref{sec:app-synthetic-distinct}.

\subsection{Non-linear sketching with the CMS-CU} \label{sec:app-synthetic-cu}

While prior research on uncertainty estimation for frequency queries focused on the CMS, the methods developed in this paper can accommodate any sketching algorithm.
Here, we begin to explore this flexibility by conducting experiments similar to those of Section~\ref{sec:app-synthetic} but with the CMS replaced by a non-linear variation known as the CMS with {\em conservative updates} (CMS-CU) \citep{estan2002new}.
We refer to Appendix~\ref{app:cms} for a review of this classical sketching algorithm.

Figures~\ref{fig:exp-zipf-marginal-cu} and \ref{fig:exp-zipf-conditional-cu} present results analogous to those of Figures~\ref{fig:exp-zipf-marginal} and \ref{fig:exp-zipf-conditional}, respectively, showing that Algorithms~\ref{alg:conformal-sketch} and~\ref{alg:conformal-sketch-frequency} still lead to shorter confidence intervals with valid coverage. Note that the benchmark approaches are not technically valid here because they were designed for the CMS and not the CMS-CU; nonetheless, their empirical performance remains qualitatively similar to that observed in Section~\ref{sec:app-synthetic}.
Unsurprisingly, our results also confirm that all methods considered here lead to shorter confidence intervals when applied with the CMS-CU instead of the CMS, consistently with the fact that the CMS-CU was designed to improve the compression efficiency by reducing the impact of random hash collisions; see Figure~\ref{fig:exp-zipf-sketch} for a direct comparison.
Thus, to provide a more practically relevant depiction of each method's performance, the experiments presented in the following sections will adopt the CMS-CU as the baseline sketch instead of the CMS.

We conclude this section by referring to Figures~\ref{fig:exp-zipf-unique-cms-cu-bins5} and~\ref{fig:exp-zipf-unique-cms-cu-bins1} in the appendix, which investigate the validity of our intervals based on the CMS-CU over distinct queries. These figures report on performance metrics analogous to those shown in Figures~\ref{fig:exp-zipf-unique-cms-bins5} and~\ref{fig:exp-zipf-unique-cms-bins1}, respectively. The results indicate that the intervals targeting marginal~\eqref{eq:marginal-coverage} or frequency-range conditional~\eqref{eq:conf-int-cond} coverage at the 95\% level tend to be valid for fewer than 95\% of all distinct queries, and that such lack of theoretical coverage is more evident now compared to when the data were sketched using the CMS. This observation motivates the experiments described in the next section, in which we apply the stronger methods presented in Section~\ref{sec:coverage-unique}.

\subsection{Coverage for distinct queries} \label{sec:app-synthetic-distinct}

This section investigates the performance of Algorithm~\ref{alg:conformal-sketch-unique}, our proposed method for constructing confidence intervals with guaranteed coverage for distinct queries.
These experiments follow the same setup as those in Section~\ref{sec:app-synthetic-cu}, simulating data from a Zipf distribution with tail parameter $a=1.5$.
The difference is that the coverage and length performance metrics are now averaged only on the distinct queries, $\textsc{Unique}(\mathcal{Z}^{\text{test}})$, from a random test set $\mathcal{Z}^{\text{test}}$ of size $M=100$.
Algorithm~\ref{alg:conformal-sketch-unique} is applied at level $1-\alpha=95\%$ using the fixed-width one-sided conformity scores described in Section~\ref{sec:conf-scores-fixed}, and varying $M'$, which controls the size of the calibration shards, as a control parameter between 1 and 100.

\begin{figure}[!htb]
\begin{center}
\includegraphics[width=0.8\linewidth]{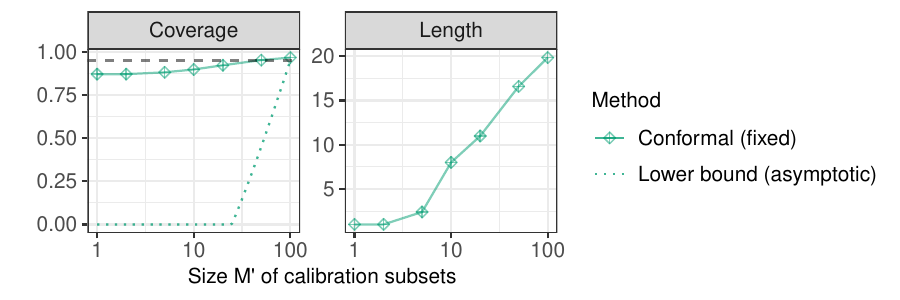}
\caption{Performance of confidence intervals for distinct queries in a test set of size 100, as a function of the parameter $M'$ of Algorithm~\ref{alg:conformal-sketch-unique}. The data are simulated from a Zipf distribution with tail parameter $a=1.5$ and sketched with the CMS-CU. Other details are as in Figure~\ref{fig:exp-zipf-marginal}.}
\label{fig:exp-zipf-unique-M}
\end{center}
\end{figure}

Figure~\ref{fig:exp-zipf-unique-M} confirms that the desired 95\% coverage for distinct queries~\eqref{eq:conf-int-unique} is achieved when Algorithm~\ref{alg:conformal-sketch-unique} is applied with $M' \approx M$, as predicted by Theorem~\ref{eq:coverage-unique}.
By contrast, the coverage for distinct queries is lower when $M'$ is small. This should not be surprising because Algorithm~\ref{alg:conformal-sketch-unique} reduces to Algorithm~\ref{alg:conformal-sketch} if $M'=1$, and the latter is designed to provide marginal coverage~\eqref{eq:marginal-coverage}, not coverage for distinct queries~\eqref{eq:conf-int-unique}. 
In fact, as shown in Figure~\ref{fig:exp-zipf-unique}, even Algorithm~\ref{alg:conformal-sketch-frequency}, which targets the relatively stronger notion of frequency-range conditional coverage~\eqref{eq:conf-int-cond}, does not always provide valid inference for distinct queries.

Finally, Figure~\ref{fig:exp-zipf-unique-M} also highlights that the distinct-query coverage practically achieved by applying on these data Algorithm~\ref{alg:conformal-sketch-unique} with smaller values of $M'$ is much higher then the worst-case asymptotic lower bound,  $\max(0,1- \alpha -  2\cdot \nu(100/M'))$, given by Theorem~\ref{eq:coverage-unique-robust}.

\subsection{Robustness to distribution shifts} \label{sec:app-synthetic-shift}

This section investigates the robustness of the confidence intervals output by Algorithms~\ref{alg:conformal-sketch} and~\ref{alg:conformal-sketch-unique} to distribution shifts in the query set.
These experiments follow the same setup as those in Section~\ref{sec:app-synthetic-distinct}, simulating data from a Zipf distribution with different values of the tail parameter.
The difference is that now the $M=100$ random test queries are sampled from a mixture distribution with two components.
The first component is the same Zipf distribution from which the sketched data are generated, while the second component is an independent continuous uniform distribution on $[0,1]$.

This setup is designed to study the effects of an extreme form of distributional shift, as objects sampled from the second component of the mixture are almost surely unique (up to rounding errors at machine precision) and are never previously observed in the integer-valued sketched data set.
The mixing proportion serves as a control parameter and it is varied from zero (no distribution shift) to one (full shift).
Note that this setup is not inconsistent with the original assumption that the data distribution has support on a discrete dictionary $\mathscr{Z}$, because even (approximately) uniform random numbers on a computer are in truth discrete.

\begin{figure}[!htb]
\begin{center}
\includegraphics[width=0.8\linewidth]{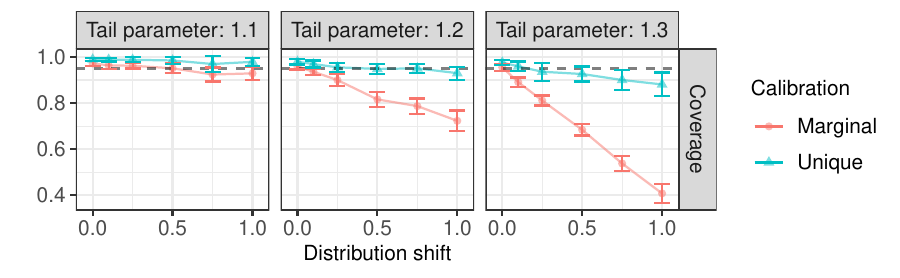}
\caption{Performance of conformal confidence intervals with marginal (Algorithm~\ref{alg:conformal-sketch}) or distinct-query (Algorithm~\ref{alg:conformal-sketch-unique}) coverage in a test set of size $M=100$ with varying degrees of distribution shift. The data are sketched with the CMS-CU instead of the CMS. Other details are as in Figure~\ref{fig:exp-zipf-marginal}.}
\label{fig:exp-zipf-unique-M-shift}
\end{center}
\end{figure}

Figure~\ref{fig:exp-zipf-unique-M-shift} reports on the results of these experiments.
The performance of the conformal confidence intervals output by Algorithm~\ref{alg:conformal-sketch}, applied with fixed conformity scores, is measured in terms of
average coverage and length over all random queries in the test set.
By contrast, the performance of the conformal confidence intervals output by Algorithm~\ref{alg:conformal-sketch-unique}, also applied with fixed conformity scores, is measured in terms of average coverage and length over the distinct queries in the test set.
Such choice facilitates the validation of Theorem~\ref{thm:shift}, which suggests Algorithm~\ref{alg:conformal-sketch} should be relatively robust to distribution shifts by these metrics.

Indeed, the empirical results confirm the distinct-query coverage guarantee provided by Algorithm~\ref{alg:conformal-sketch-unique} is more robust to distribution shifts compared to the marginal coverage property sought by Algorithm~\ref{alg:conformal-sketch}, although the performances of both methods in this setting also depend on the distribution of the sketched data. It is interesting to note that lower values of the Zipf tail parameter lead to larger numbers of unique objects in the queried data, increasing the robustness of all conformal confidence intervals to distribution shifts corresponding to unusually high proportions of new queries in the test set.

\subsection{Non-random sketching with data-driven hash functions} \label{sec:ml-sketching}

To further highlight the flexibility of conformal approach, we apply Algorithms~\ref{alg:conformal-sketch} and~\ref{alg:conformal-sketch-frequency} in combination with an alternative sketching method that departs from the CMS and the CMS-CU in that it is not based on random hash functions.
Instead, we follow the approach of \cite{bertsimas2021frequency} and fit a machine learning model to seek a compressed representation of the data that is designed to make frequency queries as efficient as possible.
In particular, a neural network model is trained to predict the relative frequency of each object, looking only at a small fraction of the data set aside during an initial warm-up phase---see below for more details about this training data set.

After the machine learning model has been fitted, any new object is assigned to one of $w$ possible buckets based on its predicted frequency, where $w$ is a width parameter that controls the memory footprint of this sketch. If the model is informative, objects with similar frequencies should be assigned to the same bin, and this is the key idea.
At the same time, a memory-efficient Bloom filter \citep{bloom1970space} is used to approximately keep track of the total number of distinct objects observed in the sketched data set. Thus, a reasonable guess for the frequency of any new query can be obtained by taking the ratio between the number of objects assigned to the same hash bucket and the approximate total number of distinct objects in the hashed data given by the Bloom filter. This procedure is outlined by Algorithm~\ref{alg:conformal-sketch-train-ml}. We also refer to \cite{bertsimas2021frequency} for a full description of this ``ML'' sketching algorithm, and to our open-source software repository for technical implementation details.

These experiments follow the same setup as those in Section~\ref{sec:app-synthetic}, simulating data from a Zipf distribution with tail parameter $a=1.1$.
Algorithm~\ref{alg:conformal-sketch-train-ml} is used to construct two-sided confidence intervals with fixed width, as explained in Section~\ref{sec:conf-scores-fixed}. 
The ML sketch is fitted on a training set collected in a data-driven way as to include exactly 500 distinct objects, following the same adaptive warm-up strategy presented in Section~\ref{sec:adaptive-warm-up}. 
Note that Algorithm~\ref{alg:conformal-sketch-train-ml} evaluates the conformity scores only on observations collected during a second independent warm-up phase. Therefore, the adaptive warm-up rule does not break the exchangeability required to obtain theoretically valid inferences.

These intervals are compared to those obtained by applying Algorithm~\ref{alg:conformal-sketch} with the CMS, the CMS-CU, or the CS \citep{charikar2002finding} as baseline sketches, varying the common width of the latter as a control parameter.
To keep the comparison fair, the ML sketch always uses the same amount of memory as the other sketches.
This is achieved by setting the number of buckets in the ML sketch equal to 50\% of the CMS, CMS-CU, and CS widths, while dedicating the remaining bits of memory to the Bloom filter.
To further facilitate the comparison with Algorithm~\ref{alg:conformal-sketch-train-ml}, the conformal confidence intervals based on the CMS, CMS-CU, and CS also utilize a similar adaptive warm-up strategy, specifically by following the heuristic approach of Algorithm~\ref{alg:conformal-sketch-train-heuristic} in Section~\ref{sec:adaptive-warm-up}.
Note that our conformal confidence interval based on the CS sketch are always two-sided because the CS sketch provides an unbiased estimate of the query frequencies \citep{cormode2020small}.

\begin{figure}[!htb]
\begin{center}
\includegraphics[width=0.8\linewidth]{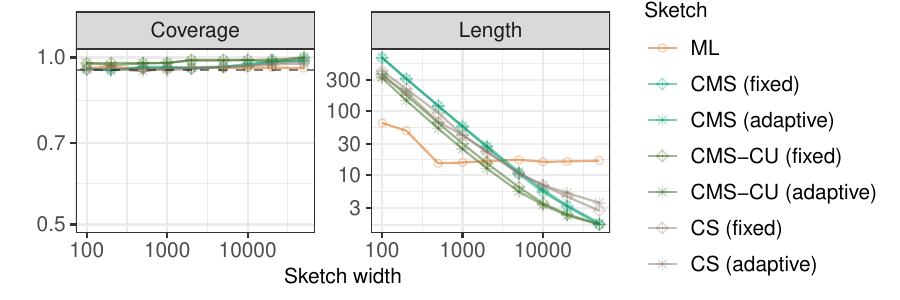}
\caption{Performance of 95\% conformal confidence intervals based on different data sketches, as a function of the sketch width. The data are simulated from a Zipf distribution with tail parameter $a=1.1$. Other details are as in Figure~\ref{fig:exp-zipf-marginal}.}
\label{fig:exp-zipf-opt-w}
\end{center}
\end{figure}

The results in Figure~\ref{fig:exp-zipf-opt-w} show that all methods achieve the desired 95\% marginal coverage, even though the  benchmark intervals based on the CMS,  CMS-CU, and CS are not known to be theoretically valid due to the heuristic nature of the adaptive warm-up involved in Algorithm~\ref{alg:conformal-sketch-train-heuristic}. 
In fact, we have observed that Algorithm~\ref{alg:conformal-sketch-train-heuristic} typically works well in practice, and that the additional data split introduced by its theoretically rigorous alternative (Algorithm~\ref{alg:conformal-sketch-train-adaptive}) may often be unnecessary.
The ML intervals produced by Algorithm~\ref{alg:conformal-sketch-train-ml} tend to be relatively more informative if the hash width is small, but they are less efficient compared to the CMS, CMS-CU, and CS as more memory becomes available.

This behavior can be explained by noting that here the performance of the ML sketch may be limited by the accuracy of the machine learning model, which is trained on a fixed number of warm-up data points that does not grow with the sketch width.
While increasing the number of training observations could further improve the performance of the ML sketch, it should be kept in mind that the training set must remain small compared to the sketched set in order to avoid excessive memory usage.

In conclusion, we note that the trade-off between random hashing and data-driven sketching may generally be affected several factors, including the amount of available memory and the ease with which a machine learning model can capture useful data patterns.
Therefore, different sketches are likely to be preferable in different applications, which highlights the advantage of having a flexible uncertainty estimation framework.

The results of additional experiments involving the ML sketch are in Appendix~\ref{app:figures}. Figure~\ref{fig:exp-zipf-opt-w-5bins} presents results similar to those in Figure~\ref{fig:exp-zipf-opt-w}, with the only difference that the conformal confidence intervals are designed to control frequency-range conditional coverage~\eqref{eq:conf-int-cond}, calibrating the conformity scores separately within $L=5$ frequency bins, instead of marginal coverage~\eqref{eq:marginal-coverage}. 
Figures~\ref{fig:exp-zipf-opt} and~\ref{fig:exp-zipf-opt-5bins} presents qualitatively similar results from experiments analogous to those in Figures~\ref{fig:exp-zipf-opt-w} and~\ref{fig:exp-zipf-opt-w-5bins}, respectively, in which the sketch width is fixed while the tail parameter of the Zipf distribution is varied.

\subsection{Additional numerical experiments} \label{sec:experiments-extra}

Appendix~\ref{app:figures} contains the results of supplementary experiments based on the CMS and CMS-CU applied to synthetic data from different distributions.
Figures~\ref{fig:exp-pyp-marginal}--\ref{fig:exp-pyp-cms} report on experiments based on data from a random probability measure distributed as the Pitman-Yor prior \citep{Pit(97)} with a standard Gaussian base distribution and parameters $\lambda>0$ and $\sigma \in [0,1)$, as explained in Appendix~\ref{app:pyp}.
We set $\lambda=5000$ and vary $\sigma$. For $\sigma=0$ the Pitman-Yor prior reduces to the Dirichlet prior \citep{Fer(73)}, while $\sigma>0$ results in heavier tails.
Figures~\ref{fig:exp-hh}--\ref{fig:exp-hh-cond} report on additional experiments in which our methods are applied in combination with the CS sketch \citep{charikar2002finding}, in order to compress data with rare high-frequency items ({\em heavy hitters}).
Concretely, these data are generated according to the following probability distribution: a heavy hitter $Z=0$ is observed with probability $1/\sqrt{m}$, where $m=100,000$; otherwise $Z \sim \text{Unif}(0,1)$ with probability $1-1/\sqrt{m}$. 
The results show that the CS leads to more informative conformal confidence intervals compared to the CMS, CMS-CU, or ML sketches. This should not be surprising given that the CS is designed to reduce the negative impact of random hash collisions in such a way as to make  frequency queries about heavy hitters relatively more accurate \citep{charikar2002finding}.
Finally, Figures~\ref{fig:exp-zipf-marginal-two-sided}--\ref{fig:exp-pyp-marginal-two-sided-cms} show the results of simulations involving two-sided confidence intervals, whose detailed setup is explained in Appendix~\ref{app:exp-two-sided}.

\section{Illustrations on empirical data} \label{sec:app}

Section~\ref{sec:app-dna} presents illustrations based on 16-mers data in SARS-CoV-2 DNA sequences, while Section~\ref{sec:app-literature} focuses on counting 2-grams in an English literature data set.

\subsection{Analysis of 16-mers in SARS-CoV-2 DNA sequences}  \label{sec:app-dna}

This illustration involves a data set of nucleotide sequences from SARS-CoV-2 viruses made publicly available by the National Center for Biotechnology Information~\citep{hatcher2017virus}. The data include 43,196 sequences, each consisting of approximately 30,000 nucleotides.
The goal is to estimate the empirical frequency of each {\it 16-mer}, a distinct sequence of 16 DNA bases in contiguous nucleotides. Given that each nucleotide has one of 4 bases, there are $4^{16} \approx$ 4.3 billion possible 16-mers.
Thus, exact tracking of all 16-mers is not unfeasible, which allows us to validate the sketch-based queries.
Sequences containing missing values are removed during pre-processing, for simplicity.

\begin{figure}[!htb]
\begin{center}
\includegraphics[width=0.75\linewidth]{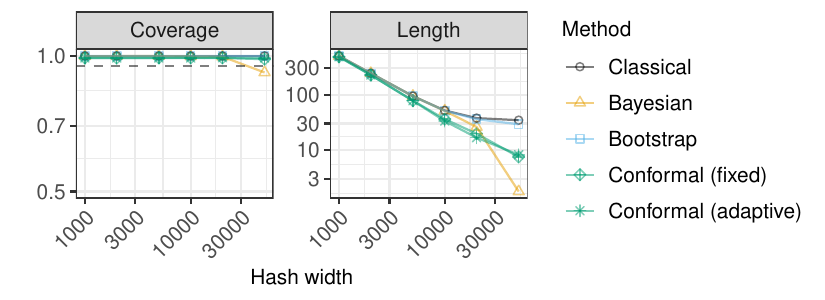}
\caption{Performance of confidence intervals based on SARS-CoV-2 sequence data. The results are shown as a function of the hash width. The data are sketched with the CMS-CU instead of the CMS. Other details are as in Figure~\ref{fig:exp-zipf-marginal}.}
\label{fig:exp-covid-marginal}
\end{center}
\end{figure}

The experiments are carried out as in Section~\ref{sec:app-synthetic}, with the difference that a larger sample of size 1,000,000 is sketched using the CMS-CU due to its higher efficiency; the width $w$ of the hash functions is varied as a control parameter.
All 16-mers are processed in a random order, which ensures their exchangeability. 
Figure~\ref{fig:exp-covid-marginal} compares the performances of all methods as a function of the hash width, in terms of marginal coverage and mean confidence interval width.

All methods achieve the desired marginal coverage, except for the Bayesian approach when $w$ is large. For small $w$, all methods return intervals of similar width, because the distribution of SARS-CoV-2 16-mers frequencies is quite concentrated with relatively narrow support (Figure~\ref{fig:exp-data-freq}), which makes it difficult to compress the data without much loss.

By contrast, the conformal methods yield noticeably shorter confidence intervals if $w$ is large.
Figure~\ref{fig:exp-covid-conditional} reports the same results stratified by the frequency of the queried objects.
Table~\ref{tab:data-w50000} lists 10 common and 10 rare queries along with their corresponding deterministic upper bounds for $w=50,000$, comparing the lower bounds obtained with each method. Table~\ref{tab:data-w5000} shows analogous results with $w=5,000$.
Figure~\ref{fig:exp-covid-sketch} confirms the advantage of sketching with the CMS-CU instead of the CMS.
Similarly, Figure~\ref{fig:exp-covid-opt} shows the CMS-CU also typically leads to more informative frequency queries compared to the ML sketch discussed in Section~\ref{sec:ml-sketching}, unless the available memory is very low.

Figure~\ref{fig:exp-covid-estimation} compares the performances of different frequency {\it point-estimates} in terms of mean absolute deviation from the true frequency. With the classical method, we take the midpoint of the 95\% confidence interval as a point estimate, although other approaches are also possible~\citep{cormode2020small}.
For the other methods, the point estimate is the lower confidence bound at level $\alpha=0.5$; in the Bayesian case, it is the posterior median.
Although a conformal lower bound with $\alpha = 0.5$ is not always a reliable estimator of conditional medians~\citep{medarametla2021distribution}, this approach outperformed the benchmarks in all of our experiments.

Figure~\ref{fig:exp-covid-unique} shows the confidence intervals reported in Figure~\ref{fig:exp-covid-marginal} approximately remain valid even if their average coverage is evaluated with respect to distinct queries only;
of course, this is not generally guaranteed and may not always be true on other data sets, as seen in Section~\ref{sec:experiments}.
Figure~\ref{fig:exp-covid-unique-M}
shows
the performance of the procedure described in Algorithm~\ref{alg:conformal-sketch-unique} for constructing conformal confidence intervals with valid coverage for distinct queries. These results show
that
Algorithm~\ref{alg:conformal-sketch-unique} leads to valid inference across a wide range of values of its parameter $M'$---the size of the calibration shards---despite the more pessimistic worst-case predictions of Theorem~\ref{eq:coverage-unique-robust}.
Finally, Figure~\ref{fig:exp-covid-unique-M-shift} investigates the robustness of the alternative types of conformal prediction intervals output by Algorithm~\ref{alg:conformal-sketch} and Algorithm~\ref{alg:conformal-sketch-unique} to distribution shifts in the test queries, similarly to Figure~\ref{fig:exp-zipf-unique-M-shift}.

\subsection{Analysis of 2-grams in an English literature data set}  \label{sec:app-literature}

This example is based on data consisting of 18 open-domain pieces of classic English literature downloaded from the Gutenberg Corpus~\citep{Gutenberg} using the NLTK Python package~\citep{bird2009natural}.
The goal is to count the frequencies of all {\it 2-grams}---consecutive pairs of English words.
After basic pre-processing to remove punctuation and unusual words (only those in a relatively small dictionary of 25,487 common English words are retained), there are approximately 1,700,000 remaining 2-grams---the total number of all {\it possible} 2-grams within this dictionary is approximately 650,000,000.

Note that such pre-processing does not remove very common words (such as ``the'', ``or'', etc.)~and it may sometimes lead to unnatural 2-grams whenever a relatively rare word is removed from an otherwise meaningful sentence (e.g., ``very uncommon for'' would become ``very for'').
Therefore, our analysis is not fully realistic from a natural language processing perspective but it is computationally efficient and still informative regarding the performance of our uncertainty estimation method. With this setup, the same experiments are then carried out as in Section~\ref{sec:app-dna}, sketching 1,000,000 randomly sampled 2-grams with the CMS-CU and querying 10,000 independent 2-grams.
As in the previous experiments, the 2-grams are processed in a random order to ensure exchangeability.

\begin{figure}[!htb]
\begin{center}
\includegraphics[width=0.75\linewidth]{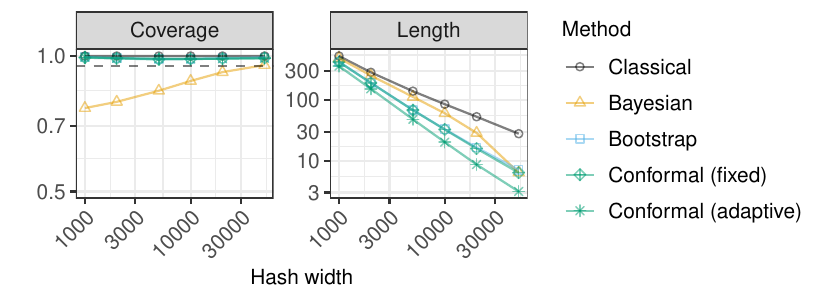}
\caption{Performance of confidence intervals for random queries, for a sketched data set of English 2-grams in classic English literature. Other details are as in Figure~\ref{fig:exp-covid-marginal}.}
\label{fig:exp-words-marginal}
\end{center}
\end{figure}

Figure~\ref{fig:exp-words-marginal} shows the conformal intervals produced by Algorithm~\ref{alg:conformal-sketch-frequency} using adaptive scores achieve the desired 95\% marginal coverage and tend to have the shortest width. By contrast, the Bayesian intervals are not valid unless the hashes are very wide.
Here, the conformal approach enjoys a larger improvement in performance compared to the other approaches because these data can be compressed efficiently due to the weaker power-law tail behavior of the frequency distribution of English 2-grams; see Figure~\ref{fig:exp-data-freq}. Further, Figures~\ref{fig:exp-words-conditional}--\ref{fig:exp-words-estimation} and Tables~\ref{tab:data-w50000}--\ref{tab:data-w5000} report additional results along the lines of those in the previous section, including empirical evidence of valid frequency-conditional coverage and a comparison of the performances of different linear and non-linear sketches.

Figure~\ref{fig:exp-words-unique} shows the confidence intervals reported in Figure~\ref{fig:exp-words-marginal} approximately remain valid even if their average coverage is evaluated with respect to distinct queries only; of course, this
is not guaranteed in general.
Figure~\ref{fig:exp-words-unique-M} illustrates the performance
of Algorithm~\ref{alg:conformal-sketch-unique}, showing that valid inference for distinct queries can be achieved with a wide range of the parameter $M'$, despite the more pessimistic worst-case predictions of Theorem~\ref{eq:coverage-unique-robust}.
Finally, Figure~\ref{fig:exp-words-unique-M-shift} investigates the robustness of the alternative types of conformal intervals output by Algorithms~\ref{alg:conformal-sketch} and~\ref{alg:conformal-sketch-unique} to distribution shifts in the test queries, similarly to Figure~\ref{fig:exp-zipf-unique-M-shift}.

\section{Discussion} \label{sec:discussion}

This work opens several opportunities for further research.
In the future one may study and compare theoretically, in some settings, the length of our conformal confidence intervals under different types of coverage guarantees. A possible approach may take inspiration from relevant work in the context of regression by \citet{lei2018distribution} and \citet{sesia2020comparison}.

Further, it would be interesting to explore the relevance of the methods and theory presented in Section~\ref{sec:coverage-unique} beyond sketching.
For example, the results of Section~\ref{sec:coverage-unique} could be repurposed to construct conformal prediction sets for regression or multi-class classification tasks that achieve valid coverage over subsets of individual test cases with certain unique attributes. In those contexts, our work may lead to an alternative framework for dealing with uncertainty estimation under algorithmic fairness constraints~\citep{Romano2020With} or stratified sampling mechanisms~\citep{dunn2018distribution,park2022pac}.

Finally,  the uncertainty estimation methods developed in this paper may also be relevant for more general forms of randomized sketching used for other numerical, statistical, and learning problems \citep{vempala2005random, halko2011finding, mahoney2011randomized, woodruff2014sketching, drineas2016randnla,martinsson2020randomized}; see e.g., \cite{dobriban2019asymptotics,liu2019ridge,lacotte2020optimal,yang2021reduce}.

\section*{Software and computations}

Accompanying software and data are available online at \url{https://github.com/msesia/conformalized-sketching}.
Experiments were carried out in parallel on a computing cluster; each experiment required less than a few hours with a standard CPU and less than 5GB of memory (20 GB are needed for the analysis of the SARS-CoV-2 DNA data).

 \subsection*{Acknowledgements}

M.~S.~is supported in part by NSF grant DMS 2210637 and by an Amazon Research Award. S.~F.~is also affiliated to IMATI-CNR ``Enrico Magenes" (Milan, Italy), and received funding from the European Research Council under the European Union's Horizon 2020 research and innovation program under grant 817257.
E.~D.~is supported in part by the NSF DMS 2046874 (CAREER) award, and ONR grant N00014-21-1-2843.
%\textcolor{red}{[Edgar: Acknowledgements?]}

\clearpage
\bibliography{biblio}

\clearpage
\appendix

% \renewcommand\cftsecaftersnum{.}
% \renewcommand\cftfigaftersnum{.}
% \renewcommand\cfttabaftersnum{.}

% %\setcounter{theorem}{0}
 \renewcommand{\thetheorem}{A\arabic{theorem}}

 \renewcommand{\thefigure}{A\arabic{figure}}

 \renewcommand{\thetable}{A\arabic{table}}

 \renewcommand{\thealgorithm}{A\arabic{algorithm}}

\section{Relevant background on the count-min sketch} \label{app:cms}

\subsection{The CMS algorithm} \label{sec:cms}

The count-min sketch (CMS) of \citet{cormode2005improved} compresses a data set by applying to each observation $d \geq 1$ different $w$-wide {\it hash} functions $h_j : \mathscr{Z} \to [w] \defeq \{1,\ldots,w\}$, for all $j \in [d]\defeq \{1,\ldots,d\}$ and some integer number of buckets $w \geq 1$.
Each hash function maps the elements of $\mathscr{Z}$ into one of $w$ buckets, so that distinct values of $z$ populate the buckets approximately uniformly.
Hash functions are typically chosen at random from a {\it pairwise independent} family $\mathcal{H}$.
This ensures the probability (over the randomness in the choice of hash functions) that two distinct objects $z_1,z_2 \in \mathscr{Z}$ are mapped by two different hash functions into the same bucket is $1/w^2$.
The data $Z_1,\ldots,Z_m$ are thus compressed into a sketch matrix $C \in \mathbb{N}^{d \times w}$ with rows summing to $m$. The element in the $j$-th row and $k$-th column of $C$ counts the data points mapped by the $j$-th hash function into the $k$-th bucket:
\begin{align} \label{eq:cms-count}
  C_{j,k} = \sum_{i=1}^{m} \I{h_j(Z_i) = k}, \qquad j \in [d],\, k\in [w].
\end{align}
One chooses $d$ and $w$ such that $d \cdot w \ll m$, and thus the matrix $C$ loses information compared to the full data set;
however, it has the advantage of requiring much less space to store.

Given a sketch $C$ from~\eqref{eq:cms-count}, we are interested in estimating the empirical frequency of an object $z \in \mathscr{Z}$, as defined in~\eqref{eq:f-true}.
A typical point estimate is the smallest count among the $d$ buckets into which $z$ is mapped:
\begin{align} \label{eq:cms-upper}
  \hat{f}_{\mathrm{up}}^{\mathrm{CMS}}(z) = \min_{j \in [d]} \left\{ C_{j,h_{j}(z)} \right\}.
\end{align}
This procedure is outlined by Algorithm~\ref{alg:CMS}.
\begin{algorithm}[H]
   \caption{CMS}
   \label{alg:CMS}
\begin{algorithmic}
  \STATE {\bfseries Input:} Data $Z_1, \ldots, Z_m$. Sketch dimensions $d,w$. Hash functions $h_1,\ldots,h_d$. Query $z$.
  \STATE {\bfseries Initialize:} $C_{j,k}=0$ for all $j \in [d], k \in [w]$.
    \FOR{$i = 1,\ldots,m$}
    \FOR{$j = 1,\ldots,d$}
    \STATE {\bfseries Increment} $C_{j,h_{j}(Z_i)} \leftarrow C_{j,h_{j}(Z_i)} + 1$
    \ENDFOR
    \ENDFOR
    \STATE {\bfseries Compute } $\hat{f}_{\mathrm{up}}^{\mathrm{CMS}}(z) = \min_{j \in [d]} \{ C_{j,h_{j}(z)} \}$.
    \STATE {\bfseries Output:} deterministic upper-bound for the frequency of $z$ in the data set: $\hat{f}_{\mathrm{up}}^{\mathrm{CMS}}(z)$.
\end{algorithmic}
\end{algorithm}

\subsection{Classical upper and lower bounds for CMS frequency queries} \label{sec:cms-classical-lower}

As $\hat{f}_{\mathrm{up}}^{\mathrm{CMS}}(z) \geq f_m(z)$, the expression in~\eqref{eq:cms-upper} always gives a deterministic upper bound for $f_m(z)$; see \cite{cormode2005improved}.
Although $\hat{f}_{\mathrm{up}}^{\mathrm{CMS}}(z)$ may be larger than $f_m(z)$ due to hash collisions, the independence of the hash functions
still enables the following classical probabilistic lower bound for $f_m(z)$.
\cite{cormode2005improved}
showed that for any $\delta, \epsilon \in (0,1)$, choosing
$d = \lceil - \log \delta \rceil$ and $w = \lceil e/\epsilon \rceil$,
for any fixed $z \in \mathscr{Z}$,
and with
$\hat{f}_{\mathrm{up}}^{\mathrm{CMS}}(z)$
from~\eqref{eq:cms-count},
\begin{equation}\label{th:cormode}
    \mathbb{P}_{\mathcal{H}}[f_m(z) \geq \hat{f}_{\mathrm{up}}^{\mathrm{CMS}}(z) - \epsilon m] \geq 1-\delta.
\end{equation}

For example, if
$\delta = 0.05$ and thus
$d = 3$, this says that $\hat{f}_{\mathrm{up}}^{\mathrm{CMS}}(z) -  m\cdot\lceil e /w \rceil$ is a lower bound on $f_m(z)$ with 95\% probability.
The subscript $\mathcal{H}$ in the bound~\eqref{th:cormode} clarifies that the randomness is with respect to the hash functions, while $Z_1, \ldots, Z_m$ and $z$ are fixed.
This bound can be useful to inform the choices of $d$ and $w$ prior to sketching, but it is not fully satisfactory as a way of quantifying the uncertainty about the true frequency of a given query.
First, it is often too conservative~\citep{ting2018count} if the data are randomly sampled from some distribution as opposed to being arbitrary and potentially worst-case.
Second, it is not flexible: $\delta$ cannot be chosen by the practitioner because it is fixed by $d$, and $\epsilon$ is uniquely determined by the hash width. Thus, the bound in~\eqref{th:cormode} does not always give practically useful confidence intervals.

\subsection{Bootstrap confidence intervals for CMS frequency queries}

An alternative approach to computing lower and upper bounds for $f_m(z)$ using the CMS was proposed by~\cite{ting2018count}, in order to address the often excessive conservativeness of the classical bounds described above.
The method of~\cite{ting2018count} is based on bootstrapping,
and departs from classical analysis of the CMS as it leverages randomness in the data instead of randomness in the hash functions.
Precisely, it assumes the data and the queried object are an independent and identically distributed (i.i.d.)~random sample from some unknown distribution.
This condition means that one is interested in the typical behavior of the algorithm over certain scenarios described by the distribution.
The condition does not always apply but, when it does, it can be extremely useful because it leads to much more informative confidence intervals.
In fact, the confidence intervals described by~\cite{ting2018count} are nearly exact
for the CMS, up to a finite-sample discrepancy between the bootstrap and population distributions.

A limitation of the bootstrap approach is that it relies on the specific {\it linear} structure of the CMS---the sketch matrix $C$ in~\eqref{eq:cms-upper} is a linear combination of the true frequencies of all objects in the data set---and is not easily extendable to other sketching algorithms that may outperform the CMS in practice.
For example, the CMS is relatively sensitive to random hash collisions, which can result in overly conservative deterministic upper bounds.
This challenge has motivated the development of alternative {\em non-linear} algorithms, such as the CMS with {\it conservative updates} (CMS-CU) of
\cite{estan2002new} which we briefly review below.

\subsection{The CMS-CU algorithm}

The difference between the CMS \citep{cormode2005improved} and the CMS-CU \citep{estan2002new} is that, whenever a new object $z$ is sketched by the latter, only the row of $C$ with the smallest value of $C_{j, h_j(z)}$ is updated, while the other counters remain unaltered.
Then, a valid deterministic upper bound for the CMS-CU can be calculated with the same formula in~\eqref{eq:cms-upper}.
This procedure is outlined in Algorithm~\ref{alg:CMS-CU}.
While the CMS-CU can lead to higher query accuracy compared to the vanilla CMS~\citep{estan2002new},
the theoretical analysis of the CMS-CU beyond a deterministic upper bound is more challenging, and it appears to be a relatively less explored topic.

\begin{algorithm}[H]
   \caption{CMS-CU}
   \label{alg:CMS-CU}
\begin{algorithmic}
  \STATE {\bfseries Input:} Data $Z_1, \ldots, Z_m$. Sketch dimensions $d,w$. Hash functions $h_1,\ldots,h_d$. Query $z$.
  \STATE {\bfseries Initialize:} $C_{j,k}=0$ for all $j \in [d], k \in [w]$.
    \FOR{$i = 1,\ldots,m$}
    \STATE {\bfseries Compute} $j^* = \arg\min_{j \in [d]} C_{j,h_{j}(Z_i)}$.
    \STATE {\bfseries Increment} $C_{j^*,h_{j^*}(Z_i)} \leftarrow C_{j^*,h_{j^*}(Z_i)} + 1$
    \ENDFOR
    \STATE {\bfseries Compute } $\hat{f}_{\mathrm{up}}^{\mathrm{CMS-CU}}(z) = \min_{j \in [d]} \{ C_{j,h_{j}(z)} \}$.
    \STATE {\bfseries Output:} deterministic upper-bound for the frequency of $z$ in the data set: $\hat{f}_{\mathrm{up}}^{\mathrm{CMS-CU}}(z)$.
\end{algorithmic}
\end{algorithm}

\FloatBarrier

\clearpage

\section{Additional methodological details} \label{sec:algoritms}

\subsection{Constructing two-sided conformal confidence intervals} \label{sec:conf-scores-twosided}

This section describes two alternatives methods for constructing two-sided conformal confidence intervals.
The first method, explained in Appendix~\ref{app:two-sided-bonferroni}, consists of separately calibrating two sequences of lower and upper one-sided confidence intervals, each adopting the significance level $\alpha/2$ instead of $\alpha$.
This is relatively easy to implement but may be less efficient than the second method, explained in Appendix~\ref{app:two-sided-chr}, which consists of directly calibrating a sequence of nested two-sided intervals.

\subsubsection{Construction based on Bonferroni correction} \label{app:two-sided-bonferroni}

One approach to building two-sided conformal confidence intervals for $f_m(X_{m+1})$ at level $1-\alpha$ consists of constructing a pair of lower and upper one-sided confidence intervals at level $1-\alpha/2$.
In particular, consider the following two nested sequences $S_t^l$ and $S_t^u$ of one-sided confidence intervals, each indexed by a scalar parameter $t$:
\begin{align*}
& S_t^l = [\hat{L}_{m,\alpha/2}(X_{m+1}; t), \hat{f}_{\mathrm{up}}^{\mathrm{CMS}}(X_{m+1})],
& S_t^u = [0, \hat{U}_{m,\alpha/2}(X_{m+1}; t)],
\end{align*}
where $\hat{f}_{\mathrm{up}}^{\mathrm{CMS}}(X_{m+1})$ is a deterministic upper bound for the unknown true empirical frequency of $X_{m+1}$; e.g., see Appendix~\ref{sec:cms}.
The sequences $S_t^l$ and $S_t^u$ can be separately calibrated using the conformal inference method described in Sections~\ref{sec:methods-marginal} and \ref{sec:frequency-coverage}, for any given choice of frequency-range partition $\mathcal{B}$, as we shall make more precise below. This gives two distinct data-adaptive thresholds $\hat{Q}^{*,l}_{n, 1-\alpha/2}$ and $\hat{Q}^{*,u}_{n, 1-\alpha/2}$, respectively, such that, $\forall B \in \mathcal{B}$,
\begin{align*}
\P{f_m(X_{m+1}) \geq \hat{L}_{m,\alpha/2}(X_{m+1}; \hat{Q}^{*,l}_{n, 1-\alpha/2}) \mid f_m(Z_{m+1}) \in B } \geq 1-\frac{\alpha}{2},
\end{align*}
and
\begin{align*}
\P{f_m(X_{m+1}) \leq \hat{U}_{m,\alpha/2}(X_{m+1}; \hat{Q}^{*,u}_{n, 1-\alpha/2}) \mid f_m(Z_{m+1}) \in B } \geq 1-\frac{\alpha}{2}.
\end{align*}
By a union bound, we obtain that the following two-sided conformal confidence interval has valid coverage, in the sense of~\eqref{eq:conf-int-cond}, at level $1-\alpha$:
\begin{align*}
[\hat{L}_{m,\alpha/2}(X_{m+1}; \hat{Q}^{*,l}_{n, 1-\alpha/2}),\hat{U}_{m,\alpha/2}(X_{m+1}; \hat{Q}^{*,u}_{n, 1-\alpha/2}) ].
\end{align*}

Different practical implementations are available to construct the sequences of candidate lower bounds $\hat{L}_{m,\alpha/2}(X_{m+1}; t)$ and upper bounds $\hat{U}_{m,\alpha/2}(X_{m+1}; t)$. Two concrete examples are explained below.

\textbf{Constant conformity scores.} A simple option to construct $\hat{L}_{m,\alpha/2}(X_{m+1}; t)$ is to directly apply the method described in Section~\ref{sec:conf-scores-fixed}, for example by shifting $\hat{f}_{\mathrm{up}}^{\mathrm{CMS}}(X_{m+1})$ downward by a constant $t$. Then, the conformalized threshold $\hat{Q}^{*,l}_{n, 1-\alpha/2}$ can be calibrated as usual.
The sequence of candidate upper bounds $\hat{U}_{m,\alpha/2}(X_{m+1}; t)$ can also be constructed similarly to $\hat{L}_{m,\alpha/2}(X_{m+1}; t)$, for example by adding a constant $t$ to the trivial lower bound of 0, up to the deterministic upper bound $\hat{f}_{\mathrm{up}}^{\mathrm{CMS}}(X_{m+1})$. The threshold $\hat{Q}^{*,u}_{n, 1-\alpha/2}$ for $\hat{U}_{m,\alpha/2}(X_{m+1}; t)$ can then be calibrated as usual with Algorithm~\ref{alg:conformal-sketch}.

\textbf{Bootstrap conformity scores.} An alternative option to construct the sequence $\hat{L}_{m,\alpha/2}(X_{m+1}; t)$ consists of shifting downward by a constant $t$ the bootstrap lower bound calculated with the method of \cite{ting2018count}, at level $\alpha/2$.
Similarly, the sequence $\hat{U}_{m,\alpha/2}(X_{m+1}; t)$ can be obtained by shifting upward by a constant $t$ the analogous bootstrap upper bound at level $1-\alpha/2$. Thus, in the special case of the vanilla CMS, our conformal confidence intervals based on these scores intuitively become very similar to the bootstrap confidence intervals of~\cite{ting2018count}. In general, however, the difference remains that the intervals of~\cite{ting2018count} rely on the linearity of the CMS, while ours are theoretically valid regardless of how the data are sketched.
We have observed this option works well in practice, at least within the scope of our numerical experiments. Therefore, this is the implementation adopted in our numerical experiments described in Section~\ref{app:exp-two-sided}.

\subsubsection{Construction based on conditional histograms} \label{app:two-sided-chr}

Two-sided conformal confidence intervals for $f_m(X_{m+1})$ can be constructed by following the general recipe outlined in Section~\ref{sec:sketch-problem}. To implement this method practically, one needs to fix an increasing sequence of candidate intervals $[\hat{L}_{m,\alpha}(\cdot; t), \hat{U}_{m,\alpha}(\cdot; t)]$, depending on $Z_{m+1}$ and $\phi(Z_{n+1}, \ldots, Z_{m})$.
Possible choices for such sequence may be directly borrowed from the existing literature on conformal inference for regression, including for example the quantile regression approach of~\cite{romano2019conformalized} or the conditional histogram approach of~\cite{sesia2021conformal}.
Here, we describe a particular implementation that combines the idea in~\cite{sesia2021conformal} with a Bayesian model, in continuity with the works of~\cite{cai2018bayesian} and~\cite{dolera2021bayesian} on Bayesian empirical frequency estimation from sketched data.
However, the same idea could easily accommodate a quantile regression model or any other machine learning algorithm instead of the Bayesian model, as explained in~\cite{sesia2021conformal}. Note that the following paragraphs largely retrace the same steps as in~\cite{sesia2021conformal}, which are however useful to recap here to make the presentation self contained.

For any $j \in [m]$, let $\hat{\varphi}_j(x)$ indicate the posterior probability of $f_m(X_{m+1}) = j$ for $X_{m+1}=x$ as estimated by any Bayesian model for frequency estimation given sketched data, such as that of~\cite{cai2018bayesian} based on a Dirichlet process prior, for example.
For convenience of notation, we will sometimes refer to the full posterior distribution of $f_m(X_{m+1})$ simply as $\hat{\varphi}$.
Note that, in general, the form of the posterior distribution $\hat{\varphi}$ may depend on $m$ as well as on the sketched data in $\phi(Z_{n+1}, \ldots, Z_{m})$.
Following in the footsteps of~\cite{sesia2021conformal}, define the following bi-valued function $\mathcal{S}$ taking as input a query $x$, the posterior distribution $\hat{\varphi}$, a scalar threshold $t \in [0,1]$, and two intervals $S^-, S^+ \subseteq \{1,\ldots,m\}$:
\begin{align} \label{eq:def-opt-problem}
  \mathcal{S}(x, \hat{\varphi}, S^-, S^+, t) \defeq \mathop{\mathrm{arg\,min}}_{(l,u) \in \{1,\ldots,m\}^2 \,:\, l \leq u} \left\{ |u-l| : \sum_{j=l}^{u} \hat{\varphi}_j(x) \geq t, \,  S^- \subseteq [l,u] \subseteq S^+  \right\}.
\end{align}
Above, it is implied that we choose the value of $(l,u)$ minimizing $\sum_{j=l}^{u} \hat{\varphi}_j(x)$ among the feasible ones with minimal $|u-l|$, whenever the optimization problem does not have a unique solution.
Therefore, we can assume without loss of generality that~\eqref{eq:def-opt-problem} has a unique solution; if that is not the case, we can break the ties at random by adding a little noise to $\hat{\varphi}$.
As explained in~\cite{sesia2021conformal}, the problem defined in~\eqref{eq:def-opt-problem} can be solved efficiently, at computational cost linear in $m$.
Note that we will sometimes refer to sub-intervals of $[m]$ as either contiguous subsets of $\{1,\ldots,m\}$ (e.g., $S^-$) or as pairs of lower and upper endpoints (e.g., $[l,u]$).

If $S^- = \emptyset$ and $S^+ = \{1,\ldots,m\}$, the expression in~\eqref{eq:def-opt-problem} computes the shortest interval with total posterior probability mass above $t$.
In general, the optimization in~\eqref{eq:def-opt-problem} involves the additional {\em nesting} constraint that the output $\mathcal{S}$ must satisfy $S^- \subseteq \mathcal{S} \subseteq S^+$, which will be needed to guarantee the resulting sequence of confidence intervals indexed by $t$ is nested.
Note that the inequality in~\eqref{eq:def-opt-problem} involving $t$ may not be binding at the optimal solution due to the discrete nature of the optimization problem.
However, the above construction could be easily modified by introducing some suitable randomization leading to confidence intervals that are even tighter on average, as explained in~\cite{sesia2021conformal}.

For any integer $T \geq 1$, consider an increasing sequence $t_\tau \in [0,1]$, for $\tau \in \{0,\ldots,T\}$.
A nested sequence of $T$ intervals indexed by $\tau \in \{0,\ldots,T\}$, which may be written as
\begin{align*}
S_t = \big[ \hat{L}_{m,\alpha}(X_{m+1}; t_\tau), \hat{U}_{m,\alpha}(X_{m+1}; t_\tau) \big],
\end{align*}
for appropriate endpoints $\hat{L}_{m,\alpha}(X_{m+1}; t_{\tau})$ and $\hat{U}_{m,\alpha}(X_{m+1}; t_{\tau})$, respectively, is then constructed from~\eqref{eq:def-opt-problem} as follows.
First, fix any {\em starting index} $\bar{\tau} \in \{0,1,\ldots,T\}$ and define $S_{\bar{\tau}}$ by applying \eqref{eq:def-opt-problem} without the nesting constraints (with $S^- = \emptyset$ and $S^+ = \{1,\ldots,m\}$):
\begin{align} \label{eq:def-s-start}
  & S_{\bar{\tau}}  \defeq \mathcal{S}(x, \hat{\varphi}, \emptyset, \{1,\ldots,m\}, t_{\bar{\tau}}),
\end{align}
Note the explicit dependence on $x$ and $\hat{\varphi}$ of the left-hand-side above is omitted for simplicity, although it is important to keep in mind that $S_{\bar{\tau}}$ does of course depend on these quantities.

Having computed the initial interval $S_{\bar{\tau}}$, we recursively extend the definition to the wider intervals indexed by $\tau = \bar{\tau} + 1, \ldots, T$ as follows:
\begin{align*}
  S_{\tau} & \defeq \mathcal{S}(x, \hat{\varphi}, S_{\tau-1}, \{1,\ldots,m\}, t_{\tau}).
\end{align*}
See~\cite{sesia2021conformal} for a schematic visualization of this step.
Similarly, the narrower intervals $S_{\tau}$ indexed by $\tau = \bar{\tau}-1, \bar{\tau}-2, \ldots, 0$ are defined recursively as:
\begin{align*}
  & S_{\tau} \defeq \mathcal{S}(x, \hat{\varphi}, \emptyset, S_{\tau+1}, t_{\tau}).
\end{align*}
See~\cite{sesia2021conformal} for a schematic visualization of this step.
As a result of this construction, the sequence of intervals $\{ S_{\tau} \}_{\tau=0}^{T}$ is nested regardless of the starting point $\bar{\tau}$ in~\eqref{eq:def-s-start}, for which a typical choice is such that $t_{\bar{\tau}} = 1-\alpha$. Then, two-sided conformal confidence intervals for $f_m(X_{m+1})$ can be obtained by applying Algorithm~\ref{alg:conformal-sketch} with this particular sequence of input nested intervals.
We refer to~\cite{sesia2021conformal} for further details on the construction of nested intervals outlined above.

\clearpage
\FloatBarrier
\subsection{Conformalized sketching with adaptive warm-up period} \label{app:adaptive-warmup}

\begin{algorithm}[!htb]
   \caption{Conformalized sketching with adaptive warm-up period (heuristic)}
   \label{alg:conformal-sketch-train-heuristic}
\begin{algorithmic}
  \STATE {\bfseries Input:} Data set $Z_1, \ldots, Z_m$. Sketching function $\phi$.
  \STATE {\bfseries \textcolor{white}{Input:}} Number $n_0 \ll m$ of unique objects to be observed during the warm-up phase.
  \STATE {\bfseries \textcolor{white}{Input:}} A (trainable) predictor to compute nested intervals $[\hat{L}_{m,\alpha}(\cdot; t), \hat{U}_{m,\alpha}(\cdot; t)]_{t \in \mathcal{T}}$.
  \STATE {\bfseries \textcolor{white}{Input:}} Number of data points $n^{\mathrm{train}} < n$ used for training $[\hat{L}_{m,\alpha}(\cdot; t), \hat{U}_{m,\alpha}(\cdot; t)]$.
%  \STATE {\bfseries \textcolor{white}{Input:}} A partition $\mathcal{B} = (B_1,\ldots,B_L)$ of $\{0,\ldots,m\}$ into $L$ intervals.
%  \STATE {\bfseries \textcolor{white}{Input:}} Random query $Z_{m+1}$. Desired coverage level $1-\alpha \in (0,1)$.
  \STATE{\bfseries Initialize} a sparse counter $f_{n}^{\mathrm{wu}}(z) = 0, \forall z \in \mathcal{Z}$.
  \FOR{$i_{\text{wp}} = 1,\ldots,m$}
  \STATE{\bfseries Increment} $f_{n}^{\mathrm{wu}}(Z_i) \leftarrow f_{n}^{\mathrm{wu}}(Z_i)  +1$.
  \IF{Number of unique observed objects $\geq n_0$}
  \STATE{\bfseries Break}
  \ENDIF
  \ENDFOR
  \STATE{\bfseries Set} $n = i_{\text{wp}}$.
  \STATE{\bfseries Initialize} a sparse counter $f_{m-n}^{\mathrm{sv}}(z) = 0, \forall z \in \mathcal{Z}$.
  \STATE{\bfseries Initialize} an empty sketch $\phi(\emptyset)$.
    \FOR{$i = n+1,\ldots,m$}
    \STATE{\bfseries Update} the sketch $\phi$ with the new observation $Z_i$.
    \IF{$f_{n}^{\mathrm{wu}}(Z_i) > 0$}
    \STATE{\bfseries Increment} $f_{m-n}^{\mathrm{sv}}(Z_i) \leftarrow f_{m-n}^{\mathrm{sv}}(Z_i)  +1$.
    \ENDIF
    \ENDFOR

  \FOR{$i = 1,\ldots,n$}
  \STATE{\bfseries Set}  $X_i = \left(Z_i, \phi(Z_{n+1}, \ldots, Z_{m}) \right)$ as in~\eqref{eq:X-def}.
  \STATE{\bfseries Set}  $Y_i = f_{m-n}^{\mathrm{sv}}(Z_i)$.
  \ENDFOR
  \STATE{\bfseries Train} $[\hat{L}_{m,\alpha}(\cdot; t), \hat{U}_{m,\alpha}(\cdot; t)]$ using the data in $\{(X_i,Y_i)\}_{i=1}^{n^{\mathrm{train}}}$.
  \FOR{$i = n^{\mathrm{train}}+1,\ldots,n$}
  \STATE{\bfseries Compute} the conformity score $E(X_i,Y_i)$ with~\eqref{eq:conf-score}, using $[\hat{L}_{m,\alpha}(\cdot; t), \hat{U}_{m,\alpha}(\cdot; t)]$.
  \ENDFOR
  \STATE {\bfseries Output:} Data sketch $\phi$;
\STATE {\color{white} \bfseries Output:} Sparse counter $f_{n}^{\mathrm{wu}}(z), \forall z \in \mathcal{Z}$;
  \STATE {\color{white} \bfseries Output:} Trained predictor $[\hat{L}_{m,\alpha}(\cdot; t), \hat{U}_{m,\alpha}(\cdot; t)]$;
  \STATE {\color{white} \bfseries Output:} Conformity scores $E(X_i,Y_i)$ for all $i \in \{n^{\mathrm{train}}+1,\ldots,n\}$.
\end{algorithmic}
\end{algorithm}

\begin{algorithm}[!htb]
   \caption{Conformalized sketching with two-step adaptive warm-up period}
   \label{alg:conformal-sketch-train-adaptive}
\begin{algorithmic}
  \STATE {\bfseries Input:} Data set $Z_1, \ldots, Z_m$. Sketching function $\phi$.
  \STATE {\bfseries \textcolor{white}{Input:}} Number $n_0 \ll m$ of unique objects to be observed during the warm-up phase.
  \STATE {\bfseries \textcolor{white}{Input:}} A (trainable) predictor to compute nested intervals $[\hat{L}_{m,\alpha}(\cdot; t), \hat{U}_{m,\alpha}(\cdot; t)]_{t \in \mathcal{T}}$.
  \STATE {\bfseries \textcolor{white}{Input:}} Number of data points $n^{\mathrm{train}} < n$ used for training $[\hat{L}_{m,\alpha}(\cdot; t), \hat{U}_{m,\alpha}(\cdot; t)]$.
%  \STATE {\bfseries \textcolor{white}{Input:}} A partition $\mathcal{B} = (B_1,\ldots,B_L)$ of $\{0,\ldots,m\}$ into $L$ intervals.
%  \STATE {\bfseries \textcolor{white}{Input:}} Random query $Z_{m+1}$. Desired coverage level $1-\alpha \in (0,1)$.
  \STATE{\bfseries Initialize} a sparse counter $f_{n}^{\mathrm{wu}}(z) = 0, \forall z \in \mathcal{Z}$.
  \FOR{$i_{\text{wp}} = 1,\ldots,m$}
  \STATE{\bfseries Increment} $f_{n}^{\mathrm{wu}}(Z_i) \leftarrow f_{n}^{\mathrm{wu}}(Z_i)  +1$.
  \IF{Number of unique observed objects $\geq n_0$}
  \STATE{\bfseries Break}
  \ENDIF
  \ENDFOR
  \STATE{\bfseries Set} $n = i_{\text{wp}}$.
  \STATE{\bfseries Initialize} a sparse counter $f_{n}^{\mathrm{wu},2}(z) = 0, \forall z \in \mathcal{Z}$.
  \FOR{$i = n+1,\ldots,2n$}
  \STATE{\bfseries Increment} $f_{n}^{\mathrm{wu}}(Z_i) \leftarrow f_{n}^{\mathrm{wu}}(Z_i)  +1$.
  \STATE{\bfseries Increment} $f_{n}^{\mathrm{wu},2}(Z_i) \leftarrow f_{n}^{\mathrm{wu},2}(Z_i)  +1$.
  \ENDFOR
  \STATE{\bfseries Initialize} a sparse counter $f_{m-n}^{\mathrm{sv}}(z) = 0, \forall z \in \mathcal{Z}$.
  \STATE{\bfseries Initialize} an empty sketch $\phi(\emptyset)$.
    \FOR{$i = 2n+1,\ldots,m$}
    \STATE{\bfseries Update} the sketch $\phi$ with the new observation $Z_i$.
    \IF{$f_{n}^{\mathrm{wu},2}(Z_i) > 0$}
    \STATE{\bfseries Increment} $f_{m-n}^{\mathrm{sv}}(Z_i) \leftarrow f_{m-n}^{\mathrm{sv}}(Z_i)  +1$.
    \ENDIF
    \ENDFOR

  \FOR{$i = n+1,\ldots,2n$}
  \STATE{\bfseries Set}  $X_i = \left(Z_i, \phi(Z_{n+1}, \ldots, Z_{m}) \right)$ as in~\eqref{eq:X-def}.
  \STATE{\bfseries Set}  $Y_i = f_{m-n}^{\mathrm{sv}}(Z_i)$.
  \ENDFOR
  \STATE{\bfseries Train} $[\hat{L}_{m,\alpha}(\cdot; t), \hat{U}_{m,\alpha}(\cdot; t)]$ using the data in $\{(X_i,Y_i)\}_{i=n + 1}^{n + n^{\mathrm{train}}}$.
  \FOR{$i = n +n^{\mathrm{train}}+1,\ldots,2n$}
  \STATE{\bfseries Compute} the conformity score $E(X_i,Y_i)$ with~\eqref{eq:conf-score}, using $[\hat{L}_{m,\alpha}(\cdot; t), \hat{U}_{m,\alpha}(\cdot; t)]$.
  \ENDFOR
  \STATE {\bfseries Output:} Data sketch $\phi$;
\STATE {\color{white} \bfseries Output:} Sparse counter $f_{n}^{\mathrm{wu}}(z), \forall z \in \mathcal{Z}$;
  \STATE {\color{white} \bfseries Output:} Trained predictor $[\hat{L}_{m,\alpha}(\cdot; t), \hat{U}_{m,\alpha}(\cdot; t)]$;
  \STATE {\color{white} \bfseries Output:} Conformity scores $E(X_i,Y_i)$ for all $i \in \{n + n^{\mathrm{train}}+1,\ldots,2n\}$.
\end{algorithmic}
\end{algorithm}

\clearpage
\FloatBarrier
\subsection{Conformalized sketching with ML algorithms} \label{app:ml}

\begin{algorithm}[!htb]
   \caption{Conformalized sketching with data-driven ML sketch}
   \label{alg:conformal-sketch-train-ml}
\begin{algorithmic}
  \STATE {\bfseries Input:} Data set $Z_1, \ldots, Z_m$.
  \STATE {\bfseries \textcolor{white}{Input:}} Number $n_0 \ll m$ of unique objects to be observed during the warm-up phase.
  \STATE {\bfseries \textcolor{white}{Input:}} A trainable model $\mathcal{M}$ to predict the relative frequency of an object.
  \STATE {\bfseries \textcolor{white}{Input:}} A Bloom filter $\mathcal{F}$.
  \STATE {\bfseries \textcolor{white}{Input:}} A (trainable) predictor to compute nested intervals $[\hat{L}_{m,\alpha}(\cdot; t), \hat{U}_{m,\alpha}(\cdot; t)]_{t \in \mathcal{T}}$.
  \STATE {\bfseries \textcolor{white}{Input:}} Number of data points $n^{\mathrm{train}} < n$ used for training $[\hat{L}_{m,\alpha}(\cdot; t), \hat{U}_{m,\alpha}(\cdot; t)]$.
%  \STATE {\bfseries \textcolor{white}{Input:}} A partition $\mathcal{B} = (B_1,\ldots,B_L)$ of $\{0,\ldots,m\}$ into $L$ intervals.
%  \STATE {\bfseries \textcolor{white}{Input:}} Random query $Z_{m+1}$. Desired coverage level $1-\alpha \in (0,1)$.
  \STATE{\bfseries Initialize} a sparse counter $f_{n}^{\mathrm{wu}}(z) = 0, \forall z \in \mathcal{Z}$.
  \FOR{$i_{\text{wp}} = 1,\ldots,m$}
  \STATE{\bfseries Increment} $f_{n}^{\mathrm{wu}}(Z_i) \leftarrow f_{n}^{\mathrm{wu}}(Z_i)  +1$.
  \IF{Number of unique observed objects $\geq n_0$}
  \STATE{\bfseries Break}
  \ENDIF
  \ENDFOR
  \STATE{\bfseries Set} $n = i_{\text{wp}}$.
  \STATE{\bfseries Train} the model $\mathcal{M}$ using the data in $\{(Z_i, f_{n}^{\mathrm{wu}}(Z_i))\}_{i \in [1,\ldots,n]}$.
  \STATE{\bfseries Initialize} the ML sketch $\phi$ based on $\mathcal{M}$ and $\mathcal{F}$, as explained in Section~\ref{sec:experiments-ML}.
  \STATE{\bfseries Initialize} a sparse counter $f_{n}^{\mathrm{wu},2}(z) = 0, \forall z \in \mathcal{Z}$.
  \FOR{$i = n+1,\ldots,2n$}
  \STATE{\bfseries Increment} $f_{n}^{\mathrm{wu}}(Z_i) \leftarrow f_{n}^{\mathrm{wu}}(Z_i)  +1$.
  \STATE{\bfseries Increment} $f_{n}^{\mathrm{wu},2}(Z_i) \leftarrow f_{n}^{\mathrm{wu},2}(Z_i)  +1$.
  \ENDFOR
  \STATE{\bfseries Initialize} a sparse counter $f_{m-n}^{\mathrm{sv}}(z) = 0, \forall z \in \mathcal{Z}$.
  \STATE{\bfseries Initialize} an empty sketch $\phi(\emptyset)$.
    \FOR{$i = 2n+1,\ldots,m$}
    \STATE{\bfseries Update} the sketch $\phi$ with the new observation $Z_i$.
    \IF{$f_{n}^{\mathrm{wu},2}(Z_i) > 0$}
    \STATE{\bfseries Increment} $f_{m-n}^{\mathrm{sv}}(Z_i) \leftarrow f_{m-n}^{\mathrm{sv}}(Z_i)  +1$.
    \ENDIF
    \ENDFOR

  \FOR{$i = n+1,\ldots,2n$}
  \STATE{\bfseries Set}  $X_i = \left(Z_i, \phi(Z_{n+1}, \ldots, Z_{m}) \right)$ as in~\eqref{eq:X-def}.
  \STATE{\bfseries Set}  $Y_i = f_{m-n}^{\mathrm{sv}}(Z_i)$.
  \ENDFOR
  \STATE{\bfseries Train} $[\hat{L}_{m,\alpha}(\cdot; t), \hat{U}_{m,\alpha}(\cdot; t)]$ using the data in $\{(X_i,Y_i)\}_{i=n + 1}^{n + n^{\mathrm{train}}}$.
  \FOR{$i = n +n^{\mathrm{train}}+1,\ldots,2n$}
  \STATE{\bfseries Compute} the conformity score $E(X_i,Y_i)$ with~\eqref{eq:conf-score}, using $[\hat{L}_{m,\alpha}(\cdot; t), \hat{U}_{m,\alpha}(\cdot; t)]$.
  \ENDFOR
  \STATE {\bfseries Output:} Data sketch $\phi$;
\STATE {\color{white} \bfseries Output:} Sparse counter $f_{n}^{\mathrm{wu}}(z), \forall z \in \mathcal{Z}$;
  \STATE {\color{white} \bfseries Output:} Trained predictor $[\hat{L}_{m,\alpha}(\cdot; t), \hat{U}_{m,\alpha}(\cdot; t)]$;
  \STATE {\color{white} \bfseries Output:} Conformity scores $E(X_i,Y_i)$ for all $i \in \{n + n^{\mathrm{train}}+1,\ldots,2n\}$.
\end{algorithmic}
\end{algorithm}

\clearpage
\FloatBarrier

\subsection{Sampling from a Pitman-Yor predictive distribution} \label{app:pyp}

The data points are sampled sequentially from the following predictive distribution, which has parameters $\lambda>0$ and $\sigma \in [0,1)$. After sampling $Z_1$ from a standard normal distribution, $\mathcal{N}(0,1)$, fix any $i\geq1$ and let $Z_{1}, \ldots, Z_{i}$ indicate the data stream observed up to that point. Denote by $k_i$ the number of distinct elements within it, and by $V_i = (V_{i,1}, \ldots, V_{i,k_i})$ the set of such distinct values. Further, let $c_{i,l}$ indicate the number of times that object $V_{i,l}$ has been observed in $Z_{1}, \ldots, Z_{i}$, for $l \in \{1,\ldots,k_i\}$. Then, $Z_{i+1}$ is generated as follows:
\begin{align*}
  Z_{i+1} \mid Z_{1}, \ldots, Z_{i} =
  \begin{cases}
    V_{i, l}, & \text{with probability } \frac{c_{i,l} - \sigma}{\lambda + i}, \text{ for } l \in \{1,\ldots,k_i\}, \\
    \mathcal{N}(0,1), & \text{with probability } \frac{\lambda + k_i \sigma}{\lambda + i}.
  \end{cases}
\end{align*}
Above, the second case which occurs with probability $(\lambda + k_i \sigma)/(\lambda + i)$ corresponds to sampling a new unique value from the standard normal distribution.

\FloatBarrier

\section{Auxiliary theoretical results} \label{sec:theory-supp}

\subsection{Probability distribution of the set of uniques}

Note that the
size of $V$ is between $1$ and $M$;
and the values taken by $V$ range over subsets
$ \{a_{j_1},\ldots, a_{j_k}\} \subseteq \mathcal{Z}$,
where  $1\le k\le M$ and $j_1,\ldots,j_k \in \mathbb{N}$ are distinct indices.

\begin{proposition}[Probability distribution of the set of uniques]\label{unipro}
Let $\mathcal{Z}^{\text{test}}$ be an i.i.d. sample of size $M$ from a discrete distribution
$P_Z = \sum_{i \in \mathbb{N}} p_j \delta_{a_j} $, where $a_j \in \mathcal{Z}$ are distinct, and $p_j\ge 0$ for all $j\in \NN$.
Let $P_Z^{[M]}$ be the probability distribution of
$\mathcal{Z}^{\text{test}}$.
Let $V = \textsc{Unique}(\mathcal{Z}^{\text{test}})$ denote the set of unique values in $\mathcal{Z}^{\text{test}}$.
For any $1\le k\le M$, and any distinct indices $j_1,\ldots,j_k \in \mathbb{N}$,
the probability mass function of $V$ at
$ \{a_{j_1},\ldots, a_{j_k}\}$
equals
\begin{align}
    P_Z^{[M]}(V = \{a_{j_1},\ldots, a_{j_k}\})
    & = \sum_{c\in C_{M,k}} \binom{M}{c_1\,c_2\,\ldots c_k}
    p_{j_1}^{c_1}\cdots p_{j_k}^{c_k}  \label{q} \\
    & = \sum_{S\subset \{j_1,\ldots, j_k\}} (-1)^{k+|S|} \left(\sum_{j\in S} p_j\right)^M. \label{q2}
\end{align}
\end{proposition}

The proof
of \eqref{q} follows directly from the definitions, while that of
\eqref{q2}---which we will use extensively later---relies on a careful combinatorial argument, pairing sets of odd and even sizes;
see Appendix \ref{pf:uni}.

To better understand \eqref{q}, consider the trivial example in which $M=1$. In this case, $P_Z^{[1]}(V = \{a_{j}\}) =p_{j}$ for all $j \in \NN$.
    Thus,
    $P_Z^{[1]}$, the distribution of uniques when sampling a single element from the distribution $P_Z$, is equal precisely to $P_Z$ itself; i.e., $P_Z^{[1]} = P_Z$.
    For $M=2$, we have that
    $P_Z^{[2]}(V = \{a_{j}\}) =p_{j}^2$ for all $j\in \NN$; this is the probability of observing $a_i$ twice in a row.
    Further,
    for all $j_1,j_2 \in \NN$ with $j_1\neq j_2$, we have that
    $P_Z^{[2]}(V = \{a_{j_1},a_{j_2}\}) =2p_{j_1}p_{j_2}$;
    this is the probability of observing $(a_{j_1},a_{j_2})$ or $(a_{j_2},a_{j_1})$, so that the set of uniques is $\{a_{j_1},a_{j_2}\}$.
 One can also verify that \eqref{q} leads to the same results.
 Continuing the above example,
 for $M=2$,
 for all  $j_1,j_2 \in \NN$ with $j_1\neq j_2$,
 \eqref{q2} leads to
  $P_Z^{[2]}(V = \{a_{j_1},a_{j_2}\}) = (p_{j_1}+p_{j_2})^2-p_{j_1}^2-p_{j_2}^2= 2p_{j_1}p_{j_2}$, agreeing with \eqref{q}.

\section{Mathematical proofs} \label{sec:proofs}

\subsection{Proof of Proposition~\ref{eq:exchangeability-XY}}
\begin{proof}
Consider $((X_{\pi(1)},Y_{\pi(1)}),\ldots,(X_{\pi(n)},Y_{\pi(n)}),(X_{\pi(m+1)},Y_{\pi(m+1)}))$ for any permutation $\pi$ of $\{1,\ldots,n,m+1\}$.
This is equal to $((X'_{1},Y'_{1}),\ldots,(X'_{n},Y'_{n}),(X'_{m+1},Y'_{m+1}))$, defined by applying the functions in~\eqref{eq:Y-def}--\eqref{eq:X-def} to a shuffled data set $Z_{\tilde{\pi}(1)},\ldots,Z_{\tilde{\pi}(m+1)}$, where $\tilde{\pi}$ indicates a permutation of $\{1,\ldots,m+1\}$ that agrees with $\pi$ on $\{1,\ldots,n,m+1\}$ and leaves $\{n+1,\ldots,m\}$ unchanged.
Therefore,
 \begin{align*}
   & \left( (X_{\pi(1)},Y_{\pi(1)}),\ldots,(X_{\pi(n)},Y_{\pi(n)}),(X_{\pi(m+1)},Y_{\pi(m+1)}) \right) \\
   & \qquad\qquad = \left( (X'_{1},Y'_{1}),\ldots,(X'_{n},Y'_{n}),(X'_{m+1},Y'_{m+1}) \right) \\
   & \qquad\qquad \overset{d}{=} \left( (X_{1},Y_{1}),\ldots,(X_{n},Y_{n}),(X_{m+1},Y_{m+1}) \right),
 \end{align*}
 where the last equality in distribution follows directly from the assumption that $Z_1,\ldots,Z_{m+1}$ are exchangeable.
% In fact, it would have been sufficient to make the weaker assumption that $Z_1,\ldots,Z_{n},Z_{m+1}$ are exchangeable.
\end{proof}

\subsection{Proof of Theorem~\ref{thm:coverage}}

\begin{proof}
We refer to the proof of the more general Theorem~\ref{thm:coverage-cond}, of which this result is a special case.
In fact, Algorithm~\ref{alg:conformal-sketch} corresponds to Algorithm~\ref{alg:conformal-sketch-frequency} applied with trivial partitions that divide the range of frequencies into a single bin: $L=1$. Further, the marginal coverage property in~\eqref{eq:marginal-coverage} is a special case of the frequency-conditional coverage property in~\eqref{eq:conf-int-cond} with the trivial partitions corresponding to $L=1$.
\end{proof}

\subsection{Proof of Theorem~\ref{thm:coverage-cond}}

The following notation will be helpful: let $B(Y_i) \in \mathcal{B}$ indicate the frequency bin into which $Y_i$ belongs, for $i \in \{1,\ldots,n,m+1\}$.
We begin by proving the result for the simpler case in which Algorithm~\ref{alg:conformal-sketch} is applied using conformity scores that do not require training, in which case $n^{\mathrm{train}}=0$.
For $i \in \{1,\ldots,n,m+1\}$, define the random variables $Y_i$ and $X_i$ as in~\eqref{eq:Y-def}--\eqref{eq:X-def}, respectively.
We already know from Proposition~\ref{eq:exchangeability-XY} that $(X_1,Y_1),\ldots,(X_{n},Y_{n}),(X_{m+1},Y_{m+1})$ are exchangeable. This implies that the conformity scores $E(X_i, Y_i)$ are exchangeable with one another, for $i \in \{1,\ldots,n,m+1\}$, because each of them only depends on $X_i, Y_i$ and on the separate data points in the sketch $\phi(Z_{n+1}, \ldots, Z_{m})$.
Therefore, $E_{m+1}$ is also exchangeable with the subset of conformity scores with indices in $\{i \in \{1,\ldots,n\} : B(Y_i) = B(Y_{m+1})\}$.

Now, fix any bin $B^* \in \mathcal{B}$ and assume $B(Y_{m+1}) = B^*$.
Now, note that the interval output by Algorithm~\ref{alg:conformal-sketch} does not cover the true frequency $f_m(Z_{m+1})$ if and only if $E_{m+1} > \hat{Q}_{n, 1-\alpha} \geq \hat{Q}_{n_l, 1-\alpha}(B^*)$. However, a standard exchangeability argument for the conformity scores in $\{i \in \{1,\ldots,n\} : B(Y_i) = B^*\}$ shows that $\mathbb{P}[E_{m+1} > \hat{Q}_{n_l, 1-\alpha}(B^*) \mid B(Y_{m+1})=B^*] \leq 1-\alpha$; for example, see Lemma~1 of~\cite{romano2019conformalized}. This completes the first part of the proof.

The second part with $n^{\mathrm{train}}>0$ follows very similarly: Proposition~\ref{eq:exchangeability-XY} implies that $(X_{n^{\mathrm{train}}+1},Y_{n^{\mathrm{train}}+1}),\ldots,(X_{n},Y_{n}),(X_{m+1},Y_{m+1})$ are exchangeable, and so must be the conformity scores $E_{i}$ for $i \in \{n^{\mathrm{train}}+1, \ldots, n, m+1\}$ because each of them only depends on the corresponding $X_i,Y_i$ and on the separate set of observations indexed by $\{1,\ldots,n^{\mathrm{train}}\}$, as well as on the sketch $\phi(Z_{n+1}, \ldots, Z_{m})$. The rest of the proof is exactly the same as in the first part because the empirical quantiles $\hat{Q}_{n_l, 1-\alpha}(B)$ are only computed on subsets of the data indexed by $\{n^{\mathrm{train}}+1, \ldots, n\}$.

\subsection{Proof of Theorem~\ref{eq:coverage-unique}}

Following the same notation as in Algorithm~\ref{alg:conformal-sketch-unique}, let $Z^*$ indicate a random object sampled uniformly from $\textsc{Unique}(\{Z_{m+1}, \ldots, Z_{m+M}\})$. Define also $X^* = \left(Z^*, \phi(Z_{n+1}, \ldots, Z_{m}) \right)$.
By construction, $Z^*$ is exchangeable with all $Z_g^*$ for $g \in [G]$, and $X^*$ is exchangeable with all $X_g^*$ for $g \in [G]$.
This implies that the conformity scores $E^*_g = E(X_g^*,Y_g^*)$ are exchangeable with one another, for all $g \in [G]$, as well as with $E^* = E(X^*,Y^*)$.
The result is then established with the same argument as in the proof of Theorem~\ref{thm:coverage-cond}.
The true frequency $f_m(Z^*)$ is not covered by the output confidence interval if and only if $E^* > \hat{Q}_{G, 1-\alpha}$, whose probability is bound from above by $1-\alpha$ according to
classical results about tolerance regions
\citep{krishnamoorthy2009statistical}, see also
Lemma~1 in~\cite{romano2019conformalized}.

\subsection{Proof of Proposition \ref{unipro}}
\label{pf:uni}

\begin{proof}
To prove \eqref{q}, note that $V = \{a_{j_1},\ldots, a_{j_k}\}$ if and only if there is a $k$-composition $c =(c_1, \ldots,c_k)$ of $M$ such that, for all $l \in[k]$, the sequence $(Z_{m+1},\ldots,Z_{m+M}) $ = $(a_{t_1},$ $\ldots,a_{t_M})$ contains exactly $c_l$ values of $a_{j_l}$.
    For a given $k$-composition  $c=(c_1, \ldots,c_k)$, there are $\binom{M}{c_1\,c_2\,\ldots c_k}$
    indices $t_1,t_2,\ldots,t_M \in \NN$ such that for all $l \in [k]$, exactly $c_l$ of them are equal to $j_l$.
    The probability that $(Z_{m+1},\ldots,Z_{m+M})$ equals any one of them is $p_{j_1}^{c_1}\cdots p_{j_k}^{c_k}$, showing \eqref{q}.

To prove~\eqref{q2}, note that, for any $S\subset \NN$,
any product arising from the expansion of  $\left(\sum_{l\in S} p_l\right)^M$ has at least one and at most $M$ distinct indices $l$.
Collecting the products
$p_{i_1}p_{i_2}\ldots p_{i_M}$
by the number
$d \in \{1,\ldots, M\}$ of distinct indices among their factors, we find
$$
\left(\sum_{l\in S} p_l\right)^M
 =
\sum_{d=1}^{M}
\sum_{\{l_1,\ldots,l_d\}\subset S,\,\,
l_i\neq l_j\textnormal{ for } i\neq j}
\,
\sum_{c\in C_{M,d}}
\binom{M}{c_1\,c_2\,\ldots c_d}
p_{l_1}^{c_1}\cdots p_{l_d}^{c_d}.
$$
Now fix any
$\{l_1,\ldots,l_d\}\subset \{j_1,\ldots, j_k\}$,
and any $c\in C_{M,d}$.
Using the previous formula
for each $S$
on the right hand side of
\eqref{q2},
the total coefficient of
$p_{l_1}^{c_1}\cdots p_{l_d}^{c_d}$
is the following sum over subsets $S$
\begin{align*}
\binom{M}{c_1\,c_2\,\ldots c_d}
\sum_{S\subset \{j_1,\ldots, j_k\}} (-1)^{k+|S|}
I\left(\{l_1,\ldots,l_d\}\subset S\right).
\end{align*}
Writing the indicator $I\left(\{l_1,\ldots,l_d\}\subset S\right)$ inside the summation constraint, and factoring out $(-1)^{k}$, this equals
\begin{align*}
(-1)^{k}
\binom{M}{c_1\,c_2\,\ldots c_d}
\sum_{\{l_1,\ldots,l_d\}\subset S\subset \{j_1,\ldots, j_k\}} (-1)^{|S|}.
\end{align*}

Now, if $\{l_1,\ldots,l_d\} = \{j_1,\ldots, j_k\}$, the above summation (after the pre-factor) has only one term---$S=\{j_1,\ldots, j_k\}$---and equals
$(-1)^{|S|}=(-1)^{k}$.

Otherwise, the above summation contains $2^{k-d}>1$ terms.
We now construct a
pairing of the sets $S$ that index of the summation, such that each pair $(S_1,S_2)$  contains an odd and even sized set.
There must be an index $j_a$, $a\in[k]$,
such that $j_a \notin \{l_1,\ldots,l_d\}$.
Suppose without loss of generality that we have
$j_k \notin \{l_1,\ldots,l_d\}$ (otherwise rename the indices $j_a$ and $j_k$).

Then, for any set $S_1$ such that
$\{l_1,\ldots,l_d\}\subset S_1\subset \{j_1,\ldots, j_k\}$
that does not contain $j_k$,
there is a corresponding set $S_2 = S_1 \cup \{j_k\}$ such that
$\{l_1,\ldots,l_d\}\subset S'\subset \{j_1,\ldots, j_k\}$.
Moreover, all sets $S$
such that
$\{l_1,\ldots,l_d\}\subset S\subset \{j_1,\ldots, j_k\}$
fall into exactly one such pair.
Further, in each pair,
there is one set of an odd size and one set of an even size.

Thus, in each pair, we have
\begin{align*}
(-1)^{|S_1|}+(-1)^{|S_2|}=0,
\end{align*}
Therefore,
when
$\{l_1,\ldots,l_d\}\neq \{j_1,\ldots, j_k\}$
\begin{align*}
\sum_{\{j_1,\ldots,j_d\}\subset S\subset \{i_1,\ldots, i_k\}} (-1)^{|S|} = 0.
\end{align*}
Hence, the coefficient of
$p_{j_1}^{c_1}\cdots p_{l_d}^{c_d}$
in the expression on the right hand side of
\eqref{q2}
is nonzero only when
$\{l_1,\ldots,l_d\}= \{j_1,\ldots, j_k\}$, in which case it equals
\begin{align*}
(-1)^{2k}\binom{M}{c_1\,c_2\,\ldots c_d}
=
\binom{M}{c_1\,c_2\,\ldots c_d}.
\end{align*}
This shows that \eqref{q} and \eqref{q2} coincide, completing the proof.

\end{proof}

\subsection{Proof of Proposition \ref{uni2}}
\label{pf:uni2}

\begin{proof}
The formula in \eqref{uz} follows directly from Equation \eqref{q} in Proposition~\ref{unipro}, because for a set $V$ of size $k$, the probability of any element being the selected unique $\zeta$ equals $1/k$. Next, $U_Z^{[1]}= P_Z$ by definition. In addition,
\begin{align*}
U_Z^{[2]}(\zeta = a_{j_1})
&=
    p_{j_1}^2+
    \frac12
    \sum_{J=\{j_1,j_2\}\subset \NN^2, |J| = 2}\,
    \sum_{c\in C_{2,2}} \binom{2}{c_1\,c_2}
    p_{j_1}^{c_1} p_{j_2}^{c_2}\\
&=
    p_{j_1}^2 +
    \frac12
    \sum_{j_2\in \NN\setminus\{j_1\}}\,
    2 p_{j_1} p_{j_2} = p_{j_1}^2 + p_{j_1} (1-p_{j_1}) =  p_{j_1}.
\end{align*}
This shows that $U_Z^{[2]} = P_Z$. Finally,
\begin{align*}
    & U_Z^{[3]}(\zeta = a_{j_1}) \\
    & =
    p_{j_1}^3+
    \frac12
    \sum_{J=\{j_1,j_2\}\subset \NN^2, |J| = 2}\,
    \sum_{c\in C_{3,2}} \binom{3}{c_1\,c_2}
    p_{j_1}^{c_1}p_{j_2}^{c_2} +
    \frac13 \binom{3}{1\,1\,1}
    \sum_{J=\{j_1, j_2, j_3\}\subset \NN^3, |J| = 3}\,
     p_{j_1} p_{j_2} p_{j_3}\\
         & =
    p_{j_1}^3+
    \frac12
    \sum_{J=\{j_1,j_2\}\subset \NN^2, |J| = 2}\,
    \sum_{c\in C_{3,2}} \binom{3}{c_1\,c_2}
    p_{j_1}^{c_1}p_{j_2}^{c_2}+
    2\sum_{J=\{j_1, j_2, j_3\}\subset \NN^3, |J| = 3}\,
     p_{j_1} p_{j_2} p_{j_3}.
\end{align*}
This further equals
\begin{align*}
    &p_{j_1}^3+
    \frac32
    \sum_{j_2\in \NN\setminus\{j_1\}}\,
    (p_{j_1}^{2}p_{j_2}+p_{j_1}p_{j_2}^{2})
    +
     2p_{j_1} \sum_{\{j_2, j_3\}\subset (\NN\setminus\{j_1\})^2,\, |J|  = 2}\,
     p_{j_2} p_{j_3}\\
    &=
    p_{j_1}^3+
    \frac32
    p_{j_1}^{2}(1-p_{j_1})
    +
    p_{j_1}
    \left(\frac32
    \sum_{j_2\in \NN\setminus\{j_1\}}\,
    p_{j_2}^{2}
    +
    2\sum_{\{j_2, j_3\}\subset (\NN\setminus\{j_1\})^2,\, |J|  = 2}\,
     p_{j_2} p_{j_3}\right).
\end{align*}
By expanding the square in $(1-p_{j_1})^2 = (\sum_{j_2\in \NN\setminus\{j_1\}} p_{j_2})^2$, this further equals
\begin{align*}
    &
    \frac
    {p_{j_1}^{2}(3-p_{j_1})}
    {2}
    +
    p_{j_1}
    \left(\frac32
    \left[
    (1-p_{j_1})^2-\sum_{\{j_2, j_3\}\subset (\NN\setminus\{j_1\})^2,\, |J|  = 2}\,
     p_{j_2} p_{j_3}
    \right] \right.\\
    &\qquad\qquad\qquad\left.+
    2\sum_{\{j_2, j_3\}\subset (\NN\setminus\{j_1\})^2,\, |J|  = 2}\,
     p_{j_2} p_{j_3}\right)\\
    &=
    \frac
    {p_{j_1}(2p_{j_1}^2-3p_{j_1}+3)}
    {2}
    +
    \frac{p_{j_1}}{2}
    \sum_{\{j_2, j_3\}\subset (\NN\setminus\{j_1\})^2,\, |J|  = 2}\,
     p_{j_2} p_{j_3}.
\end{align*}
This finishes the proof.
\end{proof}

\subsection{Proof of Proposition \ref{d2}}
\label{pf:d2}
\begin{proof}
Let the two objects be denoted by $a_1$ and $a_2$.
Then, one can verify using \eqref{q2} in Proposition~\ref{unipro} and \eqref{uz} in Proposition~\ref{uni2} that, for $j=1,2$,
    \begin{equation}\label{si}
    U_Z^{[M]}(\zeta = a_{j})
    =
    \frac{1+p_{j}^M-(1-p_{j})^M}2.
\end{equation}
Therefore,
    $$\Delta(M,M';2) =
    \frac12
    \sup_{p\in[0,1]}
    \left|
    p^M-(1-p)^M
    - \left[p^{M'}-(1-p)^{M'} \right]\right|.
    $$
   Let $\delta = (1-p)/p \ge 0$, so that $p=1/(1+\delta)$, and suppose without loss of generality that $\delta\le 1$; otherwise, change variables to $1-p \gets p$.
    Then, the term inside the absolute value above can be written as
\begin{equation}\label{A}
A(\delta)
    =
    \frac{1-\delta^{M'}}{(1+\delta)^{M'}}-
    \frac{1-\delta^M}{(1+\delta)^M}
    \ge 0.
\end{equation}
Now, denoting, for $c\ge 1$, $g(\delta,c) = \frac{1-\delta^{c}}{(1+\delta)^{c}}$, we have
\begin{align*}
    \frac{\partial g(\delta,c)}{\partial \delta}
    &=
    \frac{-c\delta^{c-1}(1+\delta)^{c}-(1-\delta^{c})\cdot c(1+\delta)^{c-1}}{(1+\delta)^{2c}}\\
    &=
    -c\frac{\delta^{c-1}(1+\delta)+(1-\delta^{c})}{(1+\delta)^{c+1}}
    =
    -c\frac{1+\delta^{c-1}}
    {(1+\delta)^{c+1}}.
\end{align*}
Hence,
    $$
    A'(\delta)
    =
    -M'\frac{1+\delta^{M'-1}}
    {(1+\delta)^{M'+1}}
    +M\frac{1+\delta^{M-1}}
    {(1+\delta)^{M+1}}.
    $$
Thus, $A'(\delta) \ge 0$ is equivalent to
    $$
    \frac{1+\delta^{M-1}}{1+\delta^{M'-1}}
    \ge \frac{M'}{M}      (1+\delta)^{M-M'},
    $$
or, with the function $h$ defined as in \eqref{h}, to
$h(\delta)\ge \ln(M'/M)$.
Now,
$$
    h'(\delta)=
    \frac{(M-1)\delta^{M-2}}{1+\delta^{M-1}}
    -
    \frac{(M'-1)\delta^{M'-2}}{1+\delta^{M'-1}}
    -\frac{M-M'}{1+\delta}.
$$
We claim that $h'(\delta)<0$ for all $\delta \in [0,1)$.
Indeed, this is equivalent to the function
$$
    \tilde{\psi}(M)=
    \frac{M}{1+\delta}-
    \frac{(M-1)\delta^{M-2}}{1+\delta^{M-1}}
$$
being increasing in $M$, for all $M\ge 2$.
Denote $x = M-1\ge 1$, $\psi (x) = \delta \cdot \tilde{\psi}(x+1)$, and $a = 1/\delta \ge 1$.
Then,
$$
    \psi(x)=
    \frac{x+1}{1+a}-
    \frac{x}{1+a^x}
$$
and
$$
    \psi'(x)=
    \frac{1}{1+a}-
    \frac{1+a^x-x a^x\ln a }{(1+a^x)^2}.
$$
Hence, $\psi'(x)> 0$ is equivalent to
$$
   (1+a)(1+a^x-x a^x\ln a )< (1+a^x)^2.
$$
Now, since $a \ge 1$ and $x\ge 1$,
we have $1+a\le 1+a^x$
and $x a^x\ln a \ge 0$.
Equality happens in both equations if and only if $x = 1$ and $a=1$.
This corresponds to $M=2$ and $\delta = 1$.
Thus, the above inequality holds for all $\delta \in [0,1)$.
This shows that $h$ is decreasing for $\delta \in [0,1)$.
Since $h(0) = 0$ and $h(1) = M'-M \le \ln(M'/M)$,
and as $h$ is continuous on $[0,1]$, there is a unique solution $\delta_* \in [0,1]$ to $h(\delta_*)=\ln(M'/M)$.
This proves the first claim.
Based on our analysis, it follows that
$A$ is maximized over $[0,1]$ at $\delta_*$.
This finishes the proof.
\end{proof}

\subsection{Proof of Corollary \ref{d3}}
\label{pf:d3}
\begin{proof}
%For $M=aM'$,
Recalling the form of the function $h$ from \eqref{h},
the equation for $\delta\in [0,1]$ from
Proposition \ref{d2} is
\begin{equation*}
    \frac{M}{M'} (\delta^{M-1}+1)
=
(1+\delta^{M'-1})(1+\delta)^{M-M'}.
\end{equation*}
For $M=aM'$, this becomes
\begin{equation}\label{dm}
    a (\delta^{aM'-1}+1)
=
(1+\delta^{M'-1})(1+\delta)^{(a-1)M'}.
\end{equation}

Now, as $a\ge 1$ and $\delta\le 1$, we have
$2 \ge (1+\delta)^{(a-1)M'}$, and thus
using the inequality $2\le (1 + 1/x)^x$ for $x\ge 1$ with $x = (a-1)M'$, we find that\footnote{All inequalities in this argument will hold for $M'$ sufficiently large, and having determined the required range, we will not repeatedly specify it.} as soon as $M' \ge 1/(a-1)$,
$$
1+\delta
\le
2^{1/[(a-1)M']}
\le
1+ \frac{1}{(a-1)M'}.
$$
Hence,
$
\delta
\le
\frac{1}{(a-1)M'};
$
and for ${M'}\ge 2/(a-1)$,
we thus find $\delta\le 1/2$.
Using this in \eqref{dm}, we obtain
\begin{equation*}
    \frac{a}{1+2^{-(M'-1)}} \le
(1+\delta)^{(a-1)M'}
\le  a\left(1+2^{-(aM'-1)}\right).
\end{equation*}

%Given two sequences $(A_{M'})_{{M'}\ge 1}$, $(B_{M'})_{{M'}\ge 1}$,
%we will write $A_{M'}\sim B_{M'}$, if as ${M'}\to \infty$, $|A_{M'}/B_{M'}-1|=O(1/{M'})$.
%Denoting by $\delta$ the unique solution of \eqref{dm},
%from the above display, we find
%(1+\delta)^{(a-1)M'} \sim a$.
Therefore,
\begin{equation*}
|(1+\delta)^{(a-1)M'}-a|
\le  a
\max\left\{2^{-(aM'-1)}, \frac{2^{-(M'-1)}}{1+2^{(M'-1)}}\right\}
\le
2^{1-M'} a.
\end{equation*}
Hence, 
using
that $x\mapsto x^{-a/(a-1)}$ is decreasing on $(0,\infty)$,
as well as the inequality $1> (1-1/x)^c\ge 1-c/x$ for all $x,c\ge 1$, applied to $x = 2^{M'-1}$ and $c = a/(a-1)$,
\begin{align*}
&\left| \frac{1}{(1+\delta)^{aM'}}
    -a^{-a/(a-1)}\right|
    \le 
    \left|\left[a(1-2^{1-M'})\right]^{-a/(a-1)} 
    -a^{-a/(a-1)}\right|\\
    &= 
    a^{-a/(a-1)} \left[\left(1-2^{1-M'}\right)^{-a/(a-1)} 
    -1\right]\\
    &\le 
    a^{-a/(a-1)} \left[\left(1-2^{1-M'}a/(a-1)\right)^{-1} 
    -1\right]
    \le 2a^{-a/(a-1)} 2^{1-M'}a/(a-1),
\end{align*}
as long as $2^{1-M'}a/(a-1) \le 1/2$, i.e., $M'\ge 2+ \log_2(a/(a-1))$.
Similarly,
\begin{align*}
    \left|\frac{1}{(1+\delta)^{M'}}
    -a^{-1/(a-1)}\right|
        \le 2a^{-1/(a-1)} 2^{1-M'}/(a-1),
\end{align*}
as long as $2^{1-M'}/(a-1) \le 1/2$, i.e., $M'\ge 2+ \log_2(1/(a-1))$.
% Therefore,
% $(1+\delta)^{M'}\sim a^{1/(a-1)}$
% and
% $(1+\delta)^{aM'}\sim a^{a/(a-1)}$.
Thus, with $A$ from \eqref{A}, using also that $\delta \le 1/[(a-1)M']$,
\begin{align*}
&|A(\delta)-a^{-1/(a-1)}(1-1/a)|
     =
    \left|\frac{1-\delta^{M'}}{(1+\delta)^{M'}}-
    \frac{1-\delta^{aM'}}{(1+\delta)^{aM'}}
    -\left(a^{-1/(a-1)}-a^{-a/(a-1)}\right)\right|\\
    &\le 
    \left|\frac{1}{(1+\delta)^{M'}}
    -a^{-1/(a-1)}\right|
    +\left|
    \frac{1}{(1+\delta)^{aM'}}
    -a^{-a/(a-1)}\right|
    +\frac{\delta^{M'}}{(1+\delta)^{M'}}+
    \frac{\delta^{aM'}}{(1+\delta)^{aM'}}\\
    &\le 
    2^{2-M'} \left[a^{-a/(a-1)}a/(a-1)+a^{-1/(a-1)}/(a-1)\right]
    +2 [(a-1)M']^{-M'}\\
    &\le 
    2^{3-M'}a^{-1/(a-1)}/(a-1)
    +2 [(a-1)M']^{-M'}.
\end{align*}
Finally,
denoting by $\delta_*$ the unique solution of \eqref{dm},
$\Delta(M,M'; 2) = A(\delta_*)$ from the proof of Proposition \ref{d2}, completing this proof.
\end{proof}

\subsection{Proof of Theorem~\ref{eq:coverage-unique-robust}}

\begin{proof}
This follows immediately by combining Corollary~\ref{d3} with \eqref{ee} and Theorem \ref{eq:coverage-unique}.
\end{proof}

\subsection{Proof of Theorem~\ref{thm:shift}} \label{pf:uni3}

\begin{proof}
First, we aim to show that, for $p_j \neq p_j'$,
\begin{equation}\label{bd}
    |U_Z^{[M]}(\{a_{j}\})-U_{Z'}^{[M]}(\{a_{j}\})| <|p_j-p'_j|.
\end{equation}
For simplicity of notation, define $p:=p_i$ and $q := p'_i$.
Define also $d:[0,1]\to \mathbb{R}$ as
$d(p)  = [p^M-(1-p)^M]/2$, for all $p\in [0,1]$.
It follows from \eqref{si} that
we need to show
\begin{equation*}
    \left| d(p) -d(q)\right| <|p-q|.
\end{equation*}
Suppose without loss of generality that $p< q$.
By the mean value theorem applied to $d$, there exists a $\omega\in [p,q]$, such that $d(p) -d(q) = d'(\omega)(p-q)$.
Therefore, it suffices to show that $|d'(\omega)|<1$ for $\omega \in (c,1-c)$.
Now,
$d'(\omega) = M[\omega^{M-1}+(1-\omega)^{M-1}]/2$.
We note here that $d'(0) = M/2>1$, $d'(1/2) = M/2^{M-1}<1$ (as $M\ge 3$), and $d'$ is strictly decreasing as a function of $\omega$ for $\omega \in [0,1/2]$.
Therefore, the equation $d'(c) = 1$
has a unique solution over $c\in [0,1/2)$.
This shows that $c$ in \eqref{c} is well defined.

Moreover, because $d'$ is strictly decreasing between $[0,1/2]$, it follows that $d'$ is maximized within the interval $[c,1-c]$ at $c$ and (by symmetry) at $1-c$.
Therefore,
$|d'(\omega)|<1$ for $\omega \in (c,1-c)$,
and \eqref{bd} follows.

Let $TV(\cdot,\cdot)$ be the total variation distance.
Then, for all $P_Z,P_{Z'}\in S_c$, with $P_Z\neq P_{Z'}$,
\begin{equation}\label{tvb}
TV(U_Z^{[M]},U_{Z'}^{[M]}) < TV(P_Z,P_{Z'}).
\end{equation}

Following \eqref{ee},
define
$e_U(Z^*) = \mathbb{P}_{\tilde Z_{1:G}\sim \left(U_Z^{[M]}\right)^{|G|} }
 [\mathcal{E}]$
 and
$e(Z^*) = \mathbb{P}_{\tilde Z_{1:G}\sim P_Z^{|G|} }
 [\mathcal{E}]$.
Let $A_U,A$ be the sets of functions over which $e_U,e$ can range, respectively.
We need to show that
\begin{align*}
 &\sup_{
 e_U \in A_U}
 \left| \mathbb{E}_{Z^*\sim U_Z^{[M]}}\, e_U(Z^*)
 -  \mathbb{E}_{Z^*\sim U_{Z'}^{[M]}}\, e_U(Z^*)\right| <
 \sup_{
 e \in A}
 \left| \mathbb{E}_{Z^*\sim P_Z}\, e(Z^*)
 -  \mathbb{E}_{Z^*\sim P_{Z'}}\, e(Z^*)\right|.
\end{align*}
Because $\mathcal{E}\subset \{a_1,a_2\}^{|G|+1}$ is arbitrary, the possible values of $e_U$ include zero and unity,
for any  value $Z^*=z$.
Hence, the above inequality is equivalent to \eqref{tvb}, completing the proof.
\end{proof}

\clearpage

\section{Supplementary figures and tables} \label{app:figures}

\subsection{Theoretical analysis of robustness to sample inflation}

\begin{figure}[!htb]
(a) \hspace*{0.45\textwidth} (b) \\
\vspace{-2.5em}
\begin{center}
\includegraphics[width=0.45\textwidth]{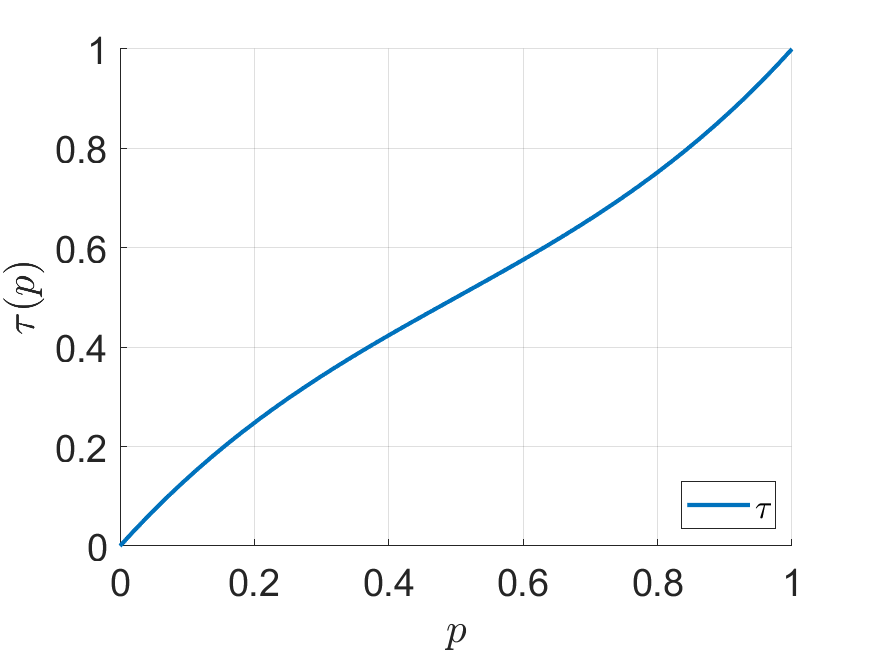}
\hfill
\includegraphics[width=0.45\textwidth]{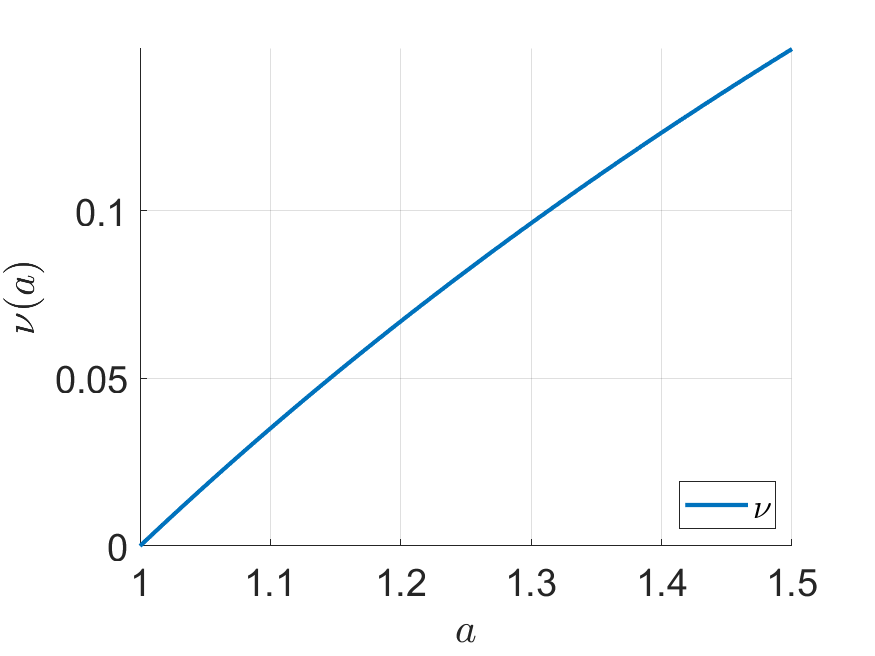}
\caption{(a) Plot of the function $\tau$ defined in \eqref{t}. (b) Plot of the function $a\mapsto \nu(a):=a^{-1/(a-1)}(1-1/a)$ defined in Corollary~\ref{d3}.}
\label{fig:t-nu}
\end{center}
\end{figure}

\FloatBarrier
\subsection{Experiments with synthetic Zipf data}

\begin{figure}[!htb]
\begin{center}
\includegraphics[width=\linewidth]{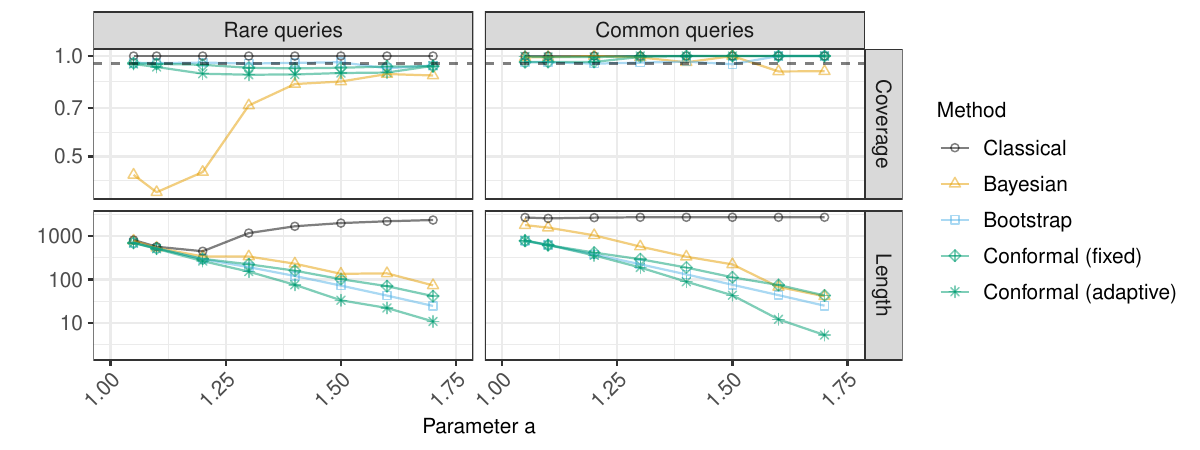}
\caption{Performance of confidence intervals stratified by the true query frequency. Left: frequency below median; right: above median. Here, the conformal confidence intervals are constructed using Algorithm~\ref{alg:conformal-sketch}, which seeks marginal coverage~\eqref{eq:marginal-coverage}, instead of Algorithm~\ref{alg:conformal-sketch-frequency}, which seeks frequency-range conditional coverage~\eqref{eq:conf-int-cond}. Other details are as in Figure~\ref{fig:exp-zipf-marginal}.}
\label{fig:exp-zipf-conditional-1bin}
\end{center}
\end{figure}

\begin{figure}[!htb]
\begin{center}
\includegraphics[width=0.75\linewidth]{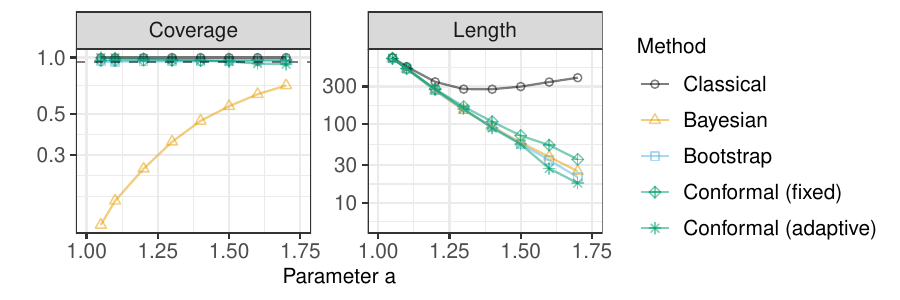} \\[-1em]
\caption{Performance of confidence intervals for random queries on synthetic Zipf data, keeping only distinct queries. The coverage is the empirical proportion of distinct queries whose frequency is covered by the output confidence intervals. The conformal confidence intervals are computed by applying Algorithm~\ref{alg:conformal-sketch-frequency} with $L=5$ frequency bins. Other details are as in Figure~\ref{fig:exp-zipf-marginal}.}
\label{fig:exp-zipf-unique-cms-bins5}
\end{center}
\end{figure}

\begin{figure}[!htb]
\begin{center}
\includegraphics[width=0.75\linewidth]{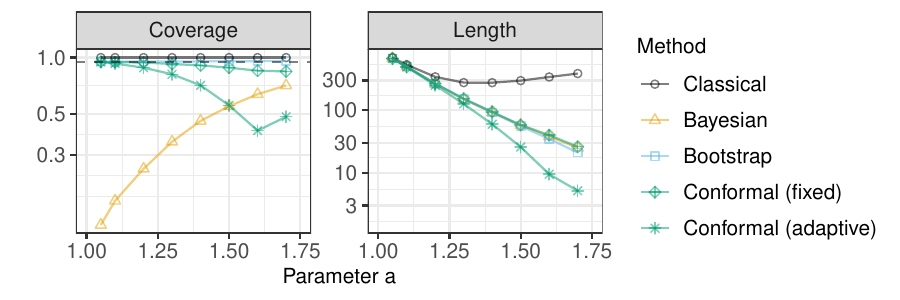} \\[-1em]
\caption{Performance of confidence intervals for random queries on synthetic Zipf data, keeping only distinct queries. The conformal confidence intervals are computed by applying Algorithm~\ref{alg:conformal-sketch}, seeking marginal coverage~\eqref{eq:marginal-coverage}, instead of Algorithm~\ref{alg:conformal-sketch-frequency}. Other details are as in Figure~\ref{fig:exp-zipf-unique-cms-bins5}.}
\label{fig:exp-zipf-unique-cms-bins1}
\end{center}
\end{figure}

\begin{figure}[!htb]
\begin{center}
\includegraphics[width=0.8\linewidth]{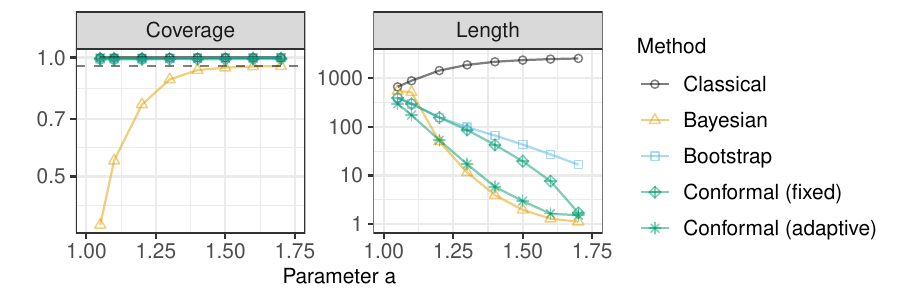}
\caption{Performance of 95\% confidence intervals with simulated Zipf data sketched with the CMS. The results are shown as a function of the Zipf tail parameter $a$. The data are sketched with the CMS-CU instead of the CMS. Other details are as in Figure~\ref{fig:exp-zipf-marginal}.}
\label{fig:exp-zipf-marginal-cu}
\end{center}
\end{figure}

\begin{figure}[!htb]
\begin{center}
\includegraphics[width=\linewidth]{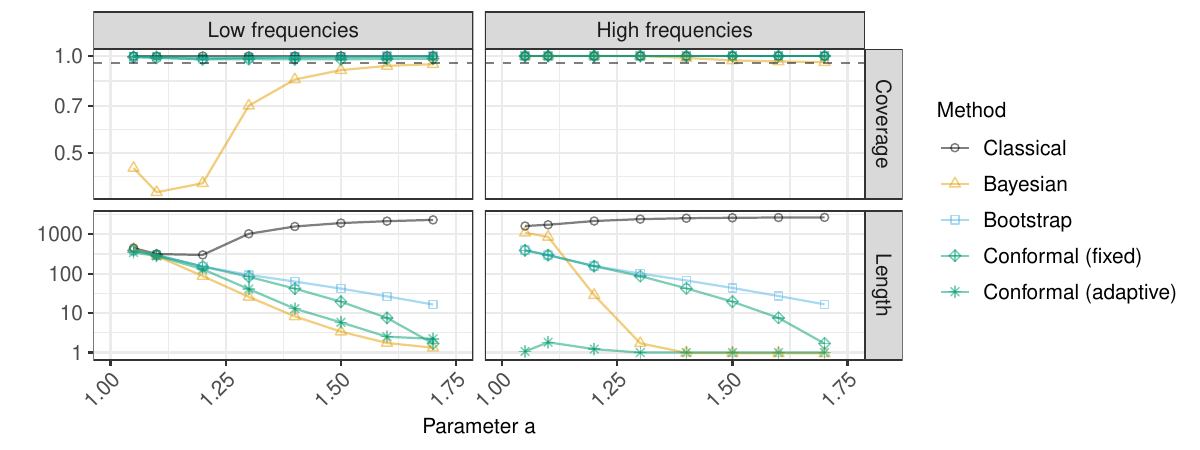}
\caption{Performance of confidence intervals stratified by the true query frequency. Left: frequency below median; right: above median. The data are sketched with the CMS-CU instead of the CMS. Other details are as in Figure~\ref{fig:exp-zipf-conditional}.}
\label{fig:exp-zipf-conditional-cu}
\end{center}
\end{figure}

\begin{figure}[!htb]
\begin{center}
\includegraphics[width=\linewidth]{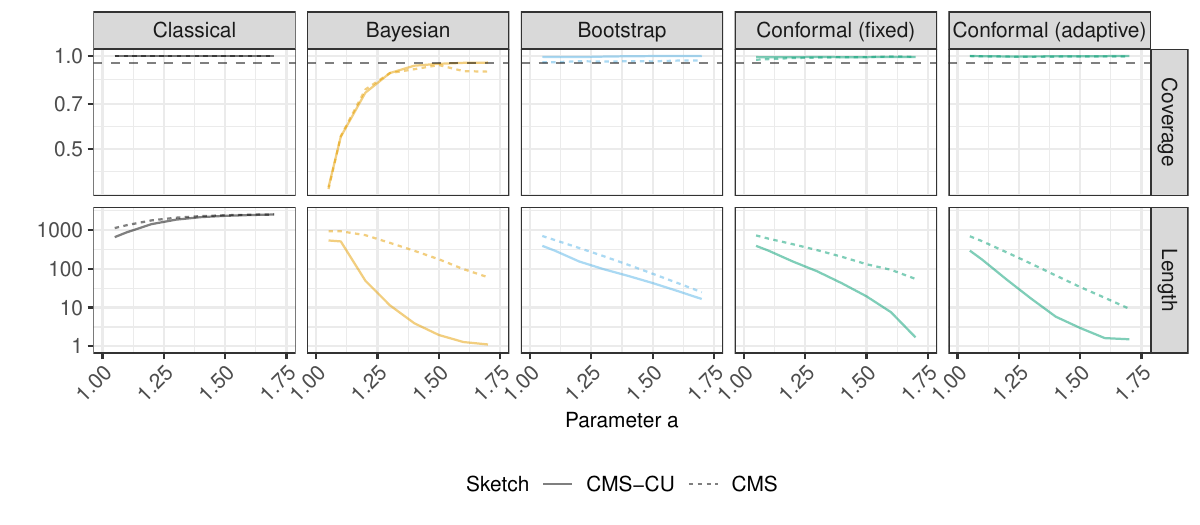} \\[-1em]
\caption{Performance of 95\% confidence intervals for random queries, based on synthetic data from a Zipf distribution. The data are sketched with either the vanilla CMS or the CMS-CU. The results are shown as a function of the Zipf tail parameter $a$.
Other details are as in Figure~\ref{fig:exp-zipf-marginal}.}
\label{fig:exp-zipf-sketch}
\end{center}
\end{figure}

\begin{figure}[!htb]
\begin{center}
\includegraphics[width=0.75\linewidth]{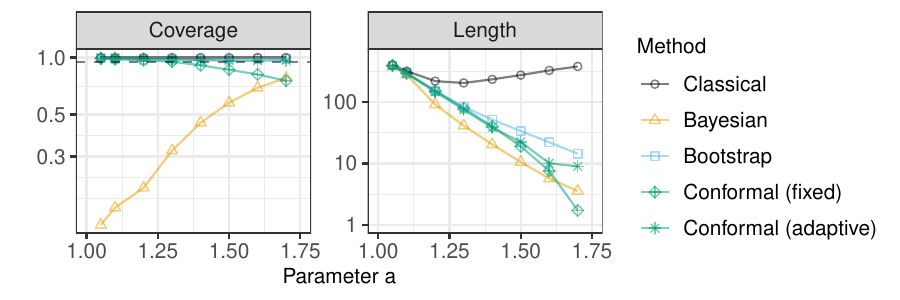} \\[-1em]
\caption{Performance of confidence intervals for random queries on synthetic Zipf data, keeping only distinct queries. The coverage is the empirical proportion of distinct queries whose frequency is covered by the output confidence intervals. The conformal confidence intervals are computed by applying Algorithm~\ref{alg:conformal-sketch-frequency} with $L=5$ frequency bins. The data are sketched with the CMS-CU instead of the CMS. Other details are as in Figure~\ref{fig:exp-zipf-unique-cms-bins5}.}
\label{fig:exp-zipf-unique-cms-cu-bins5}
\end{center}
\end{figure}

\begin{figure}[!htb]
\begin{center}
\includegraphics[width=0.75\linewidth]{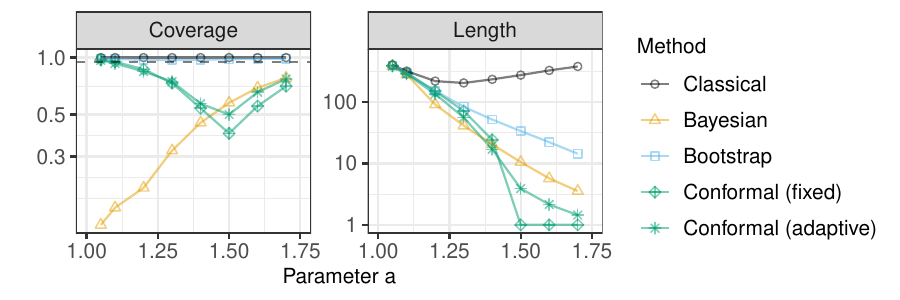} \\[-1em]
\caption{Performance of confidence intervals for random queries on synthetic Zipf data, keeping only distinct queries, after sketching the CMS-CU. The conformal confidence intervals are computed by applying Algorithm~\ref{alg:conformal-sketch}, seeking marginal coverage~\eqref{eq:marginal-coverage}, instead of Algorithm~\ref{alg:conformal-sketch-frequency}. Other details are as in Figure~\ref{fig:exp-zipf-unique-cms-cu-bins5}.}
\label{fig:exp-zipf-unique-cms-cu-bins1}
\end{center}
\end{figure}

\begin{figure}[!htb]
\begin{center}
\includegraphics[width=0.75\linewidth]{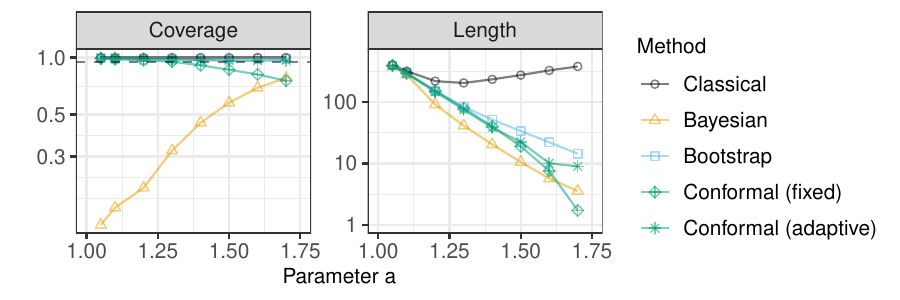} \\[-1em]
\caption{Performance of confidence intervals for random queries on synthetic Zipf data, keeping only distinct queries. The coverage is the empirical proportion of distinct queries whose frequency is covered by the output confidence intervals. The conformal confidence intervals are computed by applying Algorithm~\ref{alg:conformal-sketch-frequency} with $L=5$ frequency bins. Other details are as in Figure~\ref{fig:exp-zipf-marginal-cu}.}
\label{fig:exp-zipf-unique}
\end{center}
\end{figure}

\FloatBarrier
\subsection{Non-random sketching with data-driven hash functions} \label{sec:experiments-ML}

\begin{figure}[!htb]
\begin{center}
\includegraphics[width=0.8\linewidth]{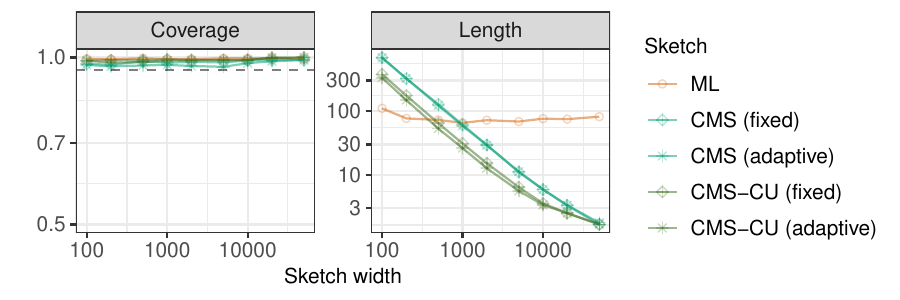}
\caption{Performance on simulated Zipf data of conformal confidence intervals based on different data sketches, as a function of the sketch width. The conformity scores are evaluated separately within $L=5$ frequency bins, seeking frequency-range conditional coverage~\eqref{eq:conf-int-cond}. Other details are as in Figure~\ref{fig:exp-zipf-opt-w}.}
\label{fig:exp-zipf-opt-w-5bins}
\end{center}
\end{figure}

\begin{figure}[!htb]
\begin{center}
\includegraphics[width=0.9\linewidth]{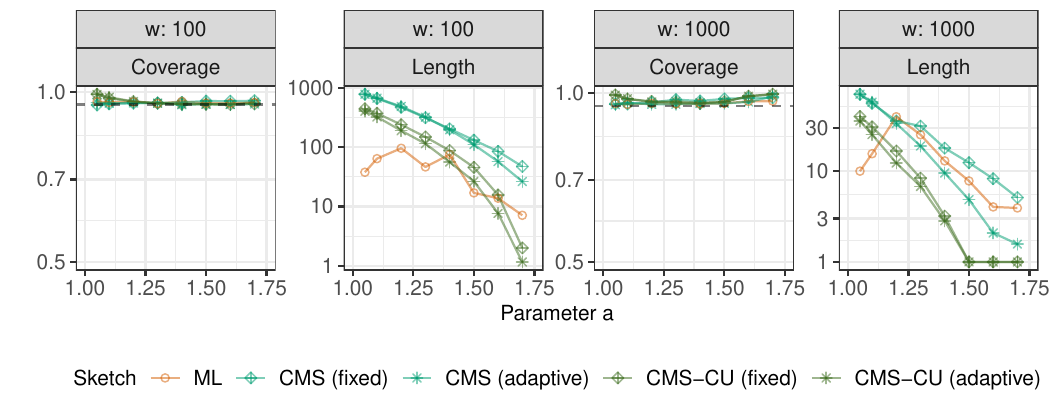}
\caption{Performance of 95\% confidence intervals based on different data sketches, either with width $w=100$ or $w=1000$. The results are shown as a function of the tail parameter of the Zipf distribution from which the data are sampled. Other details are as in Figure~\ref{fig:exp-zipf-opt-w}.}
\label{fig:exp-zipf-opt}
\end{center}
\end{figure}

\begin{figure}[!htb]
\begin{center}
\includegraphics[width=0.9\linewidth]{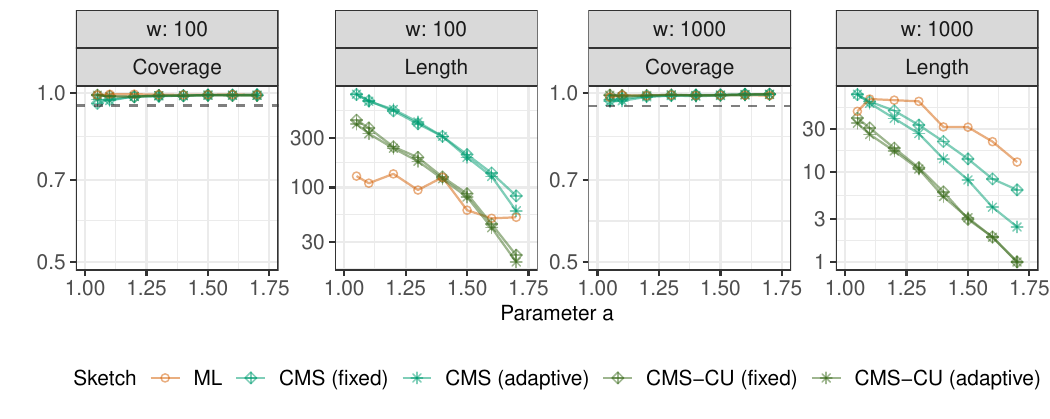}
\caption{Performance of 95\% confidence intervals based on different data sketches, either with width $w=100$ or $w=1000$. The results are shown as a function of the tail parameter of the Zipf distribution from which the data are sampled. The conformity scores are evaluated separately within $L=5$ frequency bins, seeking frequency-range conditional coverage~\eqref{eq:conf-int-cond}. Other details are as in Figure~\ref{fig:exp-zipf-opt}.}
\label{fig:exp-zipf-opt-5bins}
\end{center}
\end{figure}

\FloatBarrier
\subsection{Experiments with synthetic Pitman-Yor prior data}

\begin{figure}[!htb]
\begin{center}
\includegraphics[width=0.75\linewidth]{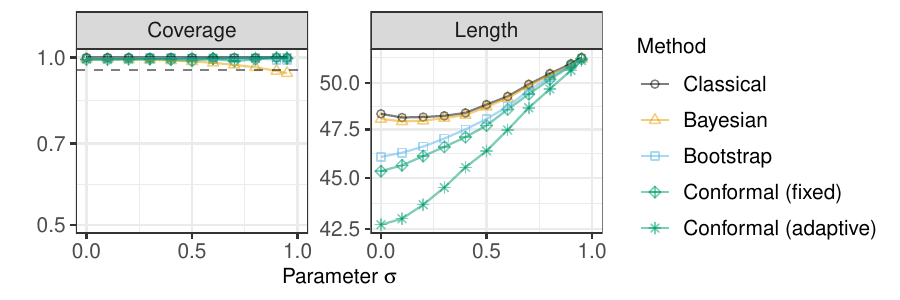} \\[-1em]
\caption{Empirical coverage and length of 95\% confidence intervals for random queries on synthetic data from the predictive distribution of a Pitman-Yor process. The data are sketched with the CMS-CU. The results are shown as a function of the Pitman-Yor process parameter $\sigma$. Other details are as in Figure~\ref{fig:exp-zipf-marginal}.}
\label{fig:exp-pyp-marginal}
\end{center}
\end{figure}

\begin{figure}[!htb]
\begin{center}
\includegraphics[width=\linewidth]{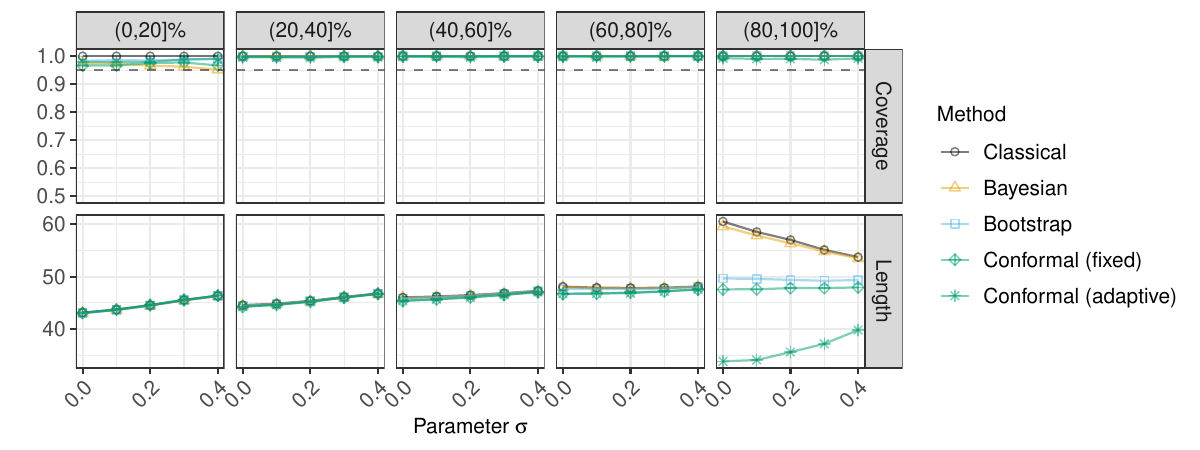} \\[-1em]
\caption{Performance of confidence intervals for random queries on synthetic data from the predictive distribution of a Pitman-Yor process. The results are stratified by the quintile of the true query frequency. Other details are as in Figure~\ref{fig:exp-pyp-marginal}.}
\label{fig:exp-pyp-conditional}
\end{center}
\end{figure}

\begin{figure}[!htb]
\begin{center}
\includegraphics[width=\linewidth]{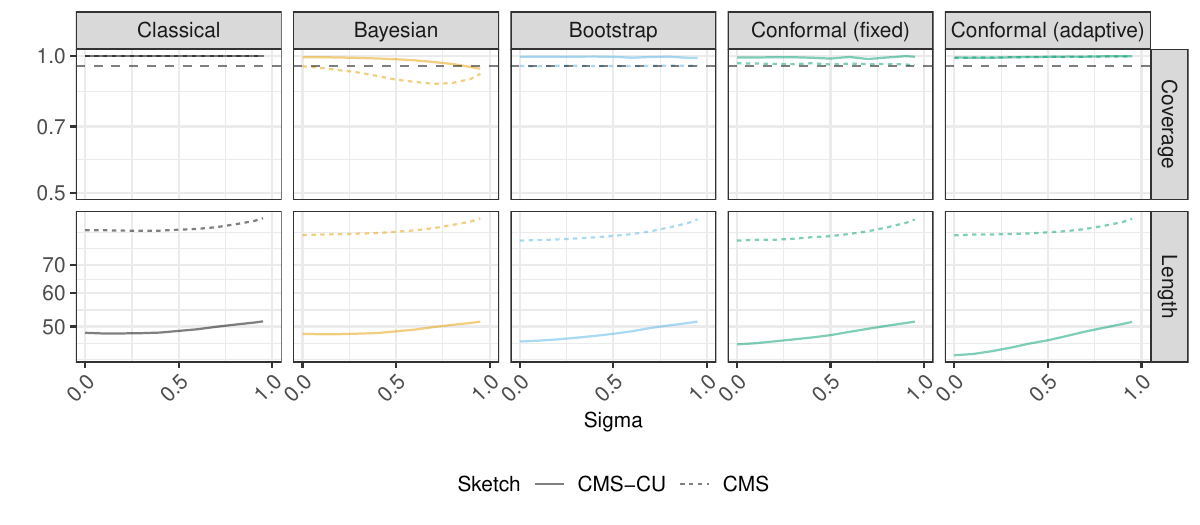} \\[-1em]
\caption{Performance of 95\% confidence intervals for random queries, based on synthetic data from the predictive distribution of a Pitman-Yor process and sketched with either the vanilla CMS or the CMS-CU. The results are shown as a function of the Pitman-Yor process parameter $\sigma$.
Other details are as in Figure~\ref{fig:exp-pyp-marginal}.}
\label{fig:exp-pyp-sketch}
\end{center}
\end{figure}

\begin{figure}[!htb]
\begin{center}
\includegraphics[width=0.75\linewidth]{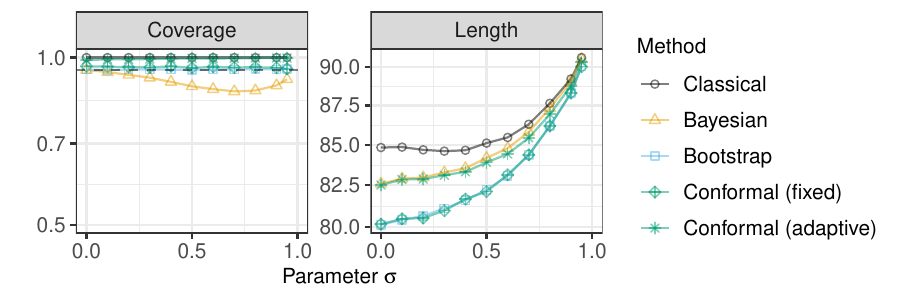} \\[-1em]
\caption{Performance of 95\% confidence intervals for random queries, based on synthetic data from the predictive distribution of a Pitman-Yor process and sketched with the vanilla CMS. The results are shown as a function of the Pitman-Yor process parameter $\sigma$.
Other details are as in Figure~\ref{fig:exp-pyp-marginal}.}
\label{fig:exp-pyp-cms}
\end{center}
\end{figure}

\FloatBarrier
\subsection{Experiments with heavy-hitter synthetic data}

\begin{figure}[!htb]
\begin{center}
\includegraphics[width=0.75\linewidth]{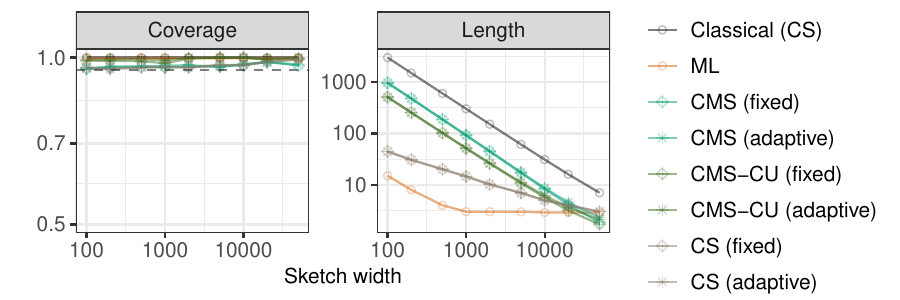} \\[-1em]
\caption{Performance of 95\% confidence intervals for random queries, based on synthetic data including heavy hitters. The data are generated according to the following probability distribution: $Z=0$ with probability $1/\sqrt{m}$, where $m=100,000$, and $Z \sim \text{Unif}(0,1)$ otherwise. The results are shown as a function of the hash width. Other details are as in Figure~\ref{fig:exp-zipf-marginal}. Note that the classical worst-case error bound for the CS is similar to that for the CMS, as described in  Appendix~\ref{sec:cms-classical-lower}. Specifically, this bound is derived by combing Markov's inequality with a Chernoff bound argument; e.g., see \cite{cormode2020small} for further details.}
\label{fig:exp-hh}
\end{center}
\end{figure}

\begin{figure}[!htb]
\begin{center}
\includegraphics[width=0.75\linewidth]{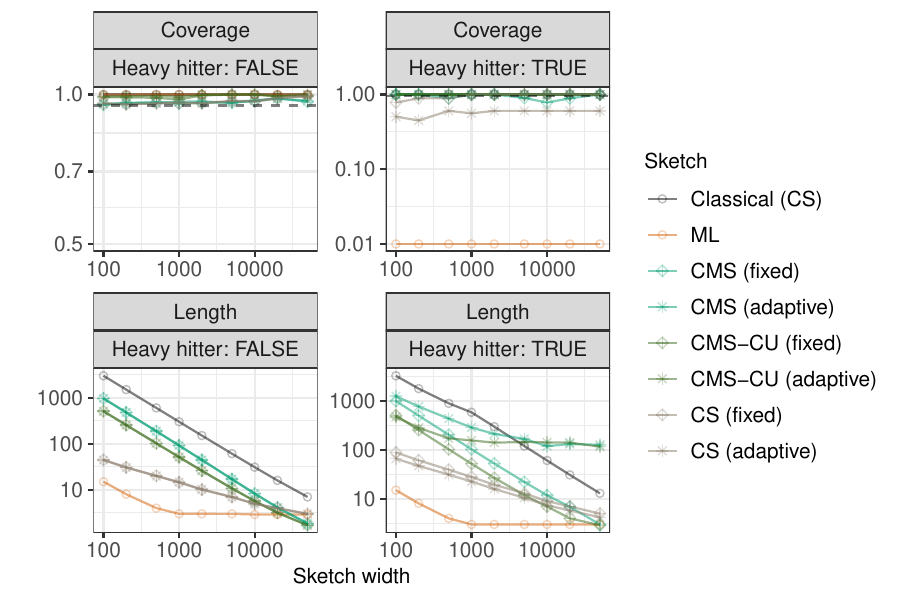} \\[-1em]
\caption{Performance of 95\% confidence intervals for random queries, based on synthetic data with heavy hitters. The coverage and average width of the conformal confidence intervals is reported separately for heavy hitters and all other objects. Other details are as in Figure~\ref{fig:exp-hh}. These results show that the CS sketch applied in combination with our method leads to the most informative confidence intervals---it achieves shorter width while ensuring valid coverage conditional on whether the queried object is a heavy hitter. Note that the true conditional coverage obtained with ML sketch for heavy hitter queries is zero, even though a truncated value of 0.01 is shown on the logarithmic scale for convenience.}
\label{fig:exp-hh-cond}
\end{center}
\end{figure}

\clearpage
\FloatBarrier
\subsection{Experiments with two-sided confidence intervals}

\begin{figure}[!htb]
\begin{center}
\includegraphics[width=0.75\linewidth]{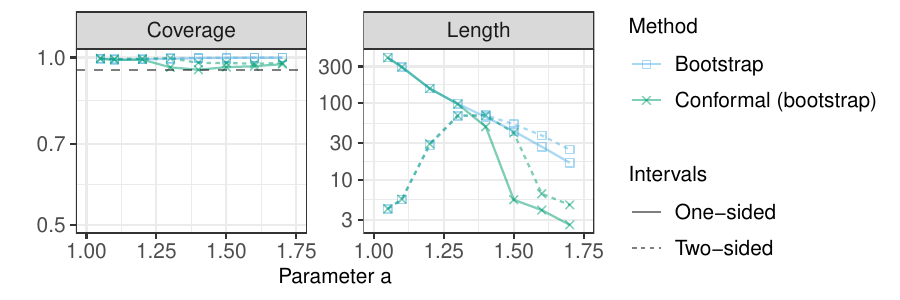}
\caption{Performance of 95\% one-sided and two-sided confidence intervals with data from a Zipf distribution, sketched with the CMS-CU. The results are shown as a function of the Zipf tail parameter $a$. Standard errors would be too mall to be clearly visible in this figure, and are hence omitted. The two dashed curves for the two-sided intervals are nearly indistinguishable from one another for $a < 1.3$.
Other details are as in Figure~\ref{fig:exp-zipf-marginal}.}
\label{fig:exp-zipf-marginal-two-sided}
\end{center}
\end{figure}

\begin{figure}[!htb]
\begin{center}
\includegraphics[width=0.75\linewidth]{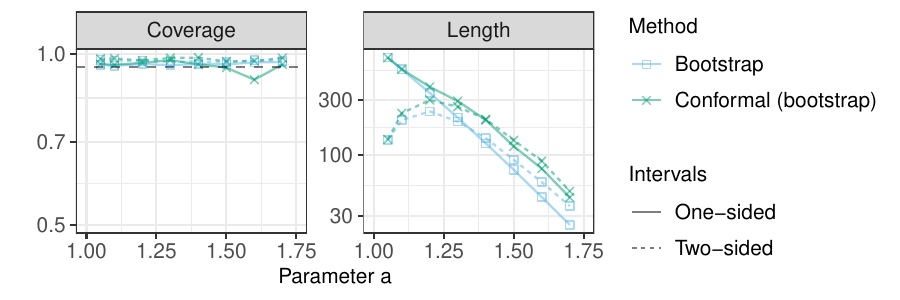}
\caption{Performance of 95\% one-sided and two-sided confidence intervals with data from a Zipf distribution, sketched with the vanilla CMS. The results are shown as a function of the Zipf tail parameter $a$. The two dashed curves for the two-sided intervals are nearly indistinguishable from one another for $a < 1.1$.
Other details are as in Figure~\ref{fig:exp-zipf-marginal-two-sided}.}
\label{fig:exp-zipf-marginal-two-sided-cms}
\end{center}
\end{figure}

%Figure~\ref{fig:exp-pyp-marginal-two-sided} reports on results based on data generated from a Pitman-Yor process prior and sketched with the CMS-CU, similarly to Figure~\ref{fig:exp-pyp-marginal}. As expected, the conformal confidence intervals are narrower than the bootstrap ones. Further, two-sided confidence intervals are much more efficient (narrower) compared to their one-sided counterparts, especially if the Pitman-Yor parameter $\sigma$ is large and the number of hash collisions is high.
%Figure~\ref{fig:exp-pyp-marginal-two-sided-cms} reports on analogous results obtained with data sketched through the vanilla CMS instead of the CMS-CU.

\begin{figure}[!htb]
\begin{center}
\includegraphics[width=0.75\linewidth]{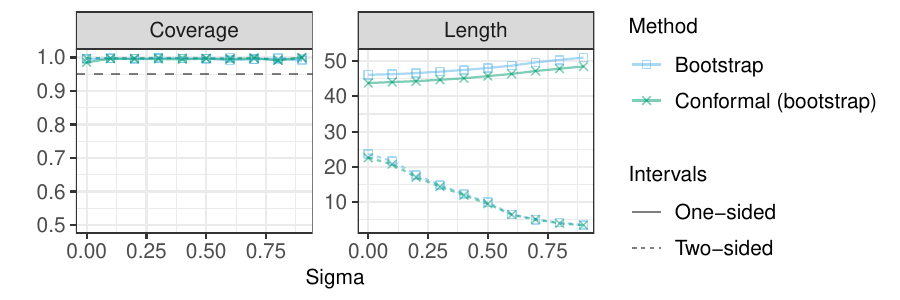}
\caption{Performance of 95\% one-sided and two-sided confidence intervals with data set sampled from the predictive distribution of a Pitman-Yor process and sketched with the CMS-CU. The results are shown as a function of the Pitman-Yor process parameter $\sigma$. The two dashed curves for the two-sided intervals are nearly indistinguishable from one another. Other details are as in Figure~\ref{fig:exp-pyp-marginal}.}
\label{fig:exp-pyp-marginal-two-sided}
\end{center}
\end{figure}

\begin{figure}[!htb]
\begin{center}
\includegraphics[width=0.75\linewidth]{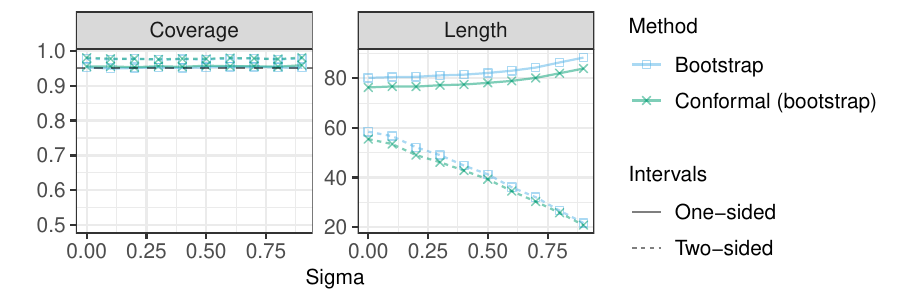}
\caption{Performance of 95\% one-sided and two-sided confidence intervals with data set sampled from the predictive distribution of a Pitman-Yor process and sketched with the vanilla CMS. The results are shown as a function of the Pitman-Yor process parameter $\sigma$. Other details are as in Figure~\ref{fig:exp-pyp-marginal-two-sided}.}
\label{fig:exp-pyp-marginal-two-sided-cms}
\end{center}
\end{figure}

\clearpage
\FloatBarrier
\subsection{Illustration on SARS-CoV-2 DNA data}

\begin{figure}[!htb]
\begin{center}
\includegraphics[width=0.75\linewidth]{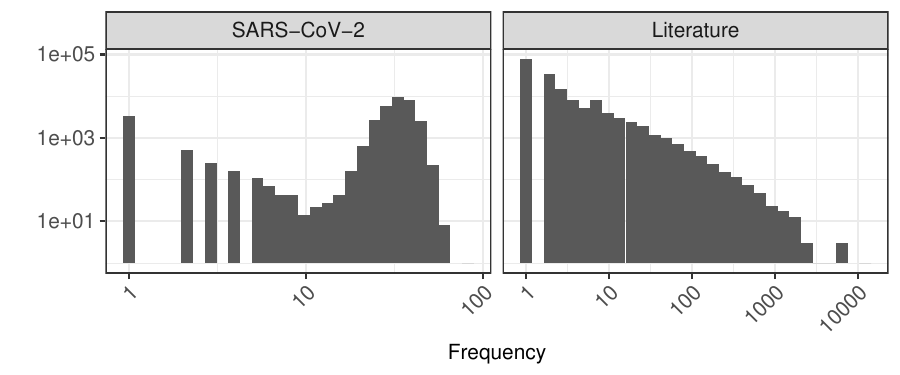} \\[-1em]
\caption{True frequency distribution of unique objects in two empirical data sets. Left: sequenced SARS-CoV-2 DNA 16-mers. Right: English 2-grams in a corpus of classic English literature.}
\label{fig:exp-data-freq}
\end{center}
\end{figure}

\begin{figure}[!htb]
\begin{center}
\includegraphics[width=\linewidth]{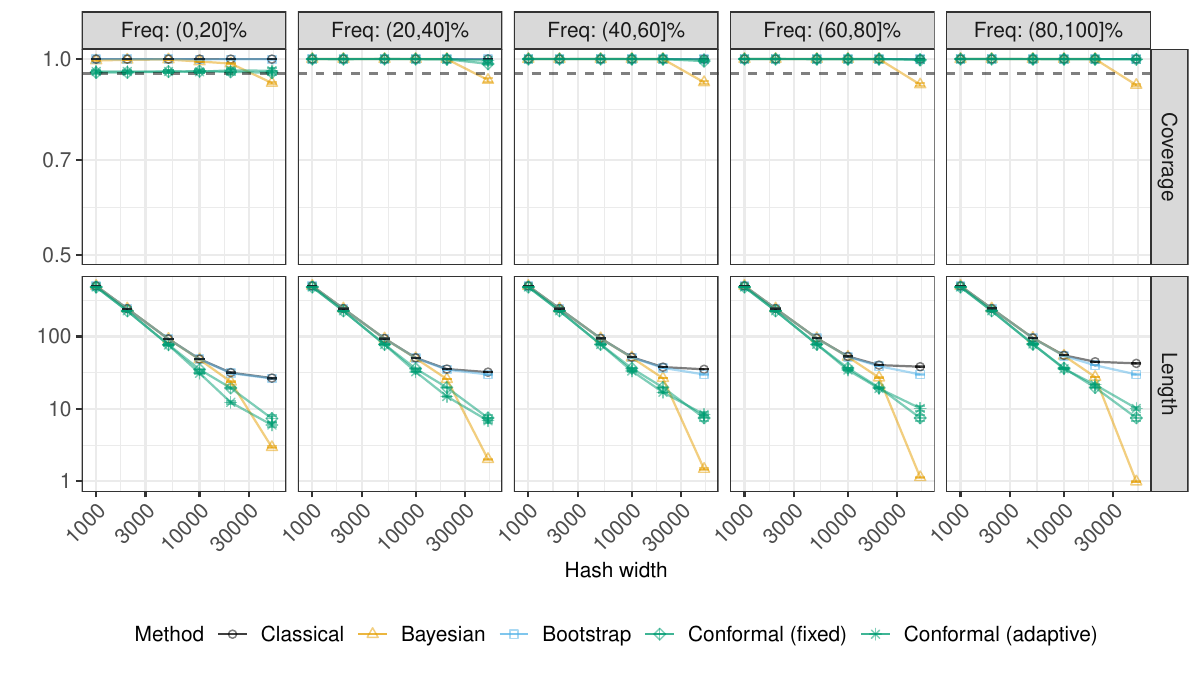} \\[-1em]
\caption{Performance of 95\% confidence intervals for random queries on SARS-CoV-2 sequence data sketched with the CMS-CU. The results are shown as a function of the hash width and stratified by the quintile of the true query frequency. Other details are as in Figure~\ref{fig:exp-covid-marginal}.}
\label{fig:exp-covid-conditional}
\end{center}
\end{figure}

\begin{figure}[!htb]
\begin{center}
\includegraphics[width=\linewidth]{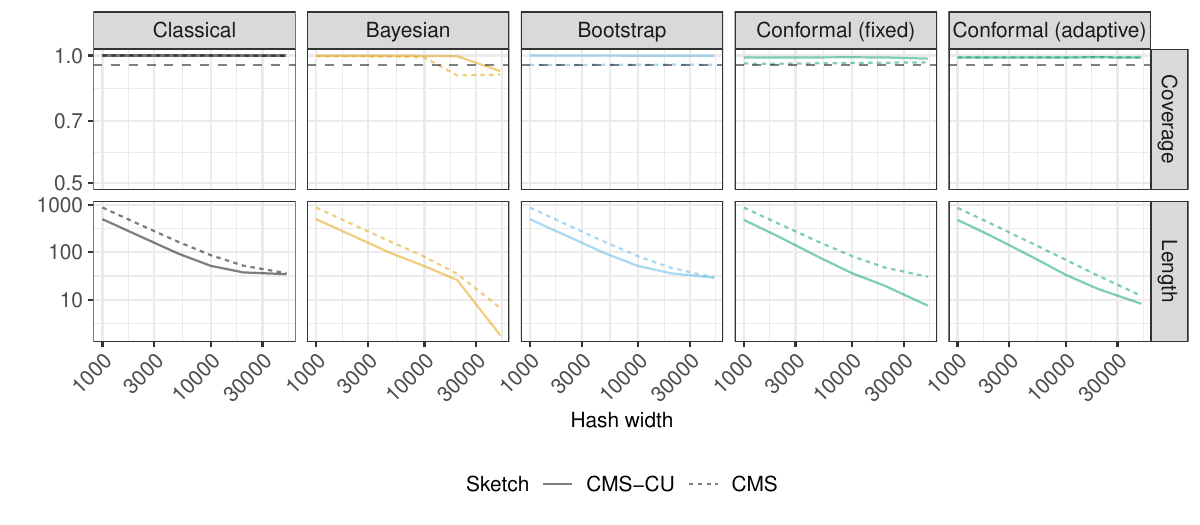} \\[-1em]
\caption{Performance of 95\% confidence intervals for random queries on SARS-CoV-2 sequence data. The data are sketched with either the vanilla CMS or the CMS-CU.
The results are shown as a function of the hash width. Other details are as in Figure~\ref{fig:exp-covid-marginal}.}
\label{fig:exp-covid-sketch}
\end{center}
\end{figure}

\begin{figure}[!htb]
\begin{center}
\includegraphics[width=0.8\linewidth]{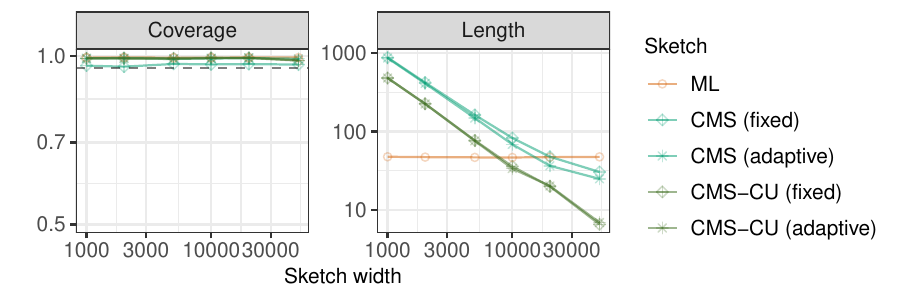} \\[-1em]
\caption{Performance of 95\% confidence intervals computed by Algorithm~\ref{alg:conformal-sketch} using different data sketches on SARS-CoV-2 sequence data, as a function of the sketch width. The results are shown as a function of the hash width. Other details are as in Figure~\ref{fig:exp-covid-marginal}.}
\label{fig:exp-covid-opt}
\end{center}
\end{figure}

\begin{figure}[!htb]
\begin{center}
\includegraphics[width=0.75\linewidth]{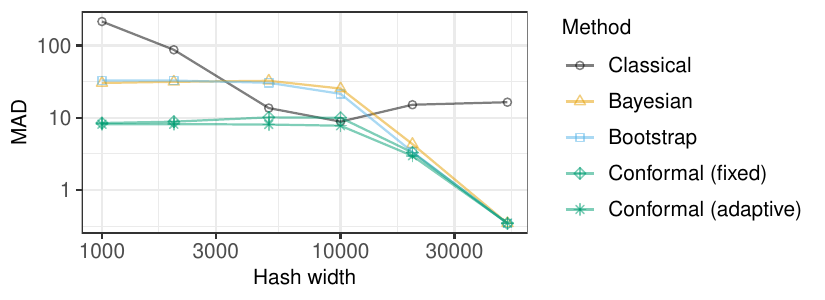} \\[-1em]
\caption{Median absolute deviation of point estimates for random queries on SARS-CoV-2 sequence data sketched with the CMS-CU. The results are shown as a function of the hash width. Other details are as in Figure~\ref{fig:exp-covid-marginal}.}
\label{fig:exp-covid-estimation}
\end{center}
\end{figure}

\clearpage
\FloatBarrier
\subsection{Illustration on  English literature data}

\begin{figure}[!htb]
\begin{center}
\includegraphics[width=0.75\linewidth]{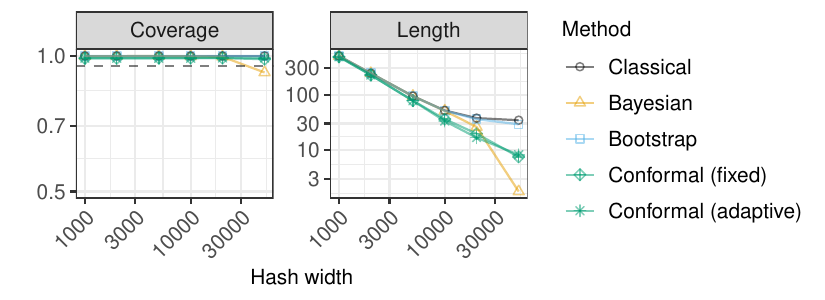} \\[-1em]
\caption{Performance of 95\% confidence intervals for random queries, on a sketched data set of 2-grams in classic English literature, keeping only distinct queries. The coverage is the empirical proportion of distinct queries whose frequency is covered by the output confidence intervals. The conformal confidence intervals are computed by applying Algorithm~\ref{alg:conformal-sketch-frequency} with $L=5$ frequency bins. The data are sketched with the CMS-CU.
Other details are as in Figure~\ref{fig:exp-covid-marginal}.}
\label{fig:exp-covid-unique}
\end{center}
\end{figure}

\begin{figure}[!htb]
\begin{center}
\includegraphics[width=0.75\linewidth]{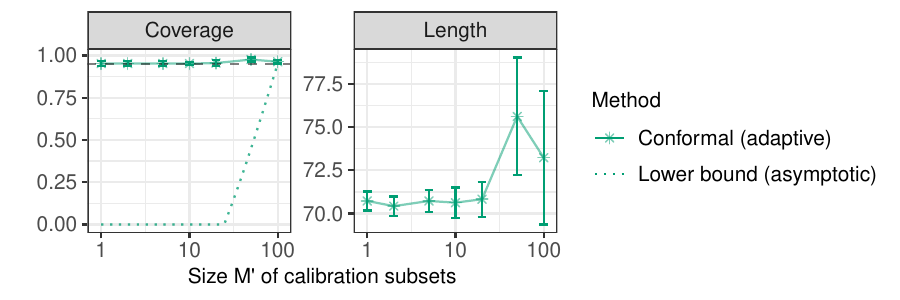}
\caption{Performance on sketched SARS-CoV-2 data of confidence intervals for distinct queries in a test set of size $M=100$, as a function of the parameter $M'$ of Algorithm~\ref{alg:conformal-sketch-unique} for constructing conformal confidence intervals satisfying~\eqref{eq:conf-int-unique}. The hash width is $w=5000$. Other details are as in Figure~\ref{fig:exp-covid-marginal}.}
\label{fig:exp-covid-unique-M}
\end{center}
\end{figure}

\begin{figure}[!htb]
\begin{center}
\includegraphics[width=0.75\linewidth]{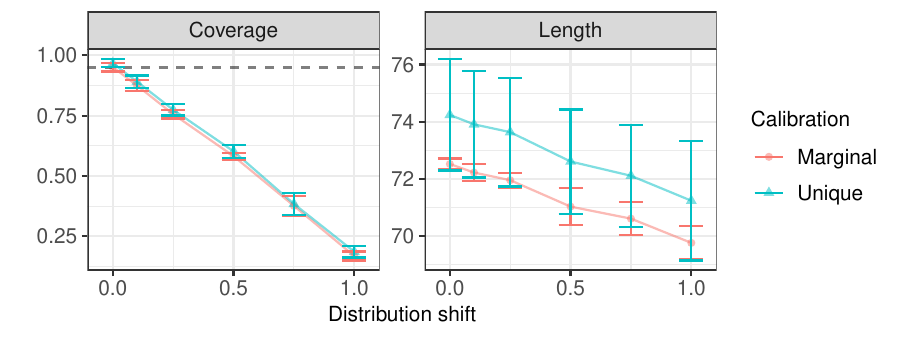}
\caption{Performance on sketched SARS-CoV-2 data of conformal confidence intervals with marginal (Algorithm~\ref{alg:conformal-sketch}) or distinct-query (Algorithm~\ref{alg:conformal-sketch-unique}) coverage in a test set of size 100 with varying degrees of distribution shift. Other details are as in Figure~\ref{fig:exp-covid-marginal}.}
\label{fig:exp-covid-unique-M-shift}
\end{center}
\end{figure}

\begin{figure}[!htb]
\begin{center}
\includegraphics[width=\linewidth]{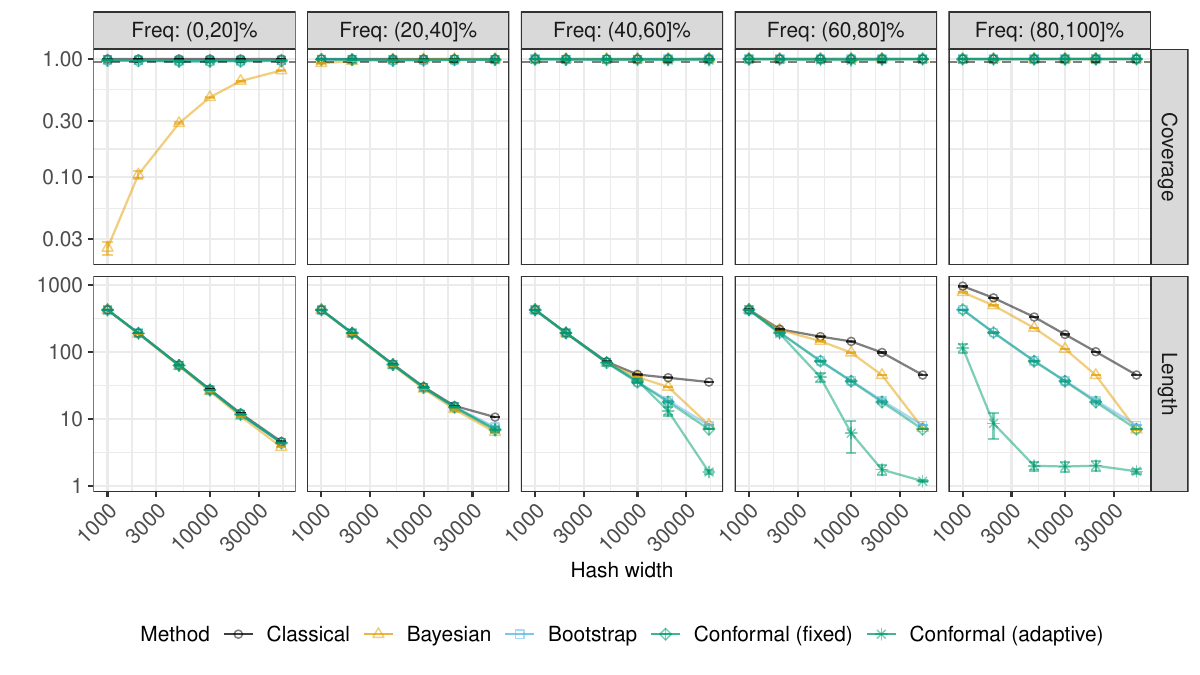} \\[-1em]
\caption{Performance of 95\% confidence intervals for random queries on a data set of 2-grams in classic English literature, sketched with the CMS-CU. The results are are shown as a function of the hash width and stratified by the quintile of the true query frequency. Other details are as in Figure~\ref{fig:exp-words-marginal}.}
\label{fig:exp-words-conditional}
\end{center}
\end{figure}

\begin{figure}[!htb]
\begin{center}
\includegraphics[width=\linewidth]{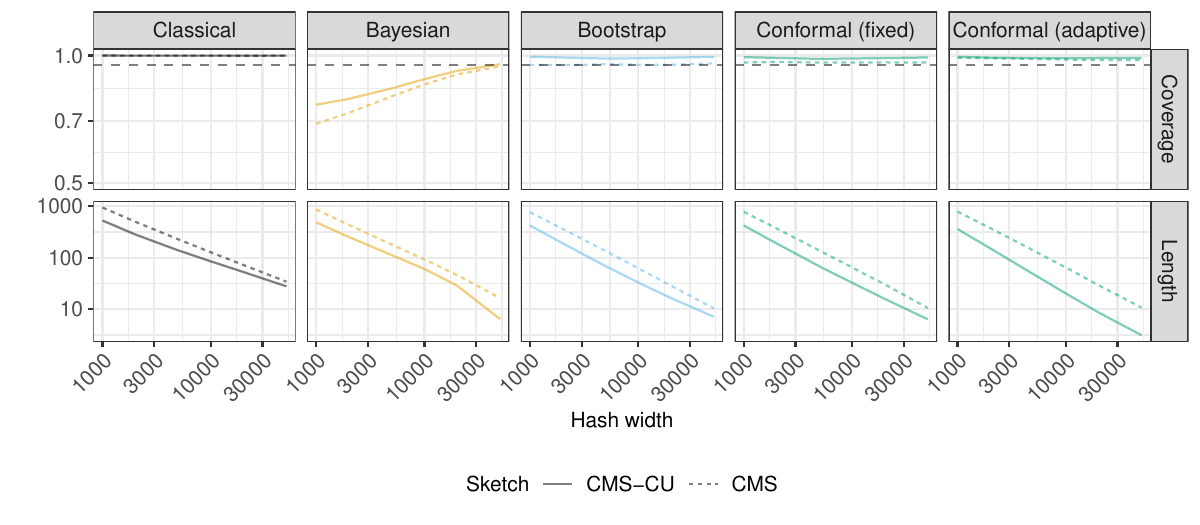} \\[-1em]
\caption{Performance of 95\% confidence intervals for random queries on a data set of 2-grams in classic English literature. The data are sketched with either the vanilla CMS or the CMS-CU. The results are shown as a function of the hash width.
Other details are as in Figure~\ref{fig:exp-words-marginal}.}
\label{fig:exp-words-sketch}
\end{center}
\end{figure}

\begin{figure}[!htb]
\begin{center}
\includegraphics[width=0.8\linewidth]{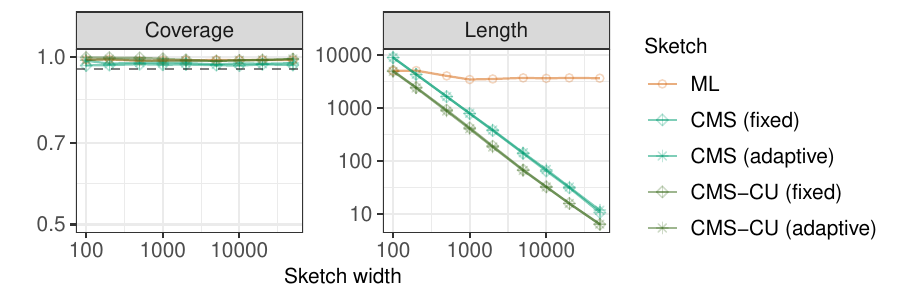} \\[-1em]
\caption{Performance of 95\% confidence intervals computed by Algorithm~\ref{alg:conformal-sketch} using different data sketches on a data set of 2-grams in classic English literature, as a function of the sketch width. The results are shown as a function of the hash width. Other details are as in Figure~\ref{fig:exp-words-marginal}.}
\label{fig:exp-words-opt}
\end{center}
\end{figure}

\begin{figure}[!htb]
\begin{center}
\includegraphics[width=0.75\linewidth]{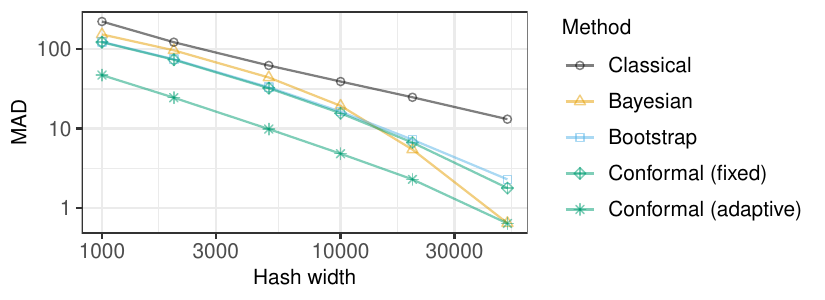} \\[-1em]
\caption{Median absolute deviation of point estimates for random queries on a data set of 2-grams in classic English literature, sketched with the CMS-CU. The results are shown as a function of the hash width. Other details are as in Figure~\ref{fig:exp-words-marginal}.}
\label{fig:exp-words-estimation}
\end{center}
\end{figure}

\begin{figure}[!htb]
\begin{center}
\includegraphics[width=0.75\linewidth]{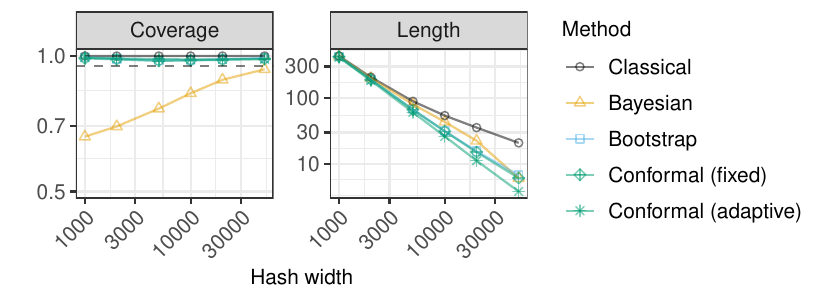} \\[-1em]
\caption{Performance of 95\% confidence intervals for random queries, on a sketched data set of 2-grams in classic English literature, keeping only distinct queries. The coverage is the empirical proportion of distinct queries whose frequency is covered by the output confidence intervals. The conformal confidence intervals are computed by applying Algorithm~\ref{alg:conformal-sketch-frequency} with $L=5$ frequency bins.
Other details are as in Figure~\ref{fig:exp-words-marginal}.}
\label{fig:exp-words-unique}
\end{center}
\end{figure}

\begin{figure}[!htb]
\begin{center}
\includegraphics[width=0.75\linewidth]{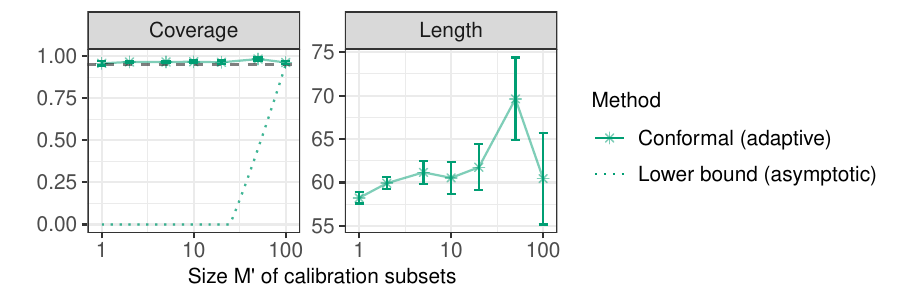}
\caption{Performance on sketched English literature data of confidence intervals for distinct queries in a test set of size $M=100$, as a function of the parameter $M'$ of Algorithm~\ref{alg:conformal-sketch-unique} for constructing conformal confidence intervals satisfying~\eqref{eq:conf-int-unique}. The hash width is $w=5000$. Other details are as in Figure~\ref{fig:exp-words-marginal}.}
\label{fig:exp-words-unique-M}
\end{center}
\end{figure}

\begin{figure}[!htb]
\begin{center}
\includegraphics[width=0.75\linewidth]{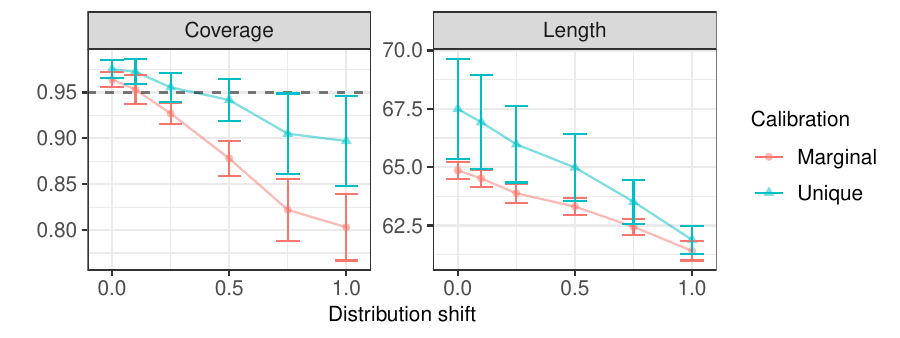}
\caption{Performance on sketched English literature data of conformal confidence intervals with marginal (Algorithm~\ref{alg:conformal-sketch}) or distinct-query (Algorithm~\ref{alg:conformal-sketch-unique}) coverage in a test set of size 100 with varying degrees of distribution shift. Other details are as in Figure~\ref{fig:exp-words-marginal}.}
\label{fig:exp-words-unique-M-shift}
\end{center}
\end{figure}

\begin{table}[!htb]
\caption{True frequencies, deterministic upper bounds, and 95\% lower bounds for 10 common (top) and 10 rare (bottom) random queries in two sketched data sets. Sketching with CMS-CU with $w=50,000$. Lower bounds written in green are below the true frequency; those in red are above. For each query, the highest lowest bound below the true frequency is highlighted in bold.} \label{tab:data-w50000}
{

\scalebox{0.78}{
\begin{tabular}[t]{lrrlllll}
\toprule
\multicolumn{3}{c}{ } & \multicolumn{5}{c}{95\% Lower bound} \\
\cmidrule(l{3pt}r{3pt}){4-8}
\multicolumn{6}{c}{ } & \multicolumn{2}{c}{Conformal} \\
\cmidrule(l{3pt}r{3pt}){7-8}
Data & Frequency & Upper bound & Classical & Bayesian & Bootstrap & Fixed & Adaptive\\
\midrule
\addlinespace[0.3em]
\multicolumn{8}{l}{\textbf{SARS-CoV-2}}\\
\hspace{1em}AATTATTATAAGAAAG & 81 & 81 & \textcolor{darkgreen}{26} & \textbf{\textcolor{darkgreen}{81}} & \textcolor{darkgreen}{52} & \textcolor{darkgreen}{50} & \textcolor{darkgreen}{36}\\
\hspace{1em}TCAGACAACTACTATT & 76 & 76 & \textcolor{darkgreen}{21} & \textbf{\textcolor{darkgreen}{55}} & \textcolor{darkgreen}{47} & \textcolor{darkgreen}{45} & \textcolor{darkgreen}{32}\\
\hspace{1em}AAAGTTGATGGTGTTG & 73 & 73 & \textcolor{darkgreen}{18} & \textbf{\textcolor{darkgreen}{59}} & \textcolor{darkgreen}{44} & \textcolor{darkgreen}{42} & \textcolor{darkgreen}{31}\\
\hspace{1em}CAATTATTATAAGAAA & 63 & 63 & \textcolor{darkgreen}{8} & \textbf{\textcolor{darkgreen}{48}} & \textcolor{darkgreen}{34} & \textcolor{darkgreen}{32} & \textcolor{darkgreen}{26}\\
\hspace{1em}ATCAGACAACTACTAT & 60 & 60 & \textcolor{darkgreen}{5} & \textbf{\textcolor{darkgreen}{44}} & \textcolor{darkgreen}{31} & \textcolor{darkgreen}{29} & \textcolor{darkgreen}{26}\\
\hspace{1em}ACCTTTGACAATCTTA & 55 & 55 & \textcolor{darkgreen}{0} & \textbf{\textcolor{darkgreen}{52}} & \textcolor{darkgreen}{26} & \textcolor{darkgreen}{24} & \textcolor{darkgreen}{27}\\
\hspace{1em}ATTTGAAGTCACCTAA & 55 & 55 & \textcolor{darkgreen}{0} & \textbf{\textcolor{darkgreen}{55}} & \textcolor{darkgreen}{26} & \textcolor{darkgreen}{24} & \textcolor{darkgreen}{27}\\
\hspace{1em}CATGCAAATTACATAT & 54 & 54 & \textcolor{darkgreen}{0} & \textbf{\textcolor{darkgreen}{54}} & \textcolor{darkgreen}{25} & \textcolor{darkgreen}{23} & \textcolor{darkgreen}{26}\\
\hspace{1em}GAATTTCACAGTATTC & 54 & 54 & \textcolor{darkgreen}{0} & \textbf{\textcolor{darkgreen}{54}} & \textcolor{darkgreen}{25} & \textcolor{darkgreen}{23} & \textcolor{darkgreen}{27}\\
\hspace{1em}TTTGTAGAAAACCCAG & 53 & 53 & \textcolor{darkgreen}{0} & \textbf{\textcolor{darkgreen}{53}} & \textcolor{darkgreen}{24} & \textcolor{darkgreen}{22} & \textcolor{darkgreen}{27}\\[0.5em]
\hspace{1em}AGTTGCAGAGTGGTTT & 24 & 24 & \textcolor{darkgreen}{0} & \textcolor{darkgreen}{13} & \textcolor{darkgreen}{0} & \textcolor{darkgreen}{0} & \textbf{\textcolor{darkgreen}{20}}\\
\hspace{1em}TCTTCACAATTGGAAC & 24 & 24 & \textcolor{darkgreen}{0} & \textcolor{darkgreen}{12} & \textcolor{darkgreen}{0} & \textcolor{darkgreen}{1} & \textbf{\textcolor{darkgreen}{20}}\\
\hspace{1em}TTCTGCTCGCATAGTG & 24 & 24 & \textcolor{darkgreen}{0} & \textcolor{darkgreen}{12} & \textcolor{darkgreen}{0} & \textcolor{darkgreen}{0} & \textbf{\textcolor{darkgreen}{20}}\\
\hspace{1em}CTACTTTAGATTCGAA & 23 & 23 & \textcolor{darkgreen}{0} & \textcolor{darkgreen}{11} & \textcolor{darkgreen}{0} & \textcolor{darkgreen}{0} & \textbf{\textcolor{darkgreen}{19}}\\
\hspace{1em}GCTGGTGTCTCTATCT & 23 & 23 & \textcolor{darkgreen}{0} & \textbf{\textcolor{darkgreen}{23}} & \textcolor{darkgreen}{0} & \textcolor{darkgreen}{1} & \textcolor{darkgreen}{19}\\
\hspace{1em}TTCTAAGAAGCCTCGG & 23 & 24 & \textcolor{darkgreen}{0} & \textcolor{darkgreen}{14} & \textcolor{darkgreen}{0} & \textcolor{darkgreen}{0} & \textbf{\textcolor{darkgreen}{20}}\\
\hspace{1em}GGGCTGTTGTTCTTGT & 22 & 24 & \textcolor{darkgreen}{0} & \textcolor{darkgreen}{12} & \textcolor{darkgreen}{0} & \textcolor{darkgreen}{0} & \textbf{\textcolor{darkgreen}{20}}\\
\hspace{1em}ACGTTCGTGTTGTTTT & 20 & 20 & \textcolor{darkgreen}{0} & \textbf{\textcolor{darkgreen}{20}} & \textcolor{darkgreen}{0} & \textcolor{darkgreen}{0} & \textcolor{darkgreen}{16}\\
\hspace{1em}GAAGTCTTTGAATGTG & 20 & 20 & \textcolor{darkgreen}{0} & \textbf{\textcolor{darkgreen}{20}} & \textcolor{darkgreen}{0} & \textcolor{darkgreen}{0} & \textcolor{darkgreen}{16}\\
\hspace{1em}CAAACCTGGTAATTTT & 3 & 3 & \textcolor{darkgreen}{0} & \textbf{\textcolor{darkgreen}{3}} & \textcolor{darkgreen}{0} & \textcolor{darkgreen}{0} & \textcolor{darkgreen}{0}\\
\addlinespace[0.3em]
\multicolumn{8}{l}{\textbf{Literature}}\\
\hspace{1em}of the & 12565 & 12568 & \textcolor{darkgreen}{12513} & \textcolor{darkgreen}{12544} & \textcolor{darkgreen}{12557} & \textcolor{darkgreen}{12556} & \textbf{\textcolor{darkgreen}{12562}}\\
\hspace{1em}in the & 6188 & 6190 & \textcolor{darkgreen}{6135} & \textcolor{darkgreen}{6169} & \textcolor{darkgreen}{6179} & \textcolor{darkgreen}{6179} & \textbf{\textcolor{darkgreen}{6180}}\\
\hspace{1em}and the & 6173 & 6175 & \textcolor{darkgreen}{6120} & \textcolor{darkgreen}{6151} & \textcolor{darkgreen}{6164} & \textcolor{darkgreen}{6164} & \textbf{\textcolor{darkgreen}{6165}}\\
\hspace{1em}the of & 6015 & 6017 & \textcolor{darkgreen}{5962} & \textcolor{darkgreen}{5990} & \textcolor{darkgreen}{6006} & \textcolor{darkgreen}{6006} & \textbf{\textcolor{darkgreen}{6007}}\\
\hspace{1em}the lord & 4186 & 4195 & \textcolor{darkgreen}{4140} & \textcolor{darkgreen}{4165} & \textcolor{darkgreen}{4184} & \textcolor{darkgreen}{4184} & \textcolor{darkgreen}{4184}\\
\hspace{1em}to the & 3465 & 3467 & \textcolor{darkgreen}{3412} & \textcolor{darkgreen}{3445} & \textcolor{darkgreen}{3456} & \textcolor{darkgreen}{3456} & \textbf{\textcolor{darkgreen}{3463}}\\
\hspace{1em}the and & 2250 & 2251 & \textcolor{darkgreen}{2196} & \textcolor{darkgreen}{2227} & \textcolor{darkgreen}{2240} & \textcolor{darkgreen}{2240} & \textbf{\textcolor{darkgreen}{2248}}\\
\hspace{1em}all the & 2226 & 2230 & \textcolor{darkgreen}{2175} & \textcolor{darkgreen}{2207} & \textcolor{darkgreen}{2219} & \textcolor{darkgreen}{2219} & \textbf{\textcolor{darkgreen}{2224}}\\
\hspace{1em}and he & 2169 & 2173 & \textcolor{darkgreen}{2118} & \textcolor{darkgreen}{2153} & \textcolor{darkgreen}{2162} & \textcolor{darkgreen}{2162} & \textbf{\textcolor{darkgreen}{2167}}\\
\hspace{1em}to be & 2062 & 2064 & \textcolor{darkgreen}{2009} & \textcolor{darkgreen}{2043} & \textcolor{darkgreen}{2053} & \textcolor{darkgreen}{2053} & \textbf{\textcolor{darkgreen}{2060}}\\[0.5em]
\hspace{1em}man on & 22 & 29 & \textcolor{darkgreen}{0} & \textcolor{darkgreen}{10} & \textcolor{darkgreen}{18} & \textcolor{darkgreen}{18} & \textcolor{darkgreen}{18}\\
\hspace{1em}their hand & 22 & 24 & \textcolor{darkgreen}{0} & \textcolor{darkgreen}{9} & \textcolor{darkgreen}{13} & \textcolor{darkgreen}{13} & \textcolor{darkgreen}{0}\\
\hspace{1em}no need & 20 & 28 & \textcolor{darkgreen}{0} & \textcolor{darkgreen}{9} & \textcolor{darkgreen}{17} & \textcolor{darkgreen}{17} & \textcolor{darkgreen}{16}\\
\hspace{1em}and brother & 12 & 14 & \textcolor{darkgreen}{0} & \textcolor{darkgreen}{2} & \textcolor{darkgreen}{3} & \textcolor{darkgreen}{3} & \textcolor{darkgreen}{0}\\
\hspace{1em}miss would & 10 & 13 & \textcolor{darkgreen}{0} & \textbf{\textcolor{darkgreen}{3}} & \textcolor{darkgreen}{2} & \textcolor{darkgreen}{2} & \textcolor{darkgreen}{0}\\
\hspace{1em}i please & 8 & 12 & \textcolor{darkgreen}{0} & \textbf{\textcolor{darkgreen}{3}} & \textcolor{darkgreen}{1} & \textcolor{darkgreen}{1} & \textcolor{darkgreen}{1}\\
\hspace{1em}also how & 3 & 13 & \textcolor{darkgreen}{0} & \textcolor{darkgreen}{2} & \textcolor{darkgreen}{2} & \textcolor{darkgreen}{2} & \textcolor{darkgreen}{0}\\
\hspace{1em}in under & 3 & 9 & \textcolor{darkgreen}{0} & \textbf{\textcolor{darkgreen}{2}} & \textcolor{darkgreen}{0} & \textcolor{darkgreen}{0} & \textcolor{darkgreen}{0}\\
\hspace{1em}ten old & 3 & 6 & \textcolor{darkgreen}{0} & \textbf{\textcolor{darkgreen}{1}} & \textcolor{darkgreen}{0} & \textcolor{darkgreen}{0} & \textcolor{darkgreen}{0}\\
\hspace{1em}fault he & 1 & 9 & \textcolor{darkgreen}{0} & \textbf{\textcolor{darkgreen}{1}} & \textcolor{darkgreen}{0} & \textcolor{darkgreen}{0} & \textcolor{darkgreen}{0}\\
\bottomrule
\end{tabular}
}

}
\end{table}

\begin{table}[!htb]

\caption{True frequencies, upper and lower bounds for 10 common (top) and 10 rare (bottom) random queries in two sketched data sets. Hash width $w=50,000$. Other details are as in Table~\ref{tab:data-w50000}.}  \label{tab:data-w5000}
{

\scalebox{0.78}{
\begin{tabular}[t]{lrrlllll}
\toprule
\multicolumn{3}{c}{ } & \multicolumn{5}{c}{95\% Lower bound} \\
\cmidrule(l{3pt}r{3pt}){4-8}
\multicolumn{6}{c}{ } & \multicolumn{2}{c}{Conformal} \\
\cmidrule(l{3pt}r{3pt}){7-8}
Data & Frequency & Upper bound & Classical & Bayesian & Bootstrap & Fixed & Adaptive\\
\midrule
\addlinespace[0.3em]
\multicolumn{8}{l}{\textbf{SARS-CoV-2}}\\
\hspace{1em}AATTATTATAAGAAAG & 81 & 209 & \textcolor{darkgreen}{0} & \textcolor{darkgreen}{4} & \textcolor{darkgreen}{0} & \textcolor{darkgreen}{0} & \textbf{\textcolor{darkgreen}{18}}\\
\hspace{1em}TCAGACAACTACTATT & 76 & 213 & \textcolor{darkgreen}{0} & \textcolor{darkgreen}{8} & \textcolor{darkgreen}{0} & \textcolor{darkgreen}{0} & \textbf{\textcolor{darkgreen}{18}}\\
\hspace{1em}AAAGTTGATGGTGTTG & 73 & 130 & \textcolor{darkgreen}{0} & \textcolor{darkgreen}{2} & \textcolor{darkgreen}{0} & \textcolor{darkgreen}{1} & \textbf{\textcolor{darkgreen}{18}}\\
\hspace{1em}CAATTATTATAAGAAA & 63 & 233 & \textcolor{darkgreen}{0} & \textcolor{darkgreen}{4} & \textcolor{darkgreen}{11} & \textcolor{darkgreen}{6} & \textbf{\textcolor{darkgreen}{19}}\\
\hspace{1em}ATCAGACAACTACTAT & 60 & 179 & \textcolor{darkgreen}{0} & \textcolor{darkgreen}{2} & \textcolor{darkgreen}{0} & \textcolor{darkgreen}{0} & \textbf{\textcolor{darkgreen}{18}}\\
\hspace{1em}ACCTTTGACAATCTTA & 55 & 292 & \textcolor{darkgreen}{0} & \textcolor{darkgreen}{15} & \textcolor{red}{70} & \textcolor{red}{67} & \textbf{\textcolor{darkgreen}{22}}\\
\hspace{1em}ATTTGAAGTCACCTAA & 55 & 258 & \textcolor{darkgreen}{0} & \textcolor{darkgreen}{11} & \textbf{\textcolor{darkgreen}{36}} & \textcolor{darkgreen}{31} & \textcolor{darkgreen}{20}\\
\hspace{1em}CATGCAAATTACATAT & 54 & 204 & \textcolor{darkgreen}{0} & \textcolor{darkgreen}{3} & \textcolor{darkgreen}{0} & \textcolor{darkgreen}{0} & \textbf{\textcolor{darkgreen}{18}}\\
\hspace{1em}GAATTTCACAGTATTC & 54 & 260 & \textcolor{darkgreen}{0} & \textcolor{darkgreen}{12} & \textbf{\textcolor{darkgreen}{38}} & \textcolor{darkgreen}{35} & \textcolor{darkgreen}{22}\\
\hspace{1em}TTTGTAGAAAACCCAG & 53 & 246 & \textcolor{darkgreen}{0} & \textcolor{darkgreen}{7} & \textbf{\textcolor{darkgreen}{24}} & \textcolor{darkgreen}{18} & \textcolor{darkgreen}{20}\\[0.5em]
\hspace{1em}ATGCTGCAATCGTGCT & 24 & 139 & \textcolor{darkgreen}{0} & \textcolor{darkgreen}{2} & \textcolor{darkgreen}{0} & \textcolor{darkgreen}{0} & \textbf{\textcolor{darkgreen}{17}}\\
\hspace{1em}ATTTCCTAATATTACA & 24 & 92 & \textcolor{darkgreen}{0} & \textcolor{darkgreen}{1} & \textcolor{darkgreen}{0} & \textcolor{darkgreen}{0} & \textbf{\textcolor{darkgreen}{17}}\\
\hspace{1em}CTCTATCATTATTGGT & 24 & 121 & \textcolor{darkgreen}{0} & \textcolor{darkgreen}{1} & \textcolor{darkgreen}{0} & \textcolor{darkgreen}{0} & \textbf{\textcolor{darkgreen}{17}}\\
\hspace{1em}TGTTTTATTCTCTACA & 24 & 199 & \textcolor{darkgreen}{0} & \textcolor{darkgreen}{3} & \textcolor{darkgreen}{0} & \textcolor{darkgreen}{1} & \textbf{\textcolor{darkgreen}{19}}\\
\hspace{1em}CAGTACATCGATATCG & 23 & 119 & \textcolor{darkgreen}{0} & \textcolor{darkgreen}{2} & \textcolor{darkgreen}{0} & \textcolor{darkgreen}{0} & \textbf{\textcolor{darkgreen}{17}}\\
\hspace{1em}TAATGGTGACTTTTTG & 23 & 92 & \textcolor{darkgreen}{0} & \textcolor{darkgreen}{1} & \textcolor{darkgreen}{0} & \textcolor{darkgreen}{0} & \textbf{\textcolor{darkgreen}{17}}\\
\hspace{1em}CAACCATAAAACCAGT & 22 & 105 & \textcolor{darkgreen}{0} & \textcolor{darkgreen}{1} & \textcolor{darkgreen}{0} & \textcolor{darkgreen}{0} & \textbf{\textcolor{darkgreen}{17}}\\
\hspace{1em}AGTTATTTGACTCCTG & 21 & 97 & \textcolor{darkgreen}{0} & \textcolor{darkgreen}{1} & \textcolor{darkgreen}{0} & \textcolor{darkgreen}{1} & \textbf{\textcolor{darkgreen}{18}}\\
\hspace{1em}ATAAAGGAGTTGCACC & 19 & 218 & \textcolor{darkgreen}{0} & \textcolor{darkgreen}{5} & \textcolor{darkgreen}{0} & \textcolor{darkgreen}{0} & \textbf{\textcolor{darkgreen}{18}}\\
\addlinespace[0.3em]
\multicolumn{8}{l}{\textbf{Literature}}\\
\hspace{1em}of the & 12565 & 12630 & \textcolor{darkgreen}{12086} & \textcolor{darkgreen}{12325} & \textcolor{darkgreen}{12463} & \textcolor{darkgreen}{12454} & \textbf{\textcolor{darkgreen}{12563}}\\
\hspace{1em}in the & 6188 & 6242 & \textcolor{darkgreen}{5698} & \textcolor{darkgreen}{5906} & \textcolor{darkgreen}{6075} & \textcolor{darkgreen}{6067} & \textbf{\textcolor{darkgreen}{6096}}\\
\hspace{1em}and the & 6173 & 6314 & \textcolor{darkgreen}{5770} & \textcolor{darkgreen}{5972} & \textcolor{darkgreen}{6147} & \textcolor{darkgreen}{6139} & \textbf{\textcolor{darkgreen}{6169}}\\
\hspace{1em}the of & 6015 & 6162 & \textcolor{darkgreen}{5618} & \textcolor{darkgreen}{5834} & \textcolor{darkgreen}{5995} & \textcolor{darkgreen}{5985} & \textbf{\textcolor{darkgreen}{6014}}\\
\hspace{1em}the lord & 4186 & 4289 & \textcolor{darkgreen}{3745} & \textcolor{darkgreen}{3975} & \textcolor{darkgreen}{4122} & \textcolor{darkgreen}{4114} & \textbf{\textcolor{darkgreen}{4185}}\\
\hspace{1em}to the & 3465 & 3558 & \textcolor{darkgreen}{3014} & \textcolor{darkgreen}{3217} & \textcolor{darkgreen}{3391} & \textcolor{darkgreen}{3380} & \textbf{\textcolor{darkgreen}{3464}}\\
\hspace{1em}the and & 2250 & 2413 & \textcolor{darkgreen}{1869} & \textcolor{darkgreen}{2081} & \textcolor{darkgreen}{2246} & \textcolor{darkgreen}{2237} & \textbf{\textcolor{darkgreen}{2249}}\\
\hspace{1em}all the & 2226 & 2346 & \textcolor{darkgreen}{1802} & \textcolor{darkgreen}{1993} & \textcolor{darkgreen}{2179} & \textcolor{darkgreen}{2170} & \textbf{\textcolor{darkgreen}{2225}}\\
\hspace{1em}and he & 2169 & 2293 & \textcolor{darkgreen}{1749} & \textcolor{darkgreen}{1937} & \textcolor{darkgreen}{2126} & \textcolor{darkgreen}{2117} & \textbf{\textcolor{darkgreen}{2168}}\\
\hspace{1em}to be & 2062 & 2121 & \textcolor{darkgreen}{1577} & \textcolor{darkgreen}{1770} & \textcolor{darkgreen}{1954} & \textcolor{darkgreen}{1945} & \textbf{\textcolor{darkgreen}{2061}}\\[0.5em]
\hspace{1em}very for & 15 & 59 & \textcolor{darkgreen}{0} & \textbf{\textcolor{darkgreen}{2}} & \textcolor{darkgreen}{0} & \textcolor{darkgreen}{0} & \textcolor{darkgreen}{0}\\
\hspace{1em}and faithful & 14 & 94 & \textcolor{darkgreen}{0} & \textbf{\textcolor{darkgreen}{3}} & \textcolor{darkgreen}{0} & \textcolor{darkgreen}{0} & \textcolor{darkgreen}{0}\\
\hspace{1em}but found & 9 & 74 & \textcolor{darkgreen}{0} & \textbf{\textcolor{darkgreen}{2}} & \textcolor{darkgreen}{0} & \textcolor{darkgreen}{0} & \textcolor{darkgreen}{0}\\
\hspace{1em}my speech & 6 & 98 & \textcolor{darkgreen}{0} & \textbf{\textcolor{darkgreen}{3}} & \textcolor{darkgreen}{0} & \textcolor{darkgreen}{0} & \textcolor{darkgreen}{0}\\
\hspace{1em}of eight & 5 & 66 & \textcolor{darkgreen}{0} & \textbf{\textcolor{darkgreen}{2}} & \textcolor{darkgreen}{0} & \textcolor{darkgreen}{0} & \textcolor{darkgreen}{0}\\
\hspace{1em}and soul & 4 & 140 & \textcolor{darkgreen}{0} & \textcolor{red}{6} & \textcolor{darkgreen}{0} & \textcolor{darkgreen}{0} & \textcolor{darkgreen}{0}\\
\hspace{1em}her prow & 3 & 79 & \textcolor{darkgreen}{0} & \textbf{\textcolor{darkgreen}{2}} & \textcolor{darkgreen}{0} & \textcolor{darkgreen}{0} & \textcolor{darkgreen}{0}\\
\hspace{1em}usual as & 2 & 56 & \textcolor{darkgreen}{0} & \textbf{\textcolor{darkgreen}{2}} & \textcolor{darkgreen}{0} & \textcolor{darkgreen}{0} & \textcolor{darkgreen}{0}\\
\hspace{1em}a invitation & 1 & 80 & \textcolor{darkgreen}{0} & \textcolor{red}{2} & \textcolor{darkgreen}{0} & \textcolor{darkgreen}{0} & \textcolor{darkgreen}{0}\\
\hspace{1em}angular log & 0 & 146 & \textcolor{darkgreen}{0} & \textcolor{red}{5} & \textcolor{darkgreen}{0} & \textcolor{darkgreen}{0} & \textcolor{darkgreen}{0}\\
\bottomrule
\end{tabular}
}

}
\end{table}

\FloatBarrier

\section{Additional experiments with two-sided confidence intervals} \label{app:exp-two-sided}

This section describes additional experiments with synthetic data similar to those described in Figures~\ref{fig:exp-zipf-marginal} (Zipf distribution) and~\ref{fig:exp-pyp-marginal} (Pitman-Yor process prior), constructing two-sided instead of one-sided confidence intervals.
For simplicity, we focus on one-sided 95\% conformalized bootstrap confidence intervals based on the simpler Bonferroni approach described in Appendix~\ref{app:two-sided-bonferroni}. The performance of these intervals
is compared to those of one and two-sided standard bootstrap confidence intervals obtained with the method of~\cite{ting2018count}.

Figure~\ref{fig:exp-zipf-marginal-two-sided} reports on results based on data generated from a Zipf distribution and sketched with the CMS-CU, similarly to Figure~\ref{fig:exp-zipf-marginal}.
Here, all methods achieve the desired 95\% marginal coverage level, but the conformal confidence intervals are shorter when the Zipf tail parameter $a$ is larger and hash collisions become rarer, consistently with Figure~\ref{fig:exp-zipf-marginal}.
It is interesting to note that the two-sided conformal confidence intervals are much narrower than their one-sided counterparts when $a$ is small and hash collisions are very common, but this is not true if $a$ is large.
The latter is likely a limitation of the specific construction we have adopted, described in Appendix~\ref{app:two-sided-bonferroni}, which may be too conservative in some cases due to the Bonferroni correction.  A suitable implementation of the more sophisticated conditional histogram~\citep{sesia2021conformal} approach described in Appendix~\ref{app:two-sided-chr} should be expected to produce two-sided intervals that are always narrower than their one-sided counterparts.
Figure~\ref{fig:exp-zipf-marginal-two-sided-cms} reports on results similar to those in Figure~\ref{fig:exp-zipf-marginal-two-sided}, with the only difference that now the data are sketched with the vanilla CMS instead of the CMS-CU.

Figure~\ref{fig:exp-pyp-marginal-two-sided} reports on results based on data generated from a Pitman-Yor process prior and sketched with the CMS-CU, similarly to Figure~\ref{fig:exp-pyp-marginal}.
Here, all methods achieve the desired 95\% marginal coverage level, and two-sided intervals are generally much shorter than their one-sided counterparts. Across all values of $\sigma$, the conformal confidence intervals tend to be shorter than the bootstrap intervals, although this difference becomes very small in the case of two-sided intervals for large values of $\sigma$.
Finally, Figure~\ref{fig:exp-pyp-marginal-two-sided-cms} reports on results similar to those in Figure~\ref{fig:exp-pyp-marginal-two-sided}, with the only difference that now the data are sketched with the vanilla CMS instead of the CMS-CU.

%%% Local Variables:
%%% mode: latex
%%% TeX-master: "cms_jmlr"
%%% End:

\end{document}